\documentclass[twocolumn,epjc3-upd,fleqn]{svjour3}

\RequirePackage[T1]{fontenc}

\smartqed  % flush right qed marks, e.g. at end of proof

\RequirePackage{graphicx,amsmath}
\RequirePackage{mathptmx}      % use Times fonts if available on your TeX system
\RequirePackage{flushend}
\RequirePackage[numbers,sort&compress]{natbib}
\RequirePackage[colorlinks,citecolor=blue,urlcolor=blue,linkcolor=blue]{hyperref}
\RequirePackage{lineno}
\usepackage{doi}
\usepackage[]{units}
\usepackage{xspace}
\usepackage{pbox}
\usepackage{bm}
\usepackage{morefloats}
\usepackage{stfloats}

\journalname{Eur. Phys. J. C}

\newcommand{\LFM}{\ensuremath{\rm Log(\mu_{p})}}
\newcommand{\Puno}{\ensuremath{\rm P_1}}
\newcommand{\PSVar}{\ensuremath{\rm PSV}}
\newcommand{\PSVarcut}{\ensuremath{\rm PSV_{cut}}}
\newcommand{\Na}{$^{22}$Na}

\newcommand{\K}{$^{40}$K}

\newcommand{\Pb}{$^{210}$Pb}

\begin{document}
%\linenumbers

\title{Performance of ANAIS-112 experiment after the first year of data taking} 

\author{J.~Amar{\'e}\thanksref{addr1,addr2},
S.~Cebri{\'an}\thanksref{addr1,addr2},
I.~Coarasa\thanksref{addr1,addr2},
C.~Cuesta\thanksref{addr1,addr4},
E.~Garc\'{\i}a\thanksref{addr1,addr2},
M.~Mart\'{\i}nez\thanksref{e1,addr1,addr2,addr3},
M.A. Oliv{\'a}n\thanksref{addr1,addr5},
Y.~Ortigoza\thanksref{addr1,addr2},
A.~Ortiz~de~Sol{\'o}rzano\thanksref{addr1,addr2},
J.~Puimed{\'o}n\thanksref{addr1,addr2},
A.~Salinas\thanksref{addr1,addr2},
M.L.~Sarsa\thanksref{e1,addr1,addr2}, 
P.~Villar\thanksref{addr1,addr2}
\and
J.A.~Villar\thanksref{e2,addr1,addr2} 
}

\thankstext{e1}{Corresponding authors: mariam@unizar.es, mlsarsa@unizar.es}
\thankstext{e2}{Deceased}
\institute{Laboratorio de F\'{\i}sica Nuclear y Astropart\'{\i}culas, Universidad de Zaragoza, C/~Pedro Cerbuna 12, 50009 Zaragoza, SPAIN \label{addr1}
           \and
Laboratorio Subterr{\'a}neo de Canfranc, Paseo de los Ayerbe s.n., 22880 Canfranc Estaci{\'o}n, Huesca, SPAIN \label{addr2}
           \and
Fundaci{\'o}n Agencia Aragonesa para la Investigaci{\'o}n y el Desarrollo (ARAID), Gobierno de Arag\'on, Avenida de Ranillas 1-D,
50018 Zaragoza, Spain\label{addr3}
           \and
\emph{Present Address: Centro de Investigaciones Energ\'eticas, Medioambientales y Tecnol\'ogicas, CIEMAT, 28040, Madrid, SPAIN} \label{addr4}
           \and
\emph{Present Address: Fundaci{\'o}n CIRCE, 50018, Zaragoza, SPAIN} \label{addr5}
}

\date{Updated on \today}
%\date{Received: date / Accepted: date}
% The correct dates will be entered by the editor

\maketitle

\begin{abstract}
ANAIS is a direct detection dark matter experiment aiming at the study of the annual modulation expected in the interaction rate. It uses same target and technique as the DAMA/LIBRA experiment, which reported a highly significant positive modulation compatible with that expected for dark matter particles distributed in the galactic halo. However, other very sensitive experiments do not find any hint of particles with the required properties, although comparison is model dependent. In 2017, ANAIS-112 experiment was installed at the Canfranc Underground Laboratory (LSC), in Spain, and after the commissioning run for calibration and general assessment, ANAIS-112 started data taking in dark matter mode on August 3$^{rd}$, 2017. It consists of nine NaI(Tl) modules, amounting 112.5 kg of mass in total. Here we report on the experimental apparatus and detector performance after the first year of data taking. Total live time available amounts to 341.72~days, being the corresponding exposure 105.32~kg~x~yr. ANAIS-112 has achieved an analysis energy threshold of 1~keVee and an average background in the region of interest, from 1 to 6~keVee, of 3.6~counts/keVee/kg/day after correcting by the event selection efficiencies. In these conditions, ANAIS-112 will be able to test the DAMA/LIBRA result at three sigma level in five years of data taking. 
\end{abstract}

\section{Introduction}
There is overwhelming evidence accumulating over the last decades supporting that a large fraction of the Universe energy-mass budget is not explained in the frame of the standard model of the particle physics if assuming the cosmological standard model~\cite{Bertone:2016nfn}. The solution to this puzzle should be searched for in any possible direction: looking for extensions of the particle physics standard model while working on new theoretical frameworks for gravity and/or dynamics. The complex nature of the dark matter (DM) and dark energy (DE) problems makes them one the main challenges for present particle physics and cosmology. DM can be understood as a new non-zero-mass particle having a very low interaction probability with baryonic matter. Axions and Weakly Interacting Massive Particles (WIMPs) are among the preferred candidates~\cite{Baudis:2016qwx}. 

Despite the wide experimental effort devoted to the understanding of the particle DM nature, no unambiguous claim has been reported by direct~\cite{Undagoitia:2015gya}, indirect~\cite{Gaskins:2016cha} or accelerator searches~\cite{Bagnaschi:2015eha} so far. However, there is a long-standing positive result on the field: the observation of a high statistically significant annual modulation in the detection rate, compatible with that expected for galactic halo dark matter particles, at the DAMA/LIBRA experiment~\cite{Bernabei:2008yi, Bernabei:2013xsa, Bernabei:2018yyw}, in operation at the Gran Sasso National Laboratory, in Italy. This result has neither been reproduced by any other experiment, nor ruled out in a model independent way, although compatibility in most conventional WIMP-DM scenarios is actually disclaimed~\cite{Savage:2009mk,Aprile:2015ade,Herrero-Garcia:2015kga,Baum:2018ekm,Kang:2018qvz,Herrero-Garcia:2018mky}. 

ANAIS stands for Annual modulation with NaI Scintillators, and has as main goal the testing of DAMA/LIBRA result 
using the same target and technique.
ANAIS-112 consists of 112.5~kg of NaI(Tl) detectors, and it was installed in 2017 at the Canfranc Underground Laboratory, LSC, in Spain. 
The dark matter run started on the 3$^{rd}$ of August, 2017, being now completed the first year of data taking. 
ANAIS-112 can test the DAMA/LIBRA result in a model independent way at 3 sigma C.L. in 5 years of data taking, 
having a significant discovery potential if the modulation signal is effectively due to dark matter 
particles~\cite{Coarasa:2017aol}. A different experiment, having different residual cosmic ray flux and 
different environmental conditions than DAMA/LIBRA has, could bring a new insight in the DAMA/LIBRA observation: 
confirmation of a modulation having the same phase and amplitude would be very difficult to 
explain as effect of backgrounds or systematics. 
In this sense, it is worth remarking here the main differences among ANAIS-112 and DAMA/LIBRA operation conditions. First, the LSC rock overburden is lower than that of Gran Sasso (800~m versus 1400~m rock), allowing for a larger expected contribution from cosmogenic related backgrounds in ANAIS-112 than in DAMA/LIBRA. On the other hand, ANAIS-112 experimental layout disposes of an active muon veto system that allows tagging and removing muon related events and monitoring the onsite muon flux. Second, the contribution from radiogenic related backgrounds is mainly independent on the laboratory overburden. Radon contribution is dependent on the local geological conditions (for production) and experimental spatial conditions, for instance ventilation of the spaces, while cosmogenically activated isotopes contribution is dependent on the exposure to cosmic rays and time spent underground. Third, PMT noise events would be monitored along the second year of operation with a blank module specifically commissioned for that goal, which is taking data since August 2018 at LSC. 

Other efforts sharing ANAIS goal in the international dark matter community are the COSINE-100 experiment, taking data also in dark matter mode at Yang-Yang Underground Laboratory, South Korea \cite{Kim:2018wcl, Adhikari:2017esn}, and in longer-term the SABRE project, aiming at installing twin detectors in Australia and Italy \cite{Tomei:2017rkg} and the COSINUS project, aiming at the R$\&$D on cryogenic detectors based on NaI \cite{Gutlein:2017iuw}. 

In this paper, we describe the most relevant experimental parameters of ANAIS-112 set-up and present the experiment 
performance along the first year of data taking. 
The structure of the paper is as follows: Section~\ref{sec:experimental} describes the ANAIS-112 experimental layout, 
including electronics and data analysis; Section~\ref{sec:calibration} focuses on the energy calibration of the experiment, 
both at low and high energy; Section~\ref{sec:eventSel} describes the procedures aiming at selecting sodium 
iodide bulk scintillation events; Section~\ref{sec:eff} summarizes the total efficiencies estimated for the 
detection and selection of events in the  Region of Interest (ROI) and Section~\ref{sec:low_energy_spectrum} briefly describes the 
ANAIS-112 background sources, referencing conveniently to other papers where this issue is more thoroughly studied. 
As far as stability of the environmental conditions is important for the implications of the annual modulation analysis 
intended, Section~\ref{sec:stability} describes the temporal variability of relevant experimental 
parameters along the first year of data taking. Finally, a brief summary and outlook are presented in Section~\ref{sec:summary}.

\section{The ANAIS-112 experiment}
\label{sec:experimental}

\begin{figure}[htbp]
\centering
\includegraphics[width=0.5\textwidth]{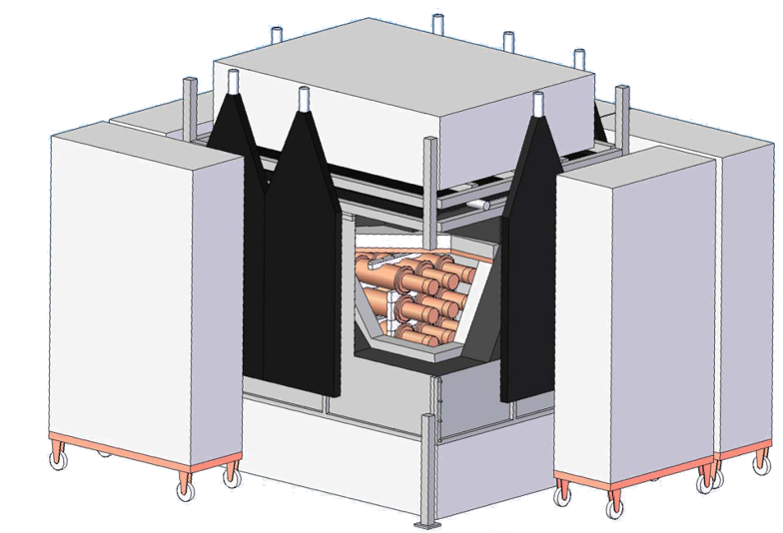}
\vspace{0.5cm}
\includegraphics[width=0.5\textwidth]{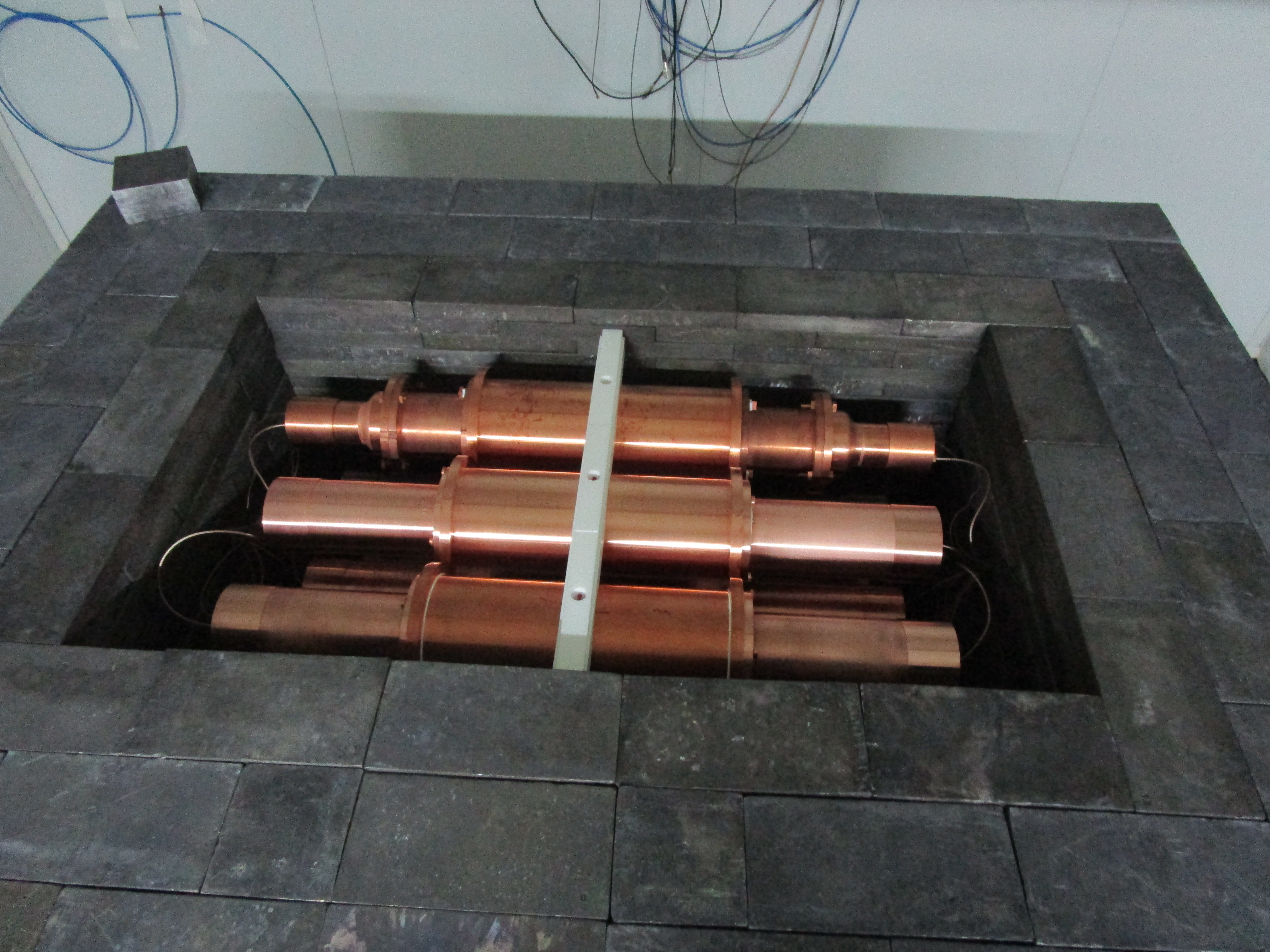}
\caption{Upper panel: schematic view of the ANAIS-112 full set-up; it consists of nine NaI modules (each crystal coupled to 2 PMT units) placed in a 3x3 array, a lead shielding (10 cm archaeological lead + 20 cm low activity lead), an anti-radon box continuously flushed with radon-free nitrogen gas, an active veto system (made of 16 plastic scintillators), and 40 cm of neutron shielding (water tanks and polyethylene blocks). Lower panel: picture taken while ANAIS-112 commissioning; detail on the calibration system for positioning the Cd sources in front of the Mylar window of each module can be also observed.
}
\label{fig:setup}
\end{figure}

ANAIS-112 experiment uses nine NaI(Tl) modules produced by Alpha Spectra Inc. (AS) in Colorado (US). These modules have been manufactured along several years and then shipped to Spain, avoiding air travel in order to prevent from cosmogenic activation, arriving at LSC the first of them at the end of 2012 and the last one by March, 2017. Each crystal is cylindrical (4.75" diameter and 11.75" length), with a mass of 12.5~kg. NaI(Tl) crystals were grown from selected ultrapure NaI powder and housed in OFE (Oxygen Free Electronic) copper; the encapsulation has a Mylar window allowing low energy
calibration using external gamma sources. Two highly efficient Hamamatsu R12669SEL2 photomultipliers (PMTs) were coupled through quartz windows to each crystal at LSC clean room. In the following we will refer as PMT-0 (PMT-1) to the units placed on the East (West) side of the shielding. 
All PMT units, as well as many of the materials used in the building of the detectors, have been screened for radiopurity using germanium detectors at the LSC low background facilities and their contribution to the experiment background has been estimated~\cite{Amare:2016rbf,Amare:2018ndh}. 
The ANAIS-112 shielding consists of 10~cm of archaeological lead, 20~cm of low activity lead, an anti-radon box (continuously flushed with radon-free nitrogen gas), 
an active muon veto system made up of 16 plastic scintillators designed to cover top and sides of the whole ANAIS set-up (see Figure~\ref{fig:setup})
and 40 cm of neutron moderator (a combination of water tanks and polyethylene blocks).
The hut housing the experiment is located at the hall B of LSC under 2450 m.w.e. rock overburden.

\subsection{ANAIS-112 DAQ System}
\label{sec:DAQ}
\begin{figure}[htbp]
\centering
\includegraphics[width=0.5\textwidth]{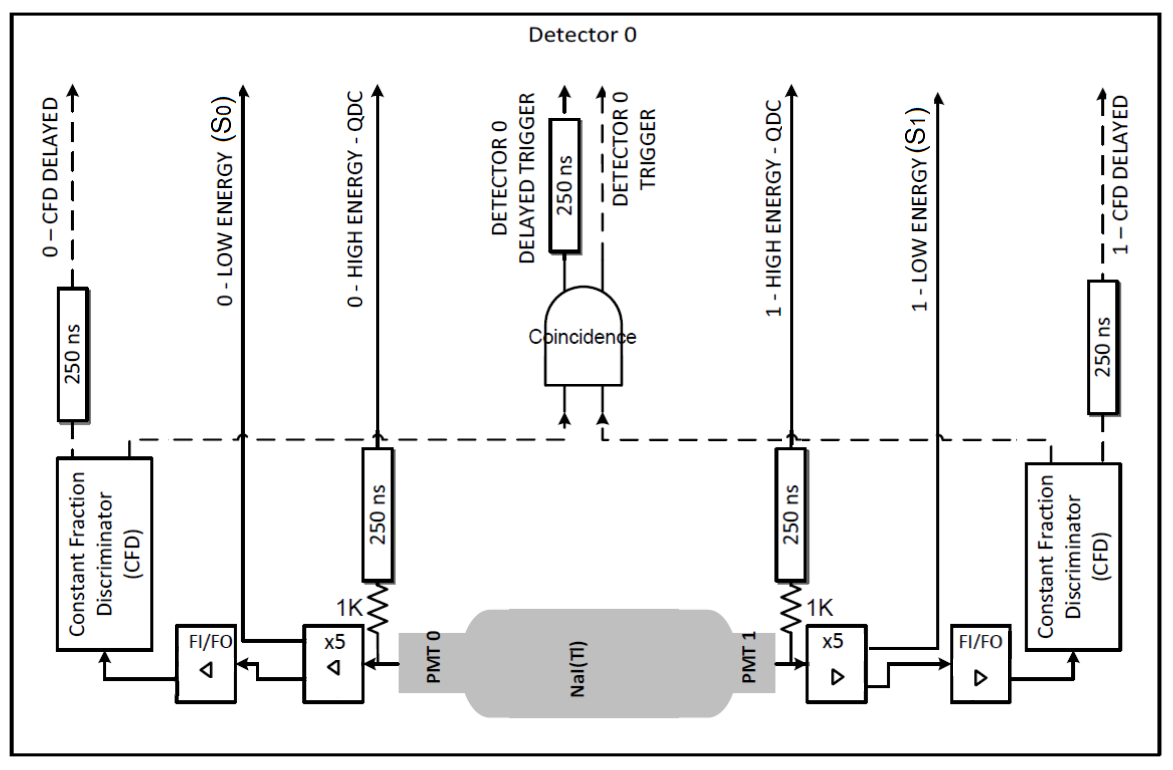}
\includegraphics[width=0.5\textwidth]{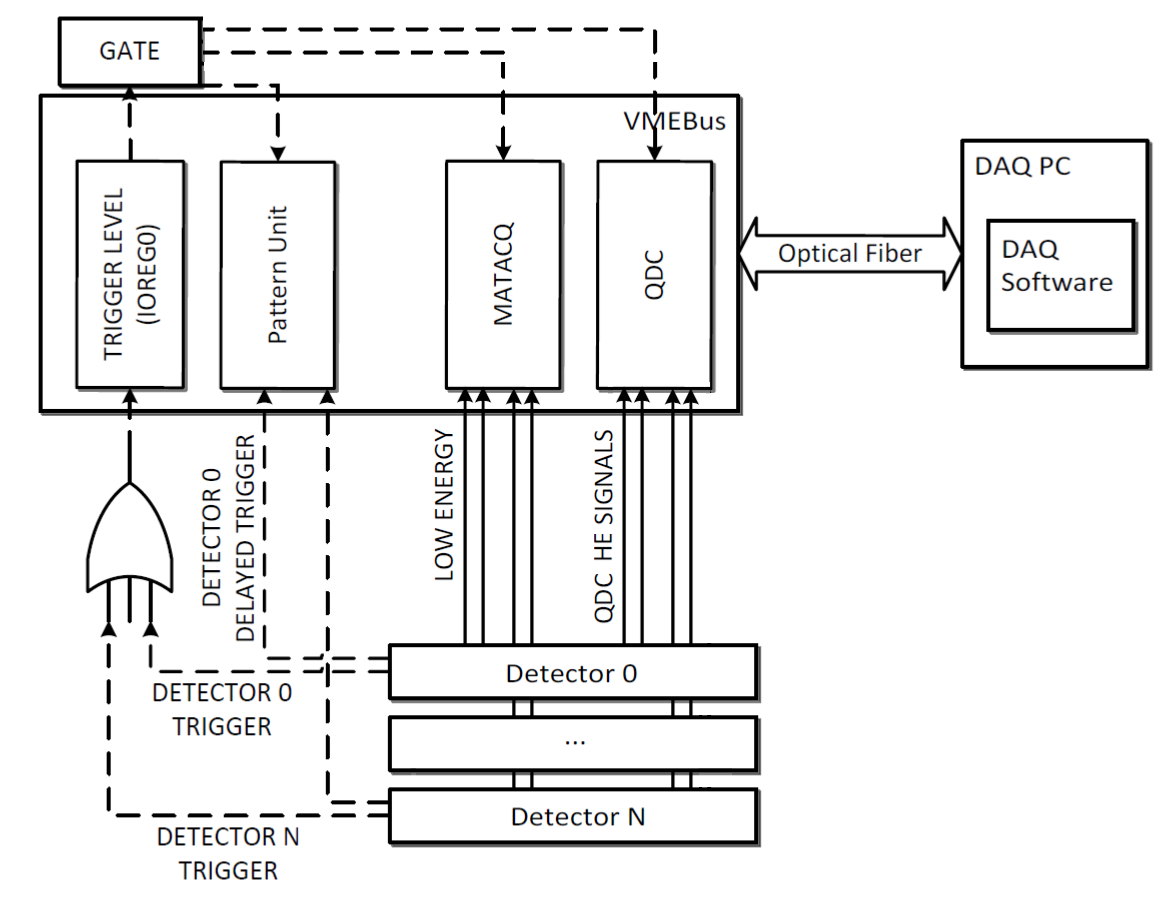}
\caption{Electronics front-end of the ANAIS-112 DAQ acquisition system for one single module (upper panel) and the complete system (lower panel).}
\label{fig:daq}
\end{figure}

The ANAIS-112 electronic chain and data acquisition system (DAQ) are designed to assure the lowest trigger level
in stable conditions over a long period of time. 
In Figure~\ref{fig:daq} we show a sketch of the main components of the electronic chain for a single module (upper panel)
and for the total system (lower panel). A more detailed description of the DAQ system can be found in \cite{MAThesis}.
\par
The first amplification stage for each PMT signal is carried out by custom-made pre-amplifiers located inside the ANAIS hut, just outside the anti-radon box, in order to minimize the length of the cables and then, optimize the signal/noise ratio. The rest of the ANAIS-112 electronics (mainly based on VME modules)
are located outside the hut, in an air-conditioned rack cabinet whose temperature is controlled and continuously monitored.
The PMT's high voltage (HV) supply is provided by a CAEN SY2527 Universal Multichannel Power Supply, 
which allows continuous monitoring of the individual voltages.
\par
The charge signal coming from every PMT is separately processed and recorded.
Each one is divided into: (1) a trigger signal; (2) a so-called low energy (LE) signal that goes to the digitizer; and 
(3) a high energy (HE) signal, conveniently attenuated for high energy events.
The trigger of each PMT signal is performed by a CAEN N843 Constant Fraction Discriminator (CFD), with a threshold level
low enough to effectively trigger at the photoelectron (phe) level,
while the single module trigger is done by the coincidence (logical AND) of the two PMT triggers in a 200~ns window. 
The HE signals are fed into CAEN V792 QDC (charge-to-digital-converter) modules, 
with 1~$\mu$s integration window and 12 bits resolution.
The LE signals (S$_0$ and S$_1$ in Figure~\ref{fig:daq}) are sent to MATACQ-CAEN V1729A digitizers, 
which continuously sample the waveforms at 2~Gs/s with high resolution (14 bits) and store them in circular buffers.

The main acquisition trigger, that is the logical OR of the nine modules individual trigger,  
is sent to a CAEN V977 I/O Register (IOREG0), whose state ON represents the DAQ {\it busy} state and 
is used for measuring the dead time, until it is reset via a VME command by the DAQ program.
It fires the QDC conversion and the 
transfer of 2520 points-length waveforms (including 500 pre-trigger points) from the MATACQ buffers to the DAQ computer memory. The I/O Register also opens 1 $\mu$s window via a gate generator (GATE) which defines the coincidence among modules. 
In addition, a Pattern Unit module registers which detectors have triggered and allows the DAQ system
to acquire only these data, reducing the dead time and saving disk space.
\par 
The building of the spectra is done by software (off-line) by integrating the S$_0$(t) and S$_1$(t) waveforms 
and adding them for each module: this variable will be called in the following {\it pulse area}. 
These waveforms also serve to perform Pulse Shape Analysis (PSA) in order to select bulk scintillation events 
in the NaI crystals from PMT-origin events (see Section~\ref{sec:eventSel})
and to distinguish alpha interactions from beta/gamma ones (Section~\ref{sec:hecal}).

The real time master clock is generated by the VME/PCI bridge, started at the
beginning of the run, and it is stored in a 64 bit scaler. 
Another VME/PCI bridge controller clock is used to measure the live time. This clock (also stored in a 64 bit scaler) is stopped
when the IOREG enters in {\it busy} state and it is started when the state is reset.
The dead time is dominated by the waveform digitization and transfer from the MATACQ board and amounts to about 5~ms per event. For the first year ANAIS-112 average trigger rate, about 5.5~Hz (see section~\ref{sec:performance}), this implies a 2.8\% dead time.

\subsection{ANAIS-112 muon veto system}
\label{sec:veto}

The role played by the veto scintillator system in ANAIS-112 data selection and annual modulation analysis is very important. It is well known that the underground muon flux has an annual modulation~\cite{Agostini:2018fnx, Agostini:2016gda, Cherwinka:2015hva, Ambrosio:2002db} which could cause muon related events in dark matter detectors to mimic the dark matter modulation. However, different phases from DAMA/LIBRA and underground muon flux modulations have been determined and DAMA/LIBRA collaboration completely discard muons as possible explanation of their signal~\cite{Bernabei:2012wp}. This is because direct effect of muon interactions in the NaI(Tl) detectors should contribute outside the ROI for the dark matter modulation analysis, at very high energies, for instance, and if present in DAMA/LIBRA ROI, they should also contribute both, to events in coincidence among several detectors and in other energy regions, which has not been observed by DAMA/LIBRA collaboration. However, contribution from other muon related events could fall into the ROI and correspond to single hit events: muons can produce secondary neutrons in the setup materials, either prompt or delayed neutrons in a wide energy range; muons can interact in the PMT, quartz optical windows, and other detectors components producing there weak scintillation signals~\cite{2014OptMa..36.1408A}; and last, because of the very slow scintillation observed in NaI(Tl)~\cite{2013OptMa..36..316C}, the tail of the scintillation pulse produced by a large energy deposition from a muon in one of the crystals could trigger several times the ANAIS-112 DAQ system (as explained in Section~\ref{sec:muonVeto}). The ANAIS-112 veto scintillator system allows to correlate muon events in the plastic scintillators with energy depositions in the NaI(Tl) crystal in order to better understand and monitor these effects. 

\begin{figure}[htbp]
\centering
\includegraphics[width=0.5\textwidth]{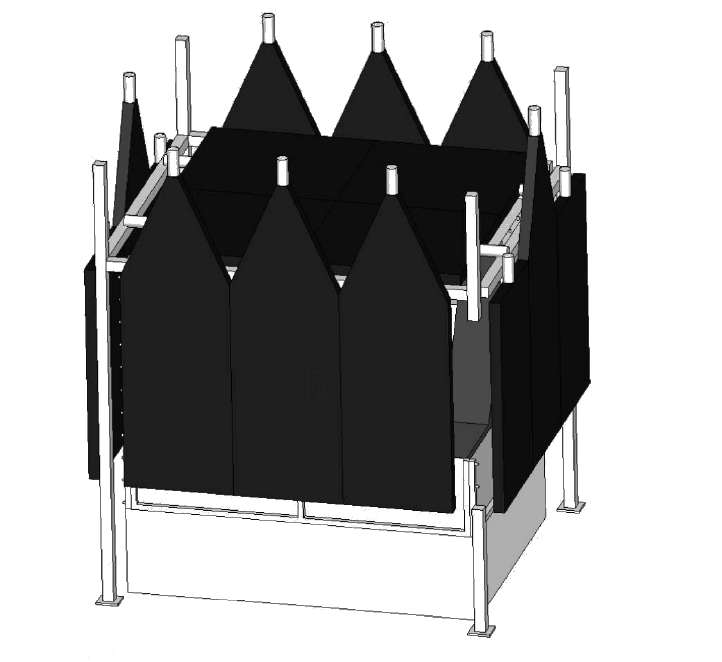}
\caption{Muon Veto System layout in ANAIS-112 setup. It consists of 16 plastic scintillators covering top and lateral sides of the shielding (see Figure~\ref{fig:setup}). It allows tagging muon related events in the NaI crystals by off-line analysis. Trigger rate of the veto system is 5.6~Hz before applying PSA techniques to select muon events from plastic scintillators background; the rate of muons finally identified is below 40~mHz.  
}
\label{fig:vetosystem}
\end{figure}

ANAIS-112 muon veto system consists of 16 plastic scintillators having different shapes which cover all but the bottom of the ANAIS-112 lead shielding (see Figure~\ref{fig:vetosystem}). A tagging strategy instead of a hardware vetoing strategy has been followed in the design of this system. This is because the goal of the muon veto system is to detect the residual muon flux at the detectors position in order to discard coincident events in the NaI(Tl) crystals, but also to study possible correlations between muon hits in the plastic scintillators and events in the NaI(Tl) crystals, specially in the ROI. In order to minimize losing muon events below plastic scintillator trigger threshold, the resulting trigger rate on the veto system was high, 5.6~Hz, being dominated by background events from the plastic scintillators. Because the plastic scintillators used in the veto system have a strong particle discrimination capability, background events are much faster than muon events, and they can be removed by Pulse Shape Analysis techniques, applied off-line. The muon detection front-end electronics were designed to provide enough information to allow this off-line particle discrimination: for each event the whole pulse area and tail pulse area are saved, being muon events discriminated by the ratio of these two parameters quite efficiently~\cite{MAThesis}. 
Muon selection parameters and operation conditions of the plastic scintillators have been chosen to minimize the rejection of real muon events.
Some other relevant information is also saved for every event: multiplicity trigger pattern and real time tagging. The DAQ software of the veto system is almost the same developed for the NaI(Tl) detectors DAQ, but it is independently run in a different computer and with a different configuration. It is launched remotely by the ANAIS-112 DAQ scripts to guarantee same start counting in clock scalers, having been the time correlation between both DAQ systems thoroughly studied while commissioning and testing, both at surface and underground LSC facilities~\cite{MAThesis}.  

The residual muon flux at the LSC has been recently determined to be 4.4x10$^{-3}$~$\mu$~s$^{-1}$~m$^{-2}$~\cite{Bandac:2017jjm}, which is over four orders of magnitude lower than the surface muon flux, but enough to have some influence in the ANAIS detectors. This flux is higher than that measured at Gran Sasso Laboratory, where the DAMA/LIBRA experiment is in operation by about one order of magnitude~\cite{Agostini:2018fnx}. The larger muon flux at LSC would produce a different (probably higher) modulation amplitude in ANAIS-112 data if muon interactions are somehow behind DAMA/LIBRA result and contribute to the ROI in an unsuspected way. 

Average muon rates observed along the first year of operation of ANAIS-112 are shown in Table~\ref{tab:muonrates} for each side of the experiment layout. Table~\ref{tab:muonratescoin} shows the average rate of coincidences in veto scintillators from two different sides. The latter is more cleary dependent on the shape of the rock overburden profile over the LSC, which is quite asymmetric, being lower on the North side. However, the angular dependence of the muon flux at LSC Hall B is not well known and then, muon flux estimation is difficult. Using only the top side information, without any correction, we get an average muon flux value of 5.6x10$^{-3}$~$\mu$~s$^{-1}$~m$^{-2}$.

\begin{table}
\caption{Average rate of muons (after PSA) in the five sides of ANAIS-112 Veto System and corresponding standard deviation of the population.}
\label{tab:muonrates}       
\centering 
\begin{tabular}{lll}
\hline\noalign{\smallskip}
Side & Average Muon rate & Standard deviation \\
     & $\mu$~s$^{-1}$ \\
\noalign{\smallskip}\hline\noalign{\smallskip}
Top &   0.01167  & 0.00055\\
North &  0.00766 & 0.00037\\
South &  0.00787  & 0.00041\\
East &   0.00598 & 0.00048\\
West &   0.00598 & 0.00035\\
\noalign{\smallskip}\hline
\end{tabular}
\end{table}

\begin{table}
\caption{Average rate of muons (after PSA) in coincidence among the different sides of ANAIS-112 Veto System and standard deviation of the population.}
\label{tab:muonratescoin}       
\centering 
\begin{tabular}{lll}
\hline\noalign{\smallskip}
Side - Side & Average Muon rate & Standard deviation\\
coincidence    & $\mu$~s$^{-1}$ \\
\noalign{\smallskip}\hline\noalign{\smallskip}
Top - North &   0.0037 & 0.0022\\
Top - South &  0.0014 & 0.0008\\
Top - East &  0.0009 & 0.0006\\
Top - West &   0.0011 & 0.0006\\
North - South &   0.0004 & 0.0002 \\
East - West &  0.00008  & 0.00005\\
\noalign{\smallskip}\hline
\end{tabular}
\end{table}

\subsection{ANAIS-112 data-taking}
\label{sec:dataTaking}
ANAIS-112 detector was assembled in early 2017 and the commissioning phase was performed from March to July.
The dark matter data taking started on the 3$^{rd}$ of August 2017. 
{\it Dark matter runs} last for about two weeks and in between we perform a $^{109}$Cd {\it calibration run}, that usually takes about 3-4 hours.
\par
\begin{figure*}[htbp]
\centering
\includegraphics[width=0.8\textwidth]{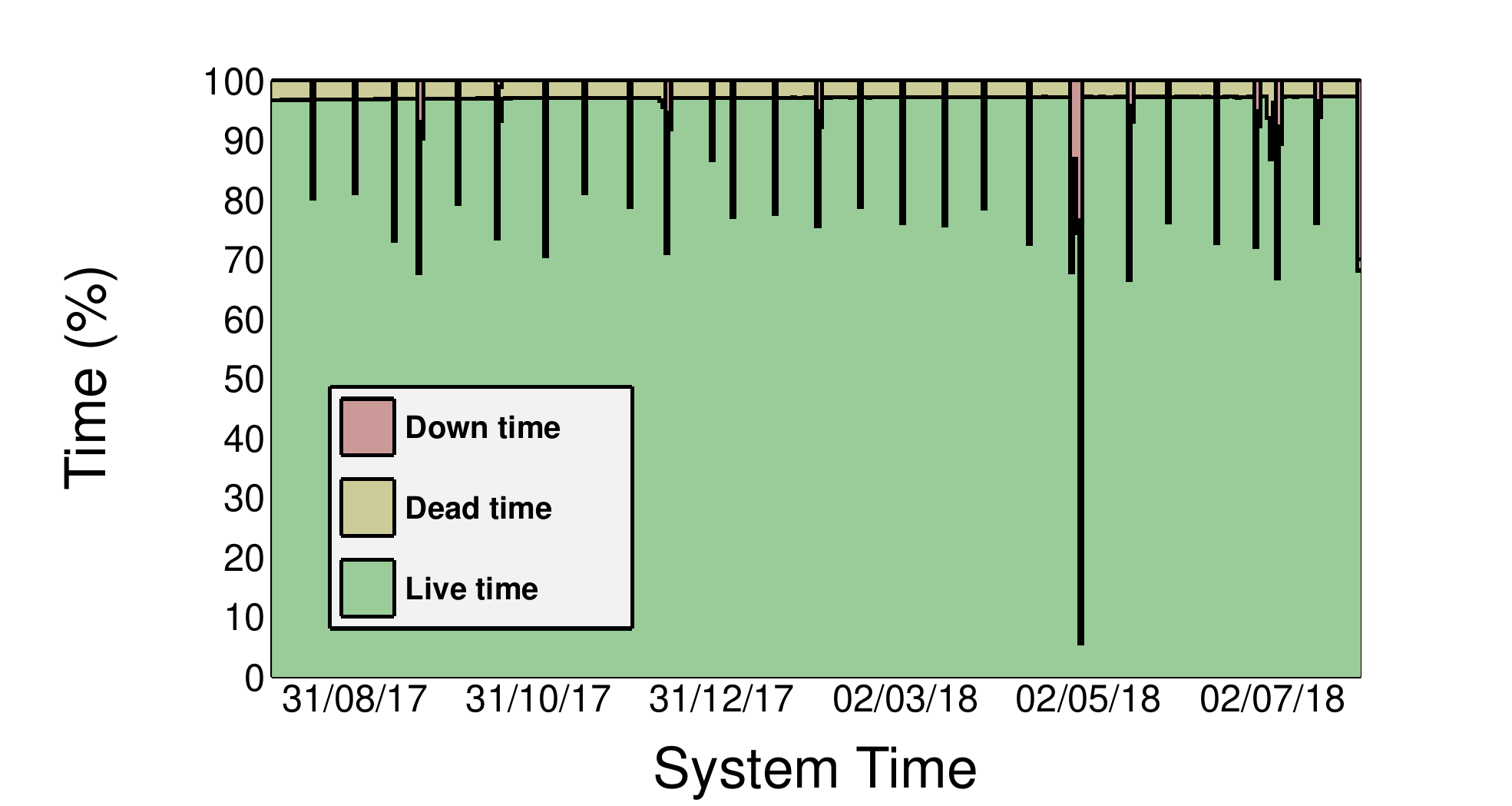}
\caption{Distribution of live, dead and down time during the first year of ANAIS-112 DM data taking. Total numbers for this first year are: 94.5\% live time, 2.9\% dead time, and 2.6\% down time. Down time periods correspond mostly to periodical calibration with low energy external sources, but some incidences are clearly observed: an electrical power failure at the LSC (end of April - beginning of May 2018) and a failure on ANAIS DAQ computer disk at the beginning of July 2018. Colors referenced are available in the online version of the paper.
}
\label{fig:duty}
\end{figure*}

One of the most important assets of ANAIS-112 in the search for the annual modulation is 
the very high duty cycle achieved. Figure~\ref{fig:duty} shows the fraction of live, dead and down time corresponding to the first year of the ANAIS-112 dark matter run. Total numbers are the following: 94.5\% live time, 2.9\% dead time, and 2.6\% down time. Excluding the down time, which is mostly due to periodical calibration of the modules using external gamma sources (seen as periodical spikes in Figure~\ref{fig:duty}), we get a live time of 97\% in DM run mode. These numbers are expected to improve for the second year of data taking, provided some small incidences affecting electrical supply of the LSC and maintenance works can be avoided. Some of those happening in this period can be clearly observed in Figure~\ref{fig:duty}: an electrical power failure at LSC (end April - beginning May 2018) and a failure on ANAIS DAQ computer disk (beginning of July 2018). From the 3$^{rd}$ of August 2017 to the 31$^{st}$ of July 2018, 341.72 days (live time) are available for the annual modulation analysis.
\par
Using this live time distribution (day by day) along the first year of 
data taking we can calculate \( \alpha\), the average of \( \cos^2(\omega t) \), 
obtaining a value of 0.46850, and \(\beta=0.00466\), the average of \( \cos(\omega t)\), 
assuming \( \omega = 2 \pi /365~d^{-1} \). These parameters should be 0.5 and 0.0, respectively, 
for a perfect coverage of the whole year-period.

\begin{figure*}[htbp]
\centering
\includegraphics[width=0.75\textwidth]{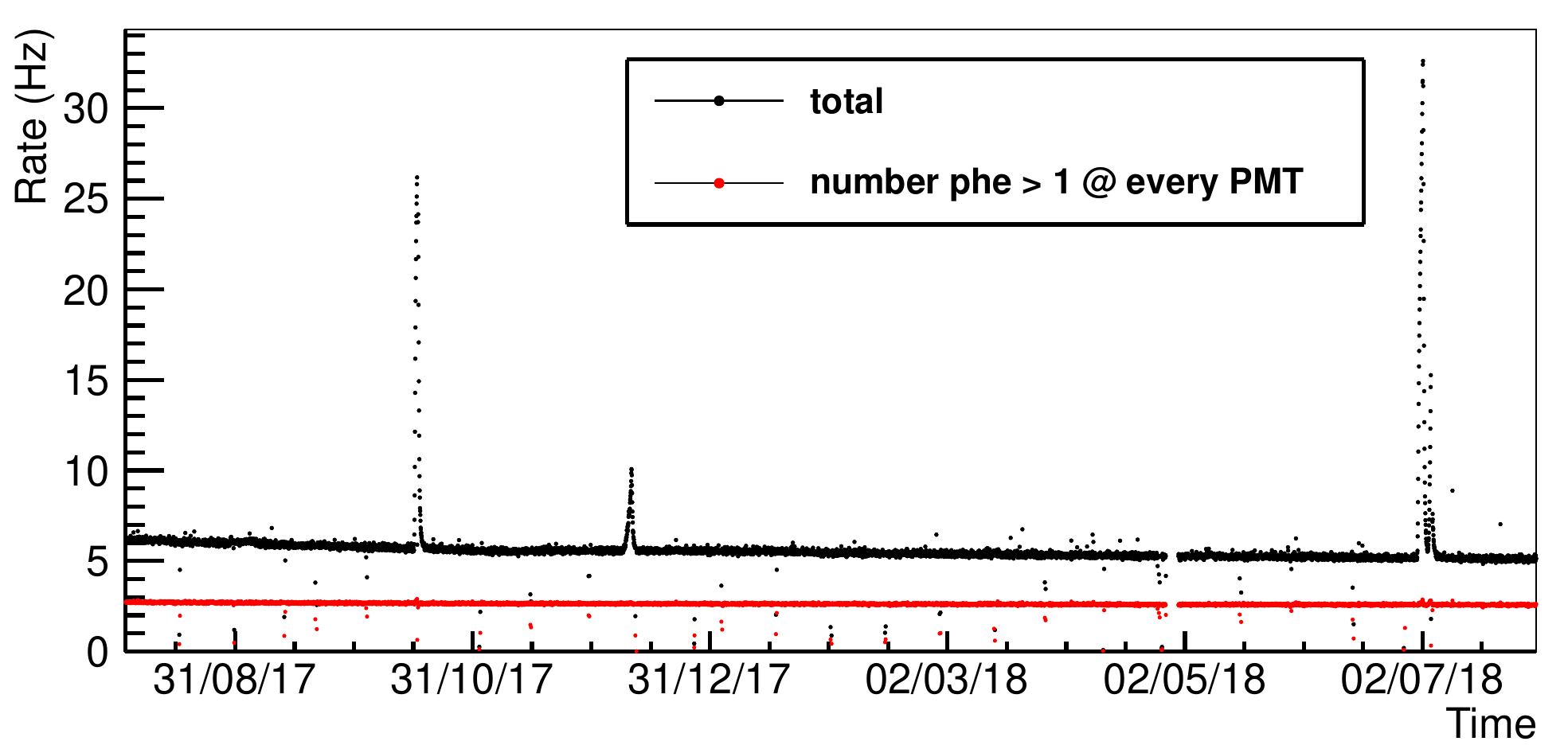}
\caption{Trigger rate (calculated in 3600~s time bins) during the first year of ANAIS-112 DM data taking: total trigger rate is shown in black, trigger rate of events having more than one phe in the pulse at each PMT is shown in red. Strong increases in the total trigger rate of the experiment are related with exhausting the liquid nitrogen dewar providing the radon free gas flux continuously coming into the shielding, but the increase is not observed in events having more than one phe in each PMT signal (red solid dots), which points to PMT related events still to be understood (see text for more information). Colors referenced are available in the online version of the paper.  
}
\label{fig:rate}
\end{figure*}
  
\subsection{ANAIS-112 modules performance}
\label{sec:performance}
Total trigger rate evolution along the first year of data 
(shown in black solid dots in Figure~\ref{fig:rate})
shows a slow but clear decrease. The liquid nitrogen supply was exhausted three times in this period, and the total trigger rate of the experiment has shown to be very sensitive to the complete exhaustion of nitrogen gas flux entering into the ANAIS-112 shielding (see Figure~\ref{fig:N2Flux_evol}). However, the rate increase in these periods is not directly explained by background events 
originating from gammas emitted by the Rn daughters: in Figure~\ref{fig:rate} it is shown as 
red solid dots the event rate after rejecting those events having only one phe in each PMT, 
which has been very stable along the whole ANAIS-112 first year data. We can conclude, most of the 
events triggering in the periods without nitrogen supply only have one phe in each PMT, 
which, by the way, usually dominate also the total trigger rate in ANAIS-112. These events are considered to be PMT-related events, but we are still working to understand their rate correlation with the nitrogen gas flux coming into the shielding using a blank module, specifically commissioned for the second year of data taking with this goal. These events are very efficiently removed by our event selection procedure (see section~\ref{sec:eventSel}) and they do not compromise the ANAIS-112 scientific goals. Moreover, in collaboration with LSC staff, radon-free gas supply has been improved for the second year of ANAIS-112 operation. 

\begin{figure*}[htbp]
\centering
\includegraphics[width=0.75\textwidth]{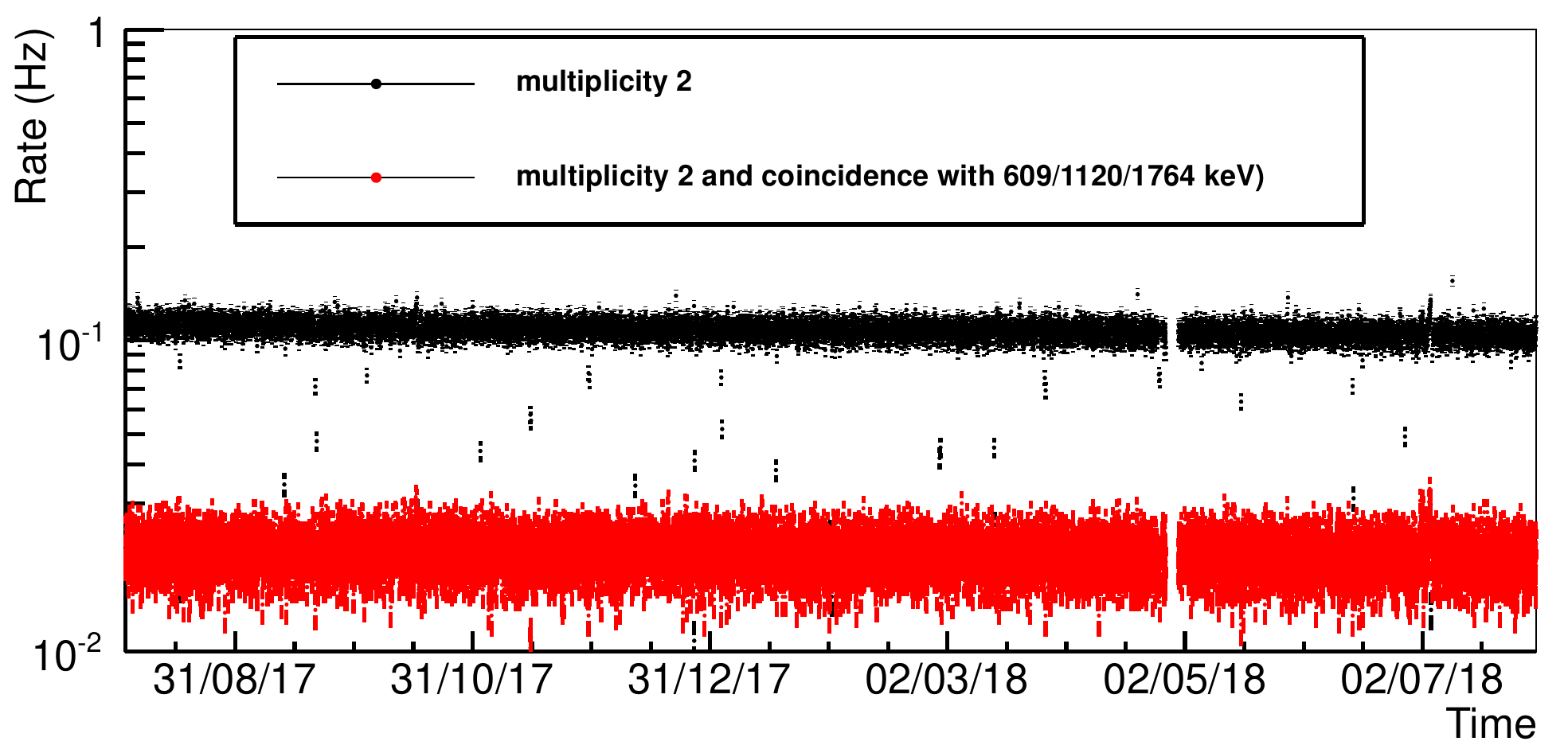}
\caption{Trigger rate (calculated in 3600~s time bins) during the first year of ANAIS-112 DM data taking: double coincidences rate (multiplicity 2) is shown in black, rate of events in coincidence with 609~keV, 
1120~keV or 1764~keV, is shown in red. Colors referenced are available in the online version of the paper.  
}
\label{fig:rate2}
\end{figure*}

We can also look at the evolution in time of the coincident events rate, for instance those having 
multiplicity 2 (two detectors with a signal within the coincidence window) 
are shown in red in Figure~\ref{fig:rate2}. These events should 
track the contribution from different background sources, as radon intrusion into the shielding or 
external gamma flux. Figure~\ref{fig:rate2} shows the rate of events in 
coincidence with 609~keV, 1120~keV or 1764~keV gamma lines as black solid dots, which are all attributable to $^{222}$Rn progeny.  
These rates are also very stable along the year of data taking.

Regarding the response of the detectors, we compute the light collection by comparing the charge collected at the PMTs 
following a known energy deposition in the crystal (in our case, the 22.6~keV line of a $^{109}$Cd source) 
with the charge distribution of a single phe, the so-called Single Electron Response (SER).
In order to obtain a population of phe not biased in amplitude by the trigger threshold, we select
very low-energy background events, for which phe can be individually counted, and take the last one in the 
waveform, integrating the charge in a narrow window around the phe maximum: [-30~ns, +60~ns].
In order to do so we use a peak identification algorithm~\cite{MAThesis}, 
based on a low pass filter and the detection of a sign change in the derivative waveform. 
It uses a very low software threshold in order to rescontruct the SER down to the baseline noise.
This routine performs very well at very low energy, being its 
efficiency limited by the phe overlap due to their non-zero width. 
Some examples of pulses with a low number of phe can be found in Section~\ref{subsec:psd}, Figure~\ref{fig:pulseExamples}.
\begin{figure*}[htbp]
\centering
\includegraphics[width=0.75\textwidth]{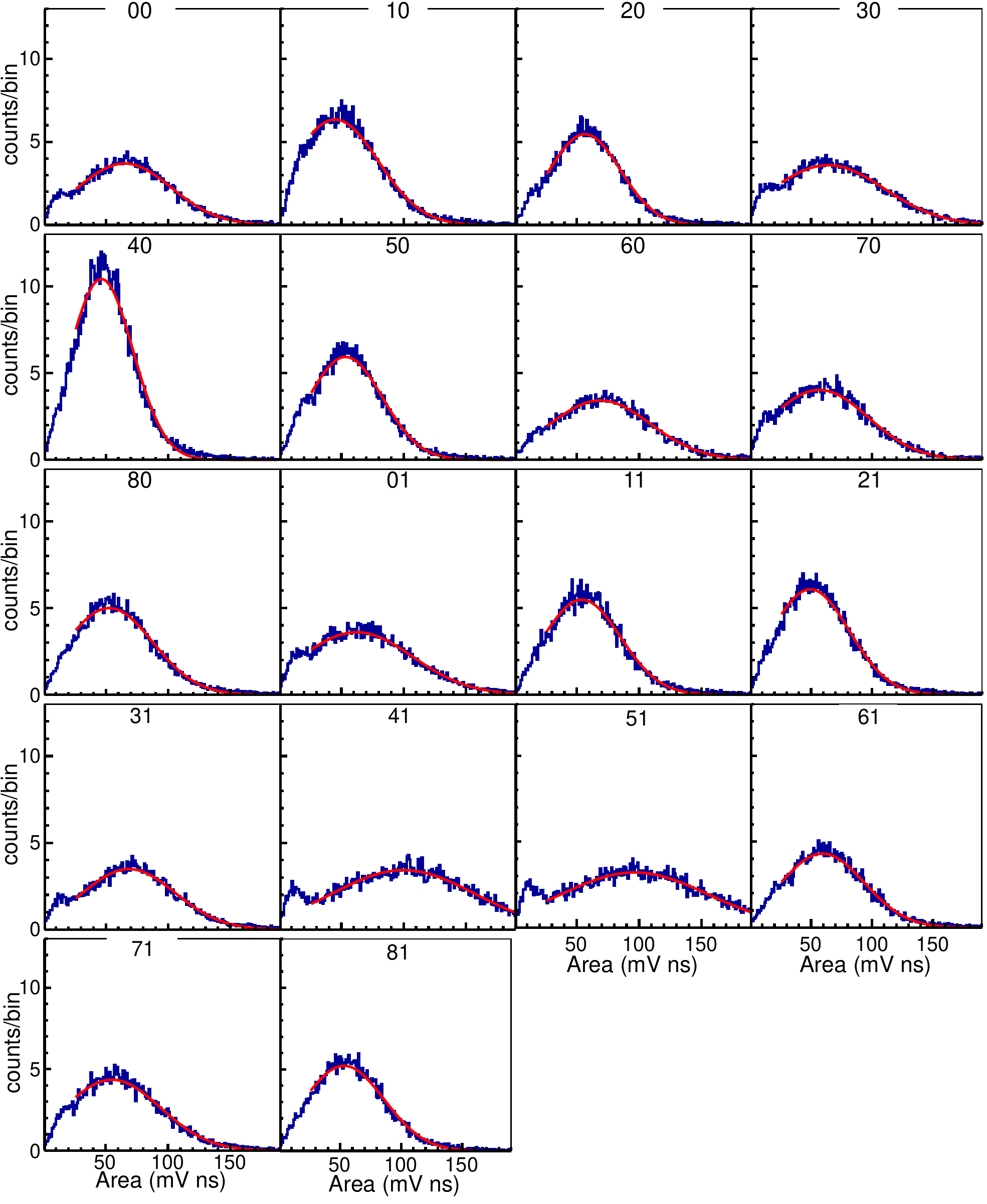}
\caption{
Example of SER area distributions for the 2~PMT$\times$9~detectors of ANAIS-112. They are obtained for background runs every two weeks, selecting the last peak in pulses having a very low number of phe, in order to not be biased by the trigger threshold (see text for more information). At each panel, the first digit
represents the detector number and the second one the PMT (PMT-0,PMT-1). The response to individual photons of a PMT unit is characteristic and strongly dependent on the HV supply applied. The units 41 and 51 (corresponding to D4 and D5 modules) were operated at higher HV, and this can be observed as a broader SER area distribution and a higher mean value, as the PMT gain is higher. 
} \label{fig:SER}
\end{figure*}

\begin{figure*}[htbp]
\centering
\includegraphics[width=0.8\textwidth]{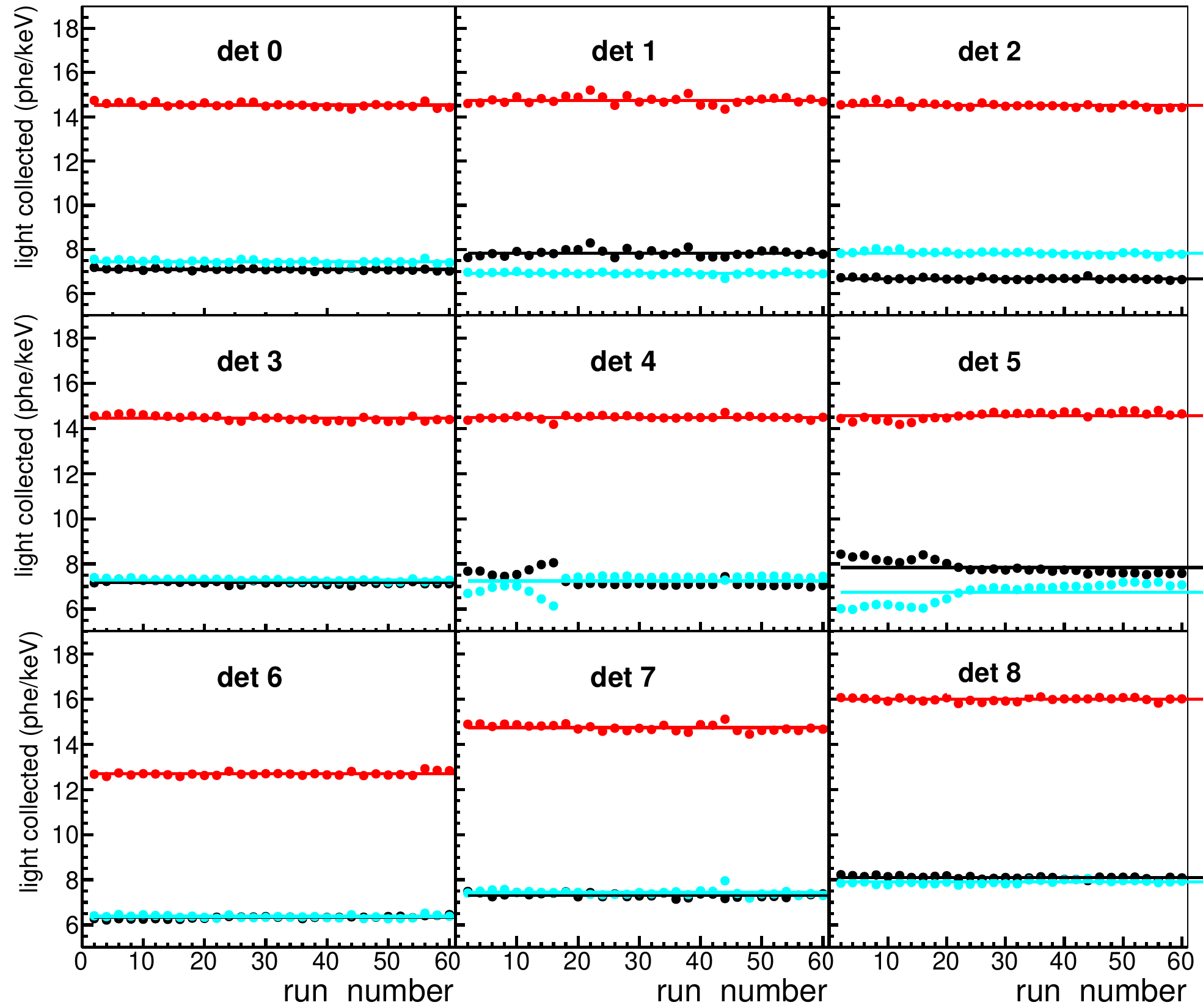}
\caption{Evolution of the light collected per unit of energy deposited for the nine NaI(Tl) modules along the first year of ANAIS-112 data taking. It is being monitored every two weeks, in average. Solid red dots correspond to the total light collected per module, black and cyan correspond to the PMT-0 and PMT-1 values, respectively. Colors referenced are available in the online version of the paper.  
}
\label{fig:lightyield}
\end{figure*}

The light collected at every PMT is then calculated as the ratio between the mean pulse area corresponding to 
the 22.6~keV calibration line and the mean of the SER pulse area distribution~\cite{Olivan:2017akd}. Figure~\ref{fig:SER} shows the distribution of the pulse area for the 18 PMTs used in ANAIS-112 set-up; this distribution is determined using background data grouped in two weeks periods, and the figure shows the distribution corresponding to one of them. 
It must be remarked the outstanding light collection measured for the nine 
AS modules, at the level of 15~phe/keV (see Table~\ref{tab:lightyield}). Only D6 module shows a 
light collection slightly below 13~phe/keV. This high light output has a direct impact in both resolution and 
energy threshold, but it also allows to improve strongly the signal vs noise filtering down to the threshold and hence, 
reduce analysis uncertainties in the search for dark matter annual modulation. 
Light collection is being continuously monitored by using periodic external calibrations using $^{109}$Cd and the SER derived from the dark matter run closer to every calibration run. 
Figure~\ref{fig:lightyield} shows the evolution of the light collected per module along the first year of 
ANAIS-112 data taking, as well as the corresponding values for the light collected at each PMT unit separately. 
It is worth remarking that all the modules show a highly stable total light collection along this first 
year of data taking, and all but D4 and D5 modules also show a highly stable light collection at each PMT. 
For D4 and D5 clear anticorrelation in the light collected at each PMT is evident in the first months of data taking. 
No explanation for this behaviour has been found up to date. 

\begin{table}
\caption{Average of the light collected per unit of deposited energy by the nine NaI(Tl) modules composing ANAIS-112 setup, derived from ANAIS-112 first year data, and standard deviation of the population.}
\label{tab:lightyield}       
\centering 
\begin{tabular}{lll}
\hline\noalign{\smallskip}
Detector & Average Light collected & Standard deviation\\
         & (phe/keV) & \\
\noalign{\smallskip}\hline\noalign{\smallskip}
D0  & 14.532 & 0.102 \\
D1  & 14.745 & 0.169 \\
D2  & 14.506 & 0.104 \\
D3  & 14.453 & 0.109 \\
D4  & 14.483 & 0.090 \\
D5  & 14.572 & 0.158\\
D6  & 12.707 & 0.104\\
D7  & 14.743 & 0.137\\
D8  & 15.994 & 0.076\\
\noalign{\smallskip}\hline
\end{tabular}
\end{table}

\subsection{ANAIS-112 data analysis and blinding strategy}
\label{sec:dataAnalysis}
The DAQ software produces as output ROOT~\cite{root} files with all the available information directly read from the 
electronic modules, mainly the PMT's waveforms, QDCs values and clocks readouts (real time and live time)
These files are then further analyzed in three subsequent levels:
\begin{itemize}
\item first level: we calculate pulse parameters, such as baseline level and RMS,
pulse area (our low energy estimator) and PSA parameters. 
We also calculate the time since the last muon veto and apply the
peak-finding algorithm and store the number and position of the identified phe for each pulse.
\item second level: in a second step we calibrate the energy response of every detector in
two different regimes (low energy (LE), and high energy (HE)), as explained in Section~\ref{sec:calibration}.
The final LE calibration is carried out every 3 dark matter runs ($\sim$1.5~months).
\item third level: we optimize the pulse shape cuts and calculate their efficiency. LE variable for single hit (multiplicity 1) events is blinded in this level of analysis, but it is kept unblinded for multiple hits events, which are used to calculate event selection efficiencies, and to check the stability of the experiment, as commented in Sections~\ref{sec:eff} and ~\ref{sec:stability}, respectively. 
\end{itemize}
We unblind $\sim$10\% of the total statistics ($\sim$34 days selected randomly from the whole first year of data taking amounting to 32.9 days live time) which we use for background assessment and tuning some of the procedures, while the rest of data remains blinded, until the moment to apply them the annual modulation analysis procedure.

\section{Energy calibration}
\label{sec:calibration}

ANAIS-112 experiment is calibrated every two weeks using external $^{109}$Cd sources. 
All the nine modules are simultaneously calibrated using a multi-source system built 
in a flexible wire in order to minimize down-time during the dark matter run. 
The system is designed to guarantee radon-free operation, with the wires moving along
a sealed tube inside the shielding up to reach their position in front of the Mylar window (see lower panel of Figure~\ref{fig:setup}).
The source plastic housing contains a certain amount of bromine, 
that under irradiation of the  $^{109}$Cd gammas produce a 
new calibration line in correspondence with the K-shell Br X-rays.
\par
\begin{figure*}[htbp]
\centering
\includegraphics[width=1.00\textwidth]{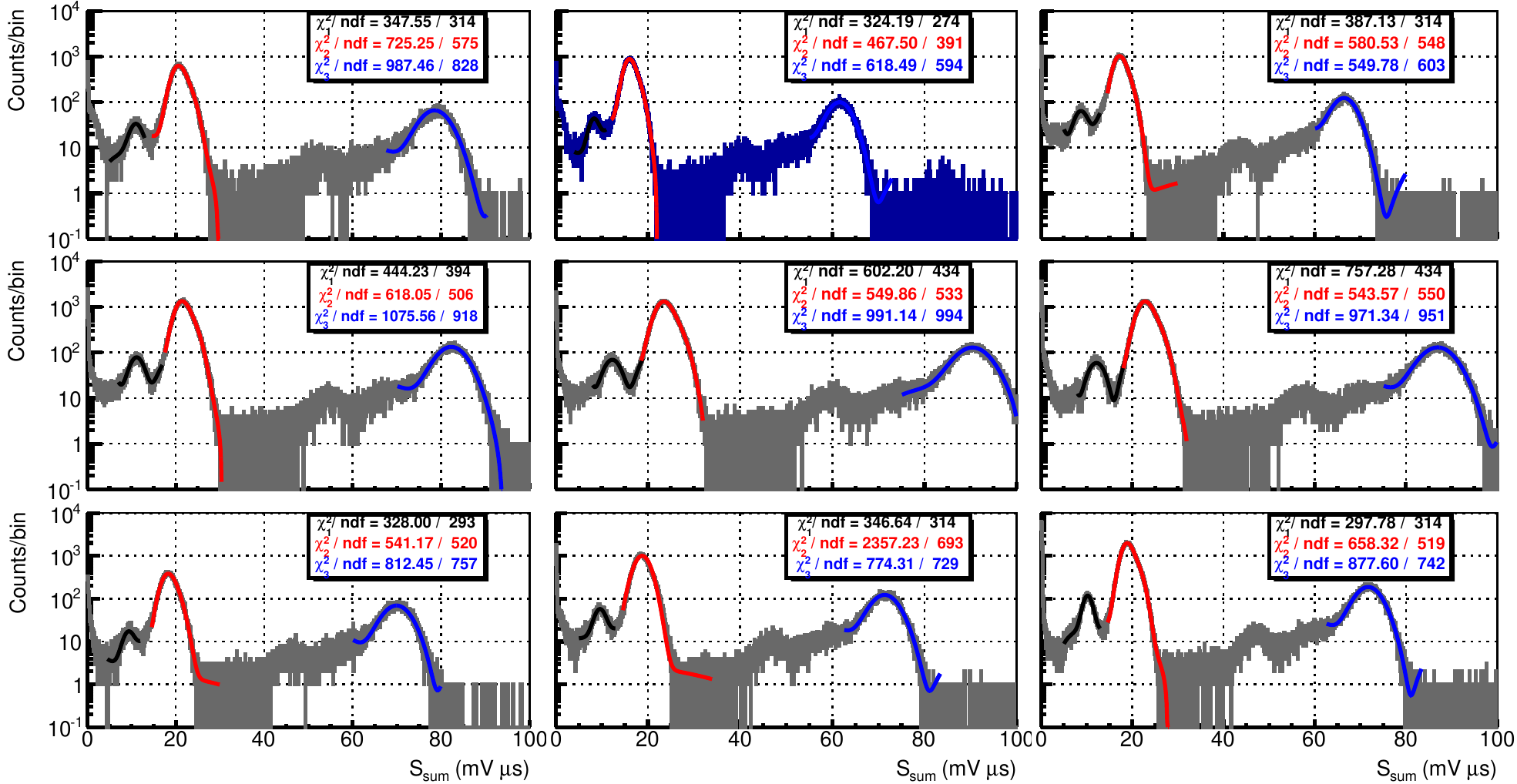}
\caption{Calibration spectra for the nine modules in one of the ANAIS-112 calibration runs. The 11.9 and 88.0~keV lines are fitted to single Gaussian lineshapes added to a second-order polynomial, while the 22.6~keV line is fitted to two Gaussian lineshapes with the same standard deviation added to a first-order polynomial. All the three fits are shown in red solid lines and chi-squared values for the three fits are given (1 - black, 2 - red, and 3 - blue labels correspond to the fits for 11.9, 22.6, and 88.0~keV lines, respectively). Colors referenced are available in the online version of the paper. 
}
\label{fig:cal9}
\end{figure*}
\begin{figure*}[htbp]
\centering
\includegraphics[width=0.75\textwidth]{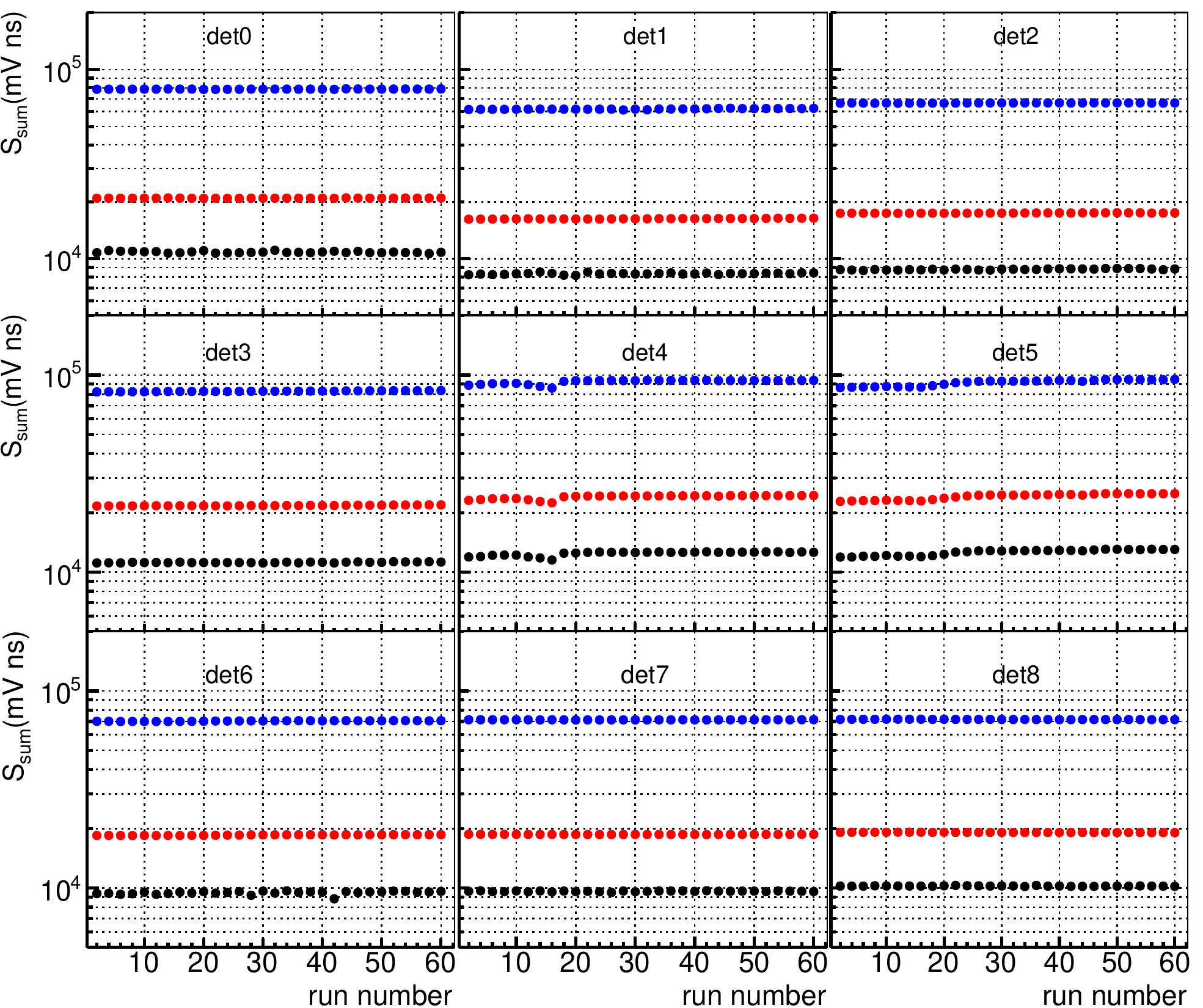}
\vspace{0.5cm}
\caption{Mean positions of the three $^{109}$Cd lines in the $S_{sum}$ variable along the first year of data taking for the nine ANAIS-112 modules. In blue 88.0~keV line, in red 22.6~keV line, and in black 11.9~keV line. Only D4 and D5 have shown clear gain drifts related with the change in the sharing of light collection between the two PMTs (see Section~\ref{sec:experimental}). Colors referenced are available in the online version of the paper. 
}
\label{fig:cal_evolution}
\end{figure*}

Every calibration run lasts for about 3-4 hours.
The energy of every event is obtained from the pulse area of the sum of the digitized
waveforms, {$S_{sum}=\sum (S_0(t)+S_1(t))$}. 
In Figure~\ref{fig:cal9} we show the calibration spectra for the nine modules corresponding 
to one of those calibration runs, arbitrarily chosen. 
The three main lines used for calibration 
(11.9, 22.6 and 88.0~keV) are fitted to gaussians and shown in red in the figure. 
The second one is in fact 
the sum of several nearby X-rays with average energy of 22.6~keV. A MC simulation showed that this average is not affected by the photon absorption in the path from the $^{109}$Cd source to the NaI crystal.
For the Br line we take as nominal energy the average of the two more intense K$_\alpha$ lines, 
resulting in 11.9~keV.
These lines are used for monitoring gain stability and to correct any possible gain drift. In Figure~\ref{fig:cal_evolution} we show the evolution of the positions of these three lines along the ANAIS-112 first year of data. Only D4 and D5 modules have shown clear gain drifts at the level of 10-15\%, related with the light sharing anomaly already shown and commented in Section~\ref{sec:experimental}.

Although the ROI for the annual modulation analysis is at very-low energy, 
below 10~keV\footnote{In the following, we will use keV for electron-equivalent energy}, a good calibration in the high energy region is mandatory 
in order to verify the background model and properly select very relevant low energy events 
populations from coincidences with high energy gammas. 
Non-linearities in the detector response have been observed in ANAIS modules, 
being due to both, intrinsic non-linearities of the scintillator itself and PMTs saturation effects. 
Because of that, specific strategies for energy calibration have been applied to ANAIS data, 
both in the high and low energy regimes, as explained in subsections~\ref{sec:hecal} and \ref{sec:lecal}. 

\subsection{High energy calibration} 
\label{sec:hecal}

\begin{figure}[htbp]
\centering
\includegraphics[width=0.5\textwidth]{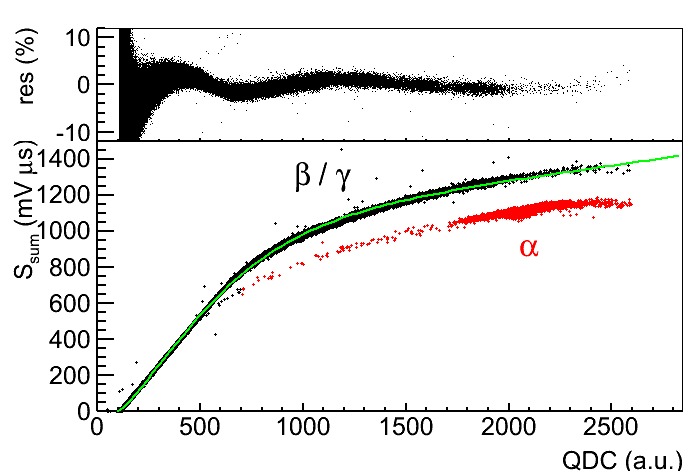}
\caption{Pulse summed-area of the digitized signals 
for one of the ANAIS-112 detectors during the two first weeks of data taking
as a function of the QDC readout.
Green line: modified logistic function used to linearize the $S_{sum}$ signal.
Red dots: $\alpha$ population, clearly separated from the $\beta$/$\gamma$ one (black dots) 
due to the different $S_{sum}$/QDC ratio. The top panel shows the residuals of the $\beta$/$\gamma$ population fitting to the modified logistic function. Colors referenced are available in the online version of the paper. }
\label{fig:heLin}
\end{figure}

The pulse amplification is set to allow a good energy reconstruction below 10~keV, 
so the events saturate above $\sim$500~keV in the digitized signal line. 
Because of that, the ANAIS-112 DAQ system (see Section~\ref{sec:DAQ}) incorporated a second signal line
conveniently attenuated, in order to keep information of the released energy of every high energy 
event using charge-to-digital converters for the signal instead of digitizers. 
In Figure~\ref{fig:heLin} the trend to saturation of the pulse area can be observed when is represented vs. the corresponding QDC value. 
These QDCs have a resolution of 4096 points, so, in order to 
reduce the quantization error in the energy signal, we do not take directly the QDC readout as estimator of the event energy in the high energy regime, but
use this information to linearize the $S_{sum}$ response using a modified logistic function, as it is shown in Figure~\ref{fig:heLin} (green line).
This linearized variable is used as estimator of the high energy assigned to each event
above $\sim$50~keV. 
We also profit from the double readout system to discriminate 
$\alpha$ events (shown in red in Figure~\ref{fig:heLin}). The integration window
of both readouts is the same, but for high energy events the digitized pulses are saturated, so
for the same QDC value
the pulse integral 
is smaller for $\alpha$ events, 
as they are faster than $\beta$/$\gamma$ (shown in black).
\par
Once the response is linearized, we calibrate the energy response above $\sim$50~keV
for each dark matter run independently using several peaks that are easily identified in the background,
because no high-energy external sources are available for this purpose in ANAIS-112. 
For every detector and run we use between 6 and 8 peaks, that we fit to Gaussian lineshapes
and use to perform a linear regression on their nominal energies against the Gaussian means using a second-order
polynomial.

\begin{figure}[htbp]
\centering
\includegraphics[width=0.5\textwidth]{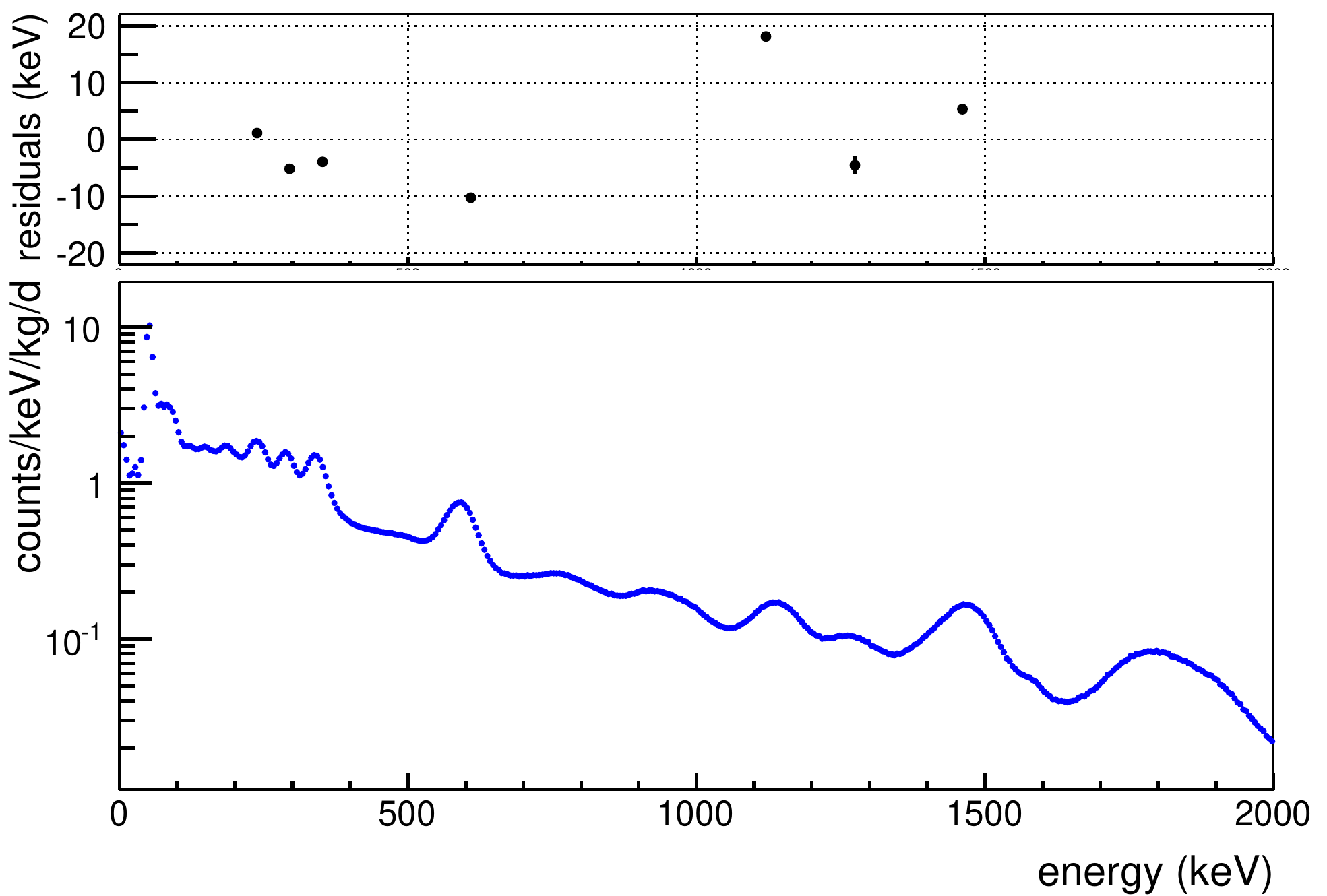}
\caption{Lower panel: high energy anticoincidence spectrum measured in the first year of ANAIS-112 
data taking (all detectors). Upper panel: residuals (fit-nominal) for the positions of the main peaks identified in the 
background.}
\label{fig:heSpc}
\end{figure}

Figure~\ref{fig:heSpc} shows the calibrated anticoincidence (i.e., only single hits) background spectrum above 50~keV for ANAIS-112 in the whole exposure corresponding to the first year of data taking. The upper panel shows the residuals for the positions of the main peaks identified in the background, being most of them below 10~keV.

\subsection{Low energy calibration} 
\label{sec:lecal}

\begin{figure}[htbp]
\centering
\includegraphics[width=0.5\textwidth]{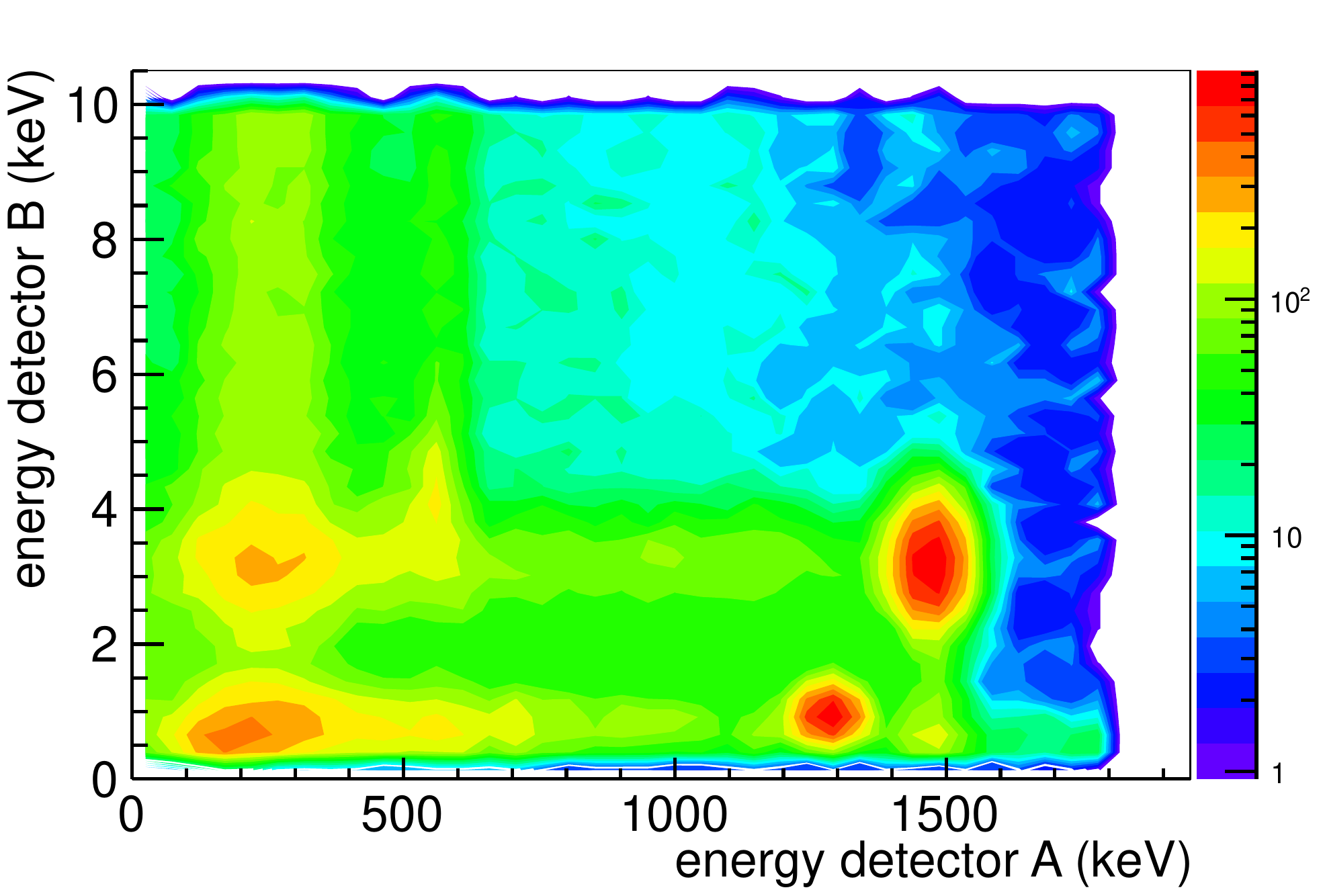}
\caption{Plot of the energies registered in two detectors triggered in coincidence when one of them has an energy deposition below 10~keV. Data correspond to the full first year of ANAIS-112. The spots at (1274.5~keV, 0.87~keV) and (1460.8~keV, 3.2~keV) 
correspond to K-shell EC decays from {\Na} and {\K} in the NaI crystal, respectively. 
}
\label{fig:coin}
\end{figure}

Thanks to the Mylar window built in the ANAIS modules we have several calibration peaks from external sources very close to the ROI, 
but we can also profit from known lines present in the background to increase the calibration accuracy in the ROI. Because the two background lines considered are actually either in the ROI or very close to the threshold, and they come from energy depositions in the bulk (as dark matter is expected to do), their inclusion in the calibration procedure improves strongly the reliability of the ANAIS-112 energy threshold estimate.
Figure~\ref{fig:coin} shows the energies registered in two detectors triggered in coincidence in the full first year of ANAIS-112. It shows those coincident events with one of the energy depositions below 10~keV.
Two spots stand out in the picture, whose origin can be ascribed to {\Na} and {\K} bulk 
contamination of the NaI crystals. 
These isotopes may decay via EC, with the emission of a $\gamma$ from the daughter nucleus de-excitation.
The atomic de-excitation energy (0.87~keV for {\Na} and 3.2~keV for {\K} for K-shell EC, which has the largest probability) 
is fully contained in the crystal where the decay occurs, 
while the high energy $\gamma$ (1274.5 and 1460.8~keV, respectively) can escape and hit another detector, thus producing a coincidence event. 
Compton-scattered $\gamma$'s are also evident in the plot, with less prominent spots  
in correspondence with the backscattered events in the lead shielding (around 200~keV).
These events can be taken as a population of background events homogeneously 
distributed in the NaI crystals array and falling in the energy window of interest for the annual modulation analysis, 
being then extremely interesting for evaluating any seasonal effect attributable to these backgrounds. 
Furthermore, they also serve as excellent calibration peaks.
We use them to calibrate the ROI, together with the two lower energy lines 
recorded during 
the periodic calibration runs (11.9 and 22.6~keV). 
We calibrate with the external $^{109}$Cd source the detectors every two weeks to control the gain stability and correct possible drifts,
but in order to increase the statistics of the 
{\Na} and {\K} peaks 
we add up one and a half months of dark matter runs to carry out the final LE calibration. 
It is worth mentioning that these peaks selected by the coincidence requirement are polluted with a 
population of very fast low-energy events in coincidence with a high-energy deposition in another crystal. 
They are presumably caused by an energy release in the PMT (likely an electron produced by a $\beta$ decaying isotope, which is able to produce
Cherenkov light) in coincidence with a $\gamma$ in one of the crystals. This hypothesis is further 
investigated in a companion paper~\cite{Amare:2018ndh}.
As we will show in Section~\ref{sec:eventSel} they can be removed with a pulse-shape cut.  

The peaks registered during calibration runs are 
fitted to Gaussian lineshapes, while for the {\Na} and {\K} peaks, with a lower number of events, 
we take the median of the distribution.
Then, we perform a linear regression on the expected energies against the peak's positions 
for every detector using a linear function and recalibrate the low energy events (below 50~keV).
\par
\begin{figure}[htbp]
\centering
\includegraphics[width=0.5\textwidth]{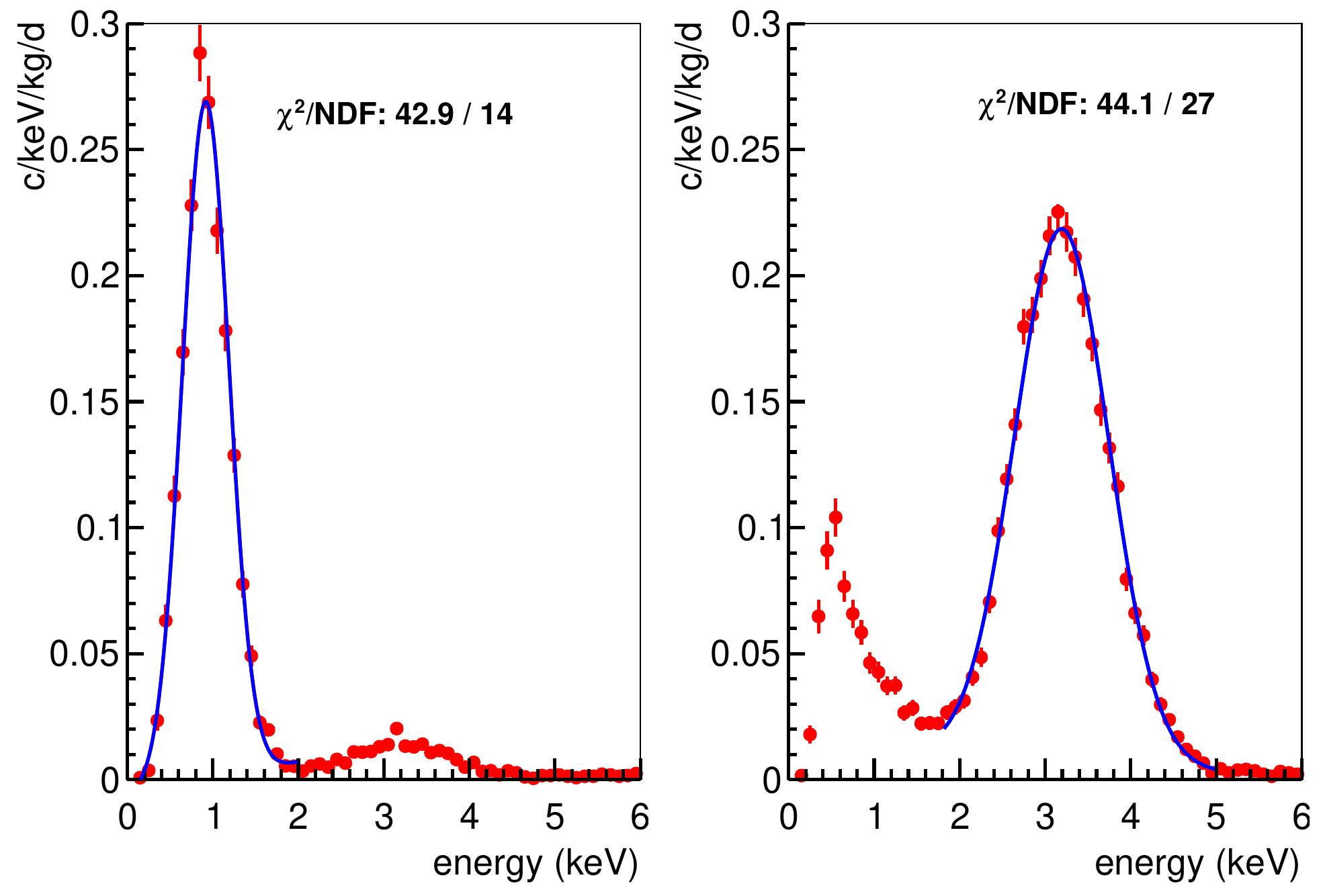}
\caption{Low energy spectra of the nine modules in coincidence 
with a high energy gamma in the range [1215-1335]~keV (left panel) and [1350-1560]~keV (right panel) 
in a second module (after recalibration described in Section~\ref{sec:lecal}). 
Data correspond to the full first year of ANAIS-112. 
Peaks at 0.87 and 3.2~keV, attributed to {\Na} and 
$^{40}$K decay in the NaI bulk, respectively, can be clearly observed. Blue line: fit to a Gaussian plus linear background lineshape. The goodness of the fits 
is also shown in the plot. Colors referenced are available in the online version of the paper. }
\label{fig:NaK}
\end{figure}

We check the goodness of this procedure and the accuracy of the final calibration at LE precisely with the lines corresponding to {\Na} and {\K} decays, found in the coincidence spectra of the whole available statistics for the first year of ANAIS-112. 
It is important to note that the 0.87~keV peak from {\Na} is actually below our analysis threshold, 
established at 1~keV after calculating the event selection efficiencies (see Section~\ref{sec:eff}). 
However, as the events corresponding to this peak
are selected by a coincidence requirement, they are almost free from noise/background except for the fast 
Cherenkov events mentioned above, so we can relax some of the cuts
described in Section~\ref{sec:eventSel} and go beyond 1~keV to fully reconstruct its energy after correcting by the trigger efficiency and that of the filter applied. More details on the selection cuts will be given in Section~\ref{sec:eff}. 
We have followed this procedure to produce the low energy spectra shown in Figure~\ref{fig:NaK}, 
in the left panel in coincidence with an energy deposition in the range [1215-1335]~keV
in another module (corresponding to the {\Na} population),  
and in the right panel requiring a coincidence with a [1350-1560]~keV event ({\K} population). 
Two peaks at 0.87 (left panel) and 3.2~keV (right panel) are clearly visible.
The selection criteria for the {\Na} population includes also 
Compton-scattered 1460.8~keV events, that produces the low-amplitude 3.2~keV peak visible in the left 
panel. On the other hand, the asymmetric population at $\sim$0.5~keV in the right panel
could correspond to the L-shell EC decays from {\K}.

\begin{figure}[htbp]
\centering
\includegraphics[width=0.5\textwidth]{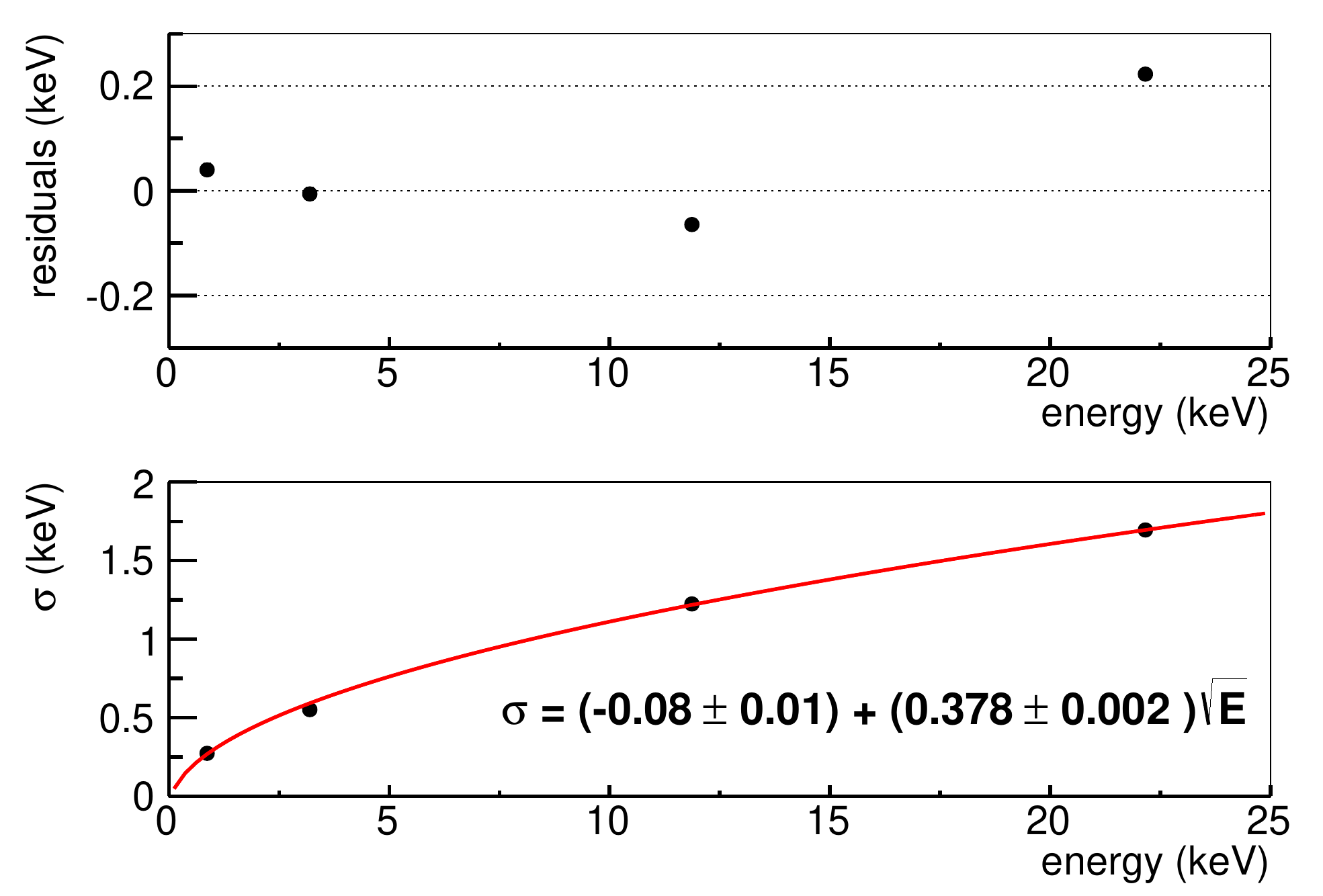}
\caption{Upper panel: residuals (fit-nominal) 
of the low energy calibration in the {\Na} and {\K} background 
peaks (efficiency corrected) and the 11.9 and 22.6~keV external calibration peaks.
Lower panel: energy resolution of the same energy lines as a function of energy. The resolution
at 22.6~keV (11.9~keV) is calculated fitting the calibration peak to a combination of four (two) Gaussian lines.
Red line corresponds to the fit to a $\alpha + \beta \sqrt{E}$ function, 
showing in the plot the values of the fitted parameters.}
\label{fig:lowEResiduals}
\end{figure}

We fit both main peaks to a Gaussian lineshape plus a linear background (blue line). 
The residuals of the fit (defined as the difference between the reconstructed mean 
position of the peak minus the nominal energy) are shown in Figure~\ref{fig:lowEResiduals}
together with those corresponding to the 11.9 and 22.6~keV calibration peaks calculated in the spectrum obtained by adding all the calibration runs. According to this, 
the calibration deviations are below 0.2~keV in the whole range and below 0.04~keV in the ROI.

In the lower panel of Figure~\ref{fig:lowEResiduals} we show also the energy resolution as a function of energy calculated for the same lines\footnote{For 
the 22.6~keV line we have used a four-Gaussian function shape,
where all the Gaussians are constrained
to have the same width and their relative intensities and positions
are fixed to the four more intense Ag X-rays~\cite{nndcAg}, while 
for the 11.9~keV line we have considered the same procedure with two Gaussians~\cite{nndcBr}.} in the coincidence spectrum and the addition of all the calibration runs, being very well reproduced by a $\alpha+\beta\sqrt{E}$ function.
The 0.87~keV peak, with a resolution of $\sigma$=0.27~keV, is in good agreement 
with the expected value if the resolution is dominated by the phe counting statistics,
considering an average light collection of 14.5~phe/keV (see Table~\ref{tab:lightyield}).

\section{Event selection}
\label{sec:eventSel}

DM particles should interact with normal matter through processes having a very 
low cross-section. In general, they are typically searched for in the WIMP-nucleon cross-section range from 
10$^{-45}$ to 10$^{-40}$~cm$^2$~\cite{Undagoitia:2015gya}. This implies that the probability of a dark matter particle 
interacting twice while crossing ANAIS-112 setup is negligible, and then we should only select 
events compatible with single-hit interactions, so %and for instance, 
coincidences among different modules are rejected. 
This very low interaction cross-section also means that the probability of interacting in the 
PMT or other setup components having low mass is also negligible, and that we should expect 
the interaction with the Na and I atoms as being dominant. Depending on the DM candidate, this interaction could occur only with the nuclei or also with the electrons in the target material. Because of this, we should remove from our analysis those events arising at the PMTs or being related with a muon crossing ANAIS-112 setup (see Section~\ref{sec:experimental}). 
In the following we summarize the filter procedures designed to select bulk scintillation events compatible with a dark matter signal and in next section the corresponding acceptance efficiencies for such filters will be calculated. 

\subsection{Rejection of muon related events}
\label{sec:muonVeto}

\begin{figure}[htbp]
\centering
\includegraphics[width=0.5\textwidth]{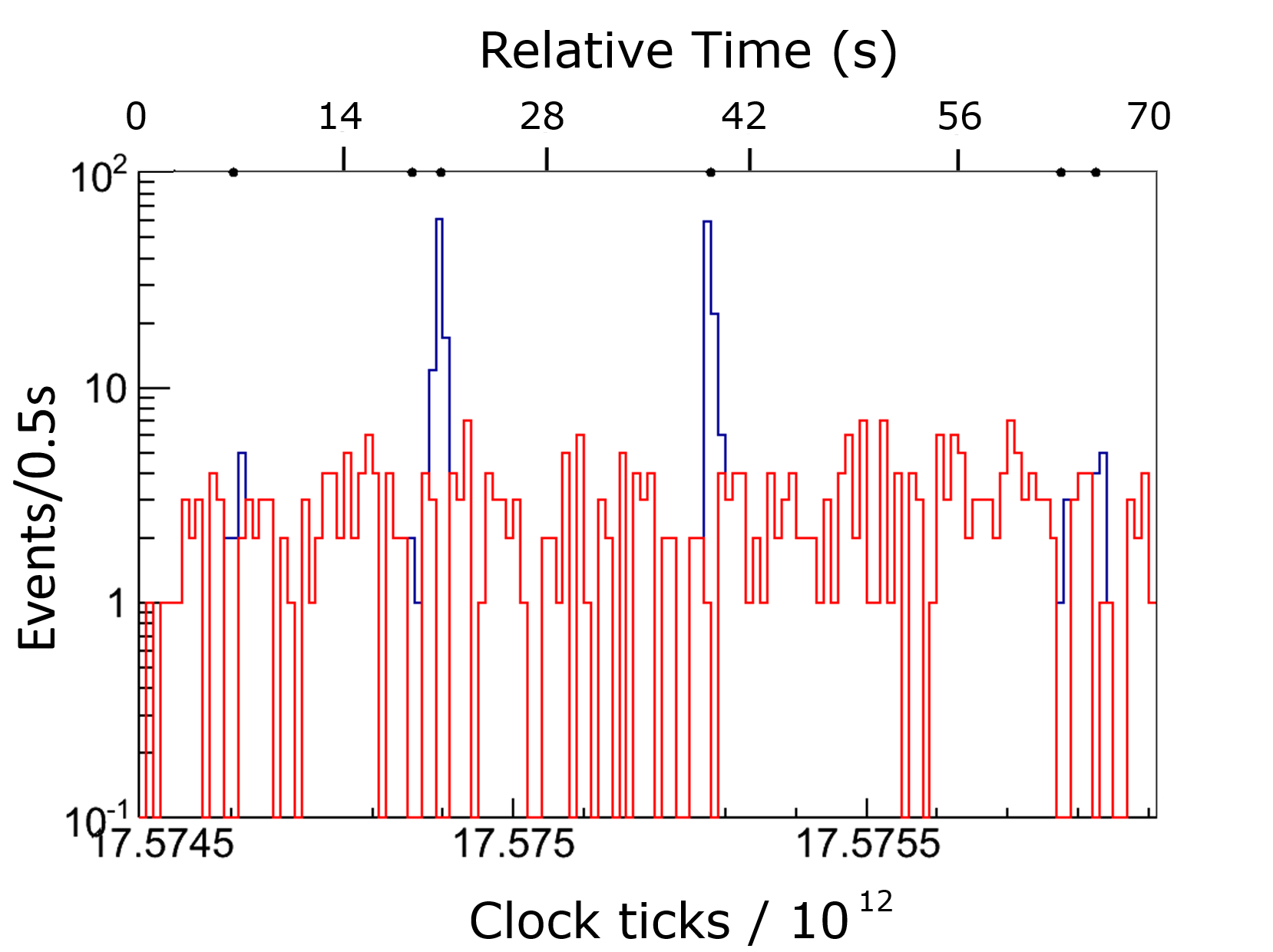}
\caption{Total trigger rate in ANAIS-112 in a period of 70~s %75 s 
is shown in blue. Events in the veto scintillators are marked as dots in the upper axis, pointing at a clear correlation between muons crossing ANAIS-112 setup and total trigger rate increases. We show in red the trigger rate of events passing the filter of being more than one second after a trigger in the veto scintillators. Top axis time scale (given in seconds) is relative to the beginning of the period shown, bottom axis scale is relative to the beginning of the run and it is given in clock ticks (50 ns/tick). Colors referenced are available in the online version of the paper. 
}
\label{fig:muon1}
\end{figure}

Figure~\ref{fig:muon1} shows a detail on the trigger rate of ANAIS-112 dark matter run for a period of 70~s. %75~s. 
The time of every event triggering any of the ANAIS-112 crystals is precisely determined using clock counters on a time basis of 50~ns. Precise synchronization among clock counters in ANAIS-112 DAQ and Veto Systems is carried out periodically. 
Veto scintillators trigger timing is marked in Figure~\ref{fig:muon1} as dots on the upper axis, and it can be clearly seen the correlation with some of the remarkable increases in the total trigger rate of 
ANAIS-112 modules. These high rate periods can then be related to a direct muon interaction in at least one of the NaI(Tl) crystals, which is producing subsequent triggers of the DAQ system. 
We remove most of these total trigger rate increases by rejecting events arriving less than one second after a veto scintillation trigger (see Figure~\ref{fig:muon1}: blue line is total trigger rate, red line is the trigger rate after this filtering), and we correct off-line the live time of our measurement by subtracting to the live time one second for every muon veto trigger. 
Several experiments have measured a clear seasonal modulation in the muon rate underground, 
and we need to be sure that muon related events are conveniently rejected before the annual modulation analysis is applied. The fraction of live time rejected after the application of this filtering amounts only to a 3\%.

\begin{figure}[htbp]
\centering
\includegraphics[width=0.5\textwidth]{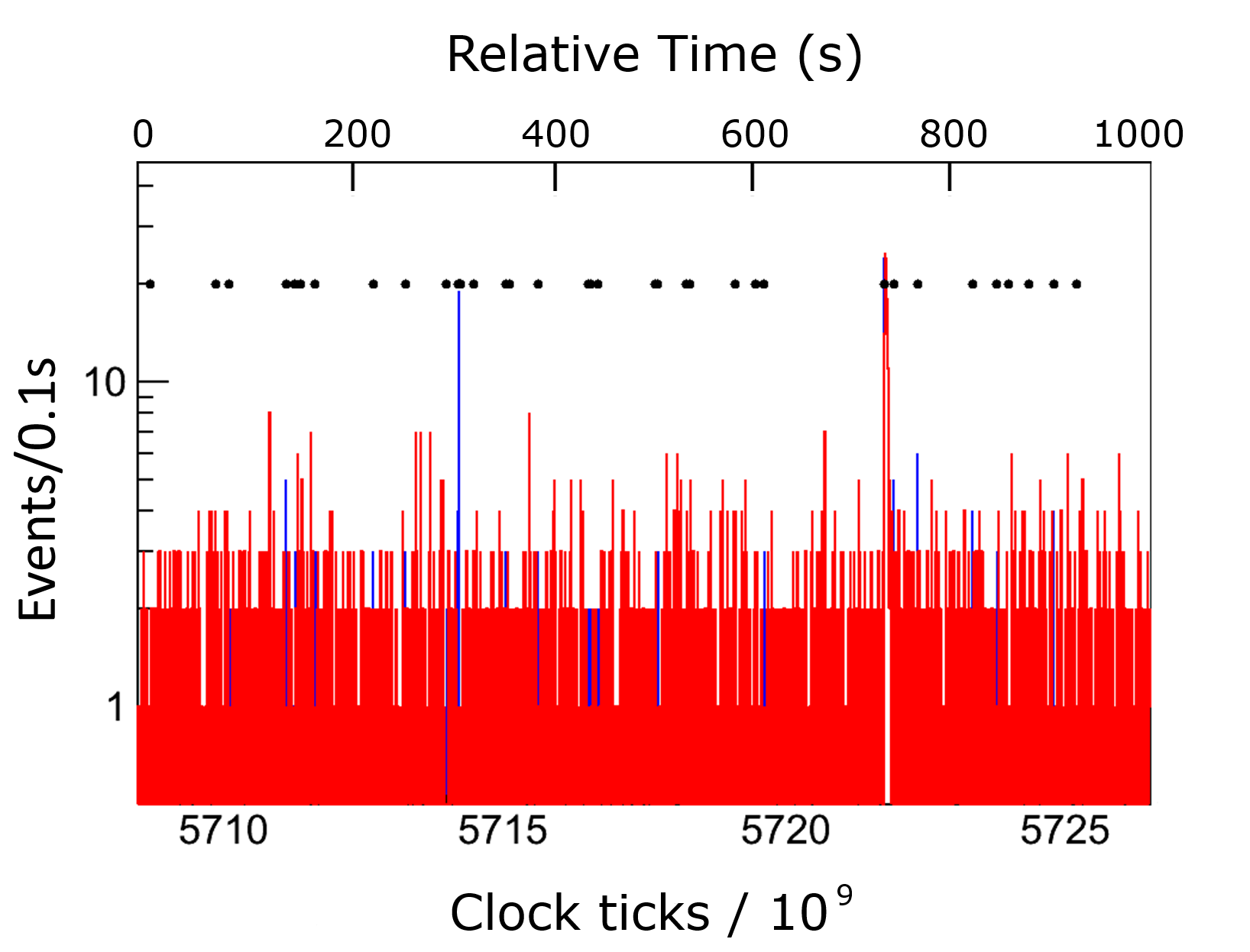}
\includegraphics[width=0.5\textwidth]{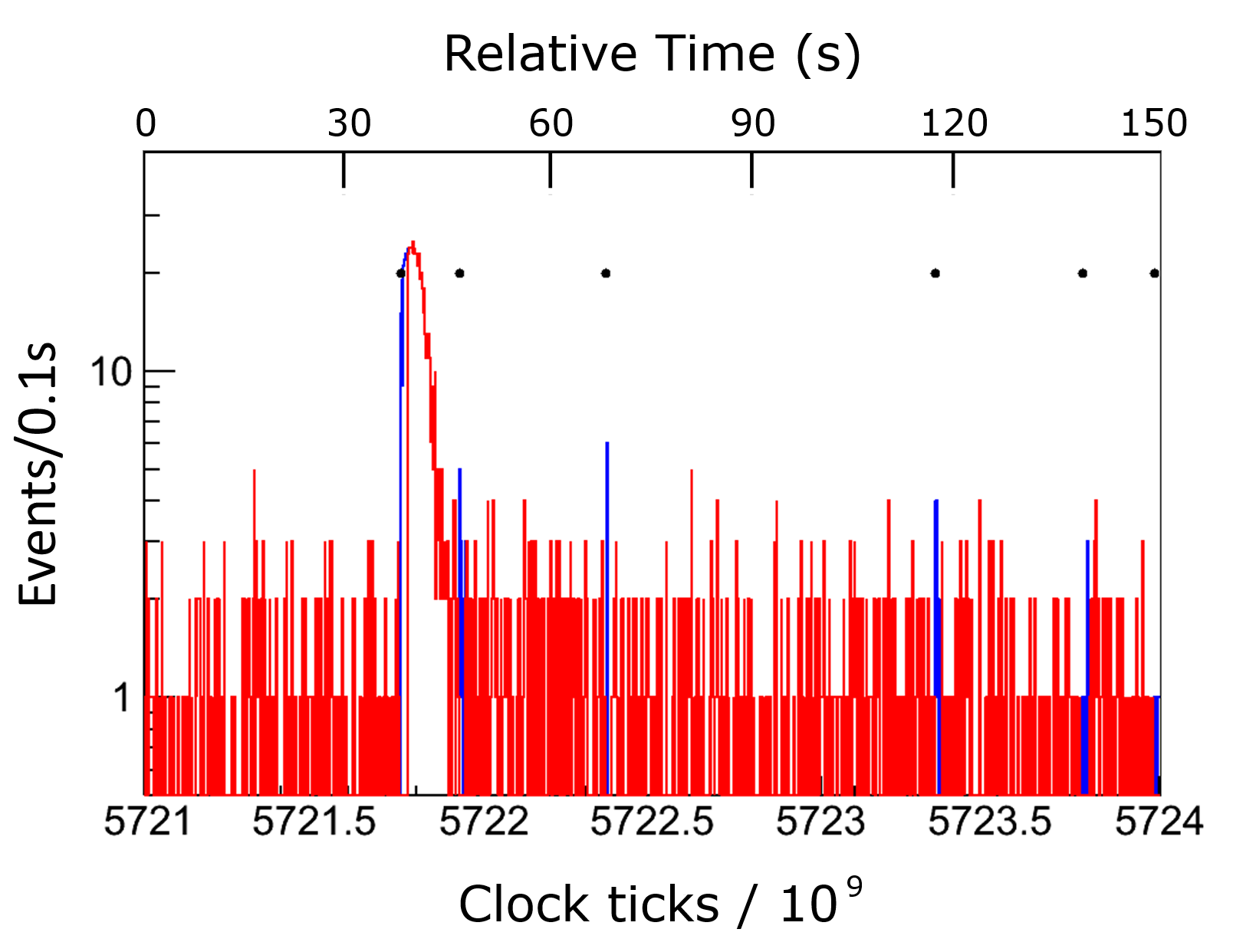}
\caption{Total trigger rate in ANAIS-112 in a 1000~s %250 s 
period is shown in blue and in red the trigger rate after removing 1 s following a veto trigger (top). A zoomed view around a very important muon related event is shown in the bottom panel. Some muon related events in the NaI(Tl) detectors are not removed by the 1 s delay filter, but those events have a very low number of phe in the pulses and they can be rejected too (see text). Top axis time scale (given in seconds) is relative to the beginning of the period shown, bottom axis scale is relative to the beginning of the run and it is given in clock ticks (50 ns/tick). Colors referenced are available in the online version of the paper. 
}
\label{fig:muon2}
\end{figure}

\begin{figure}[htbp]
\centering
\includegraphics[width=0.5\textwidth]{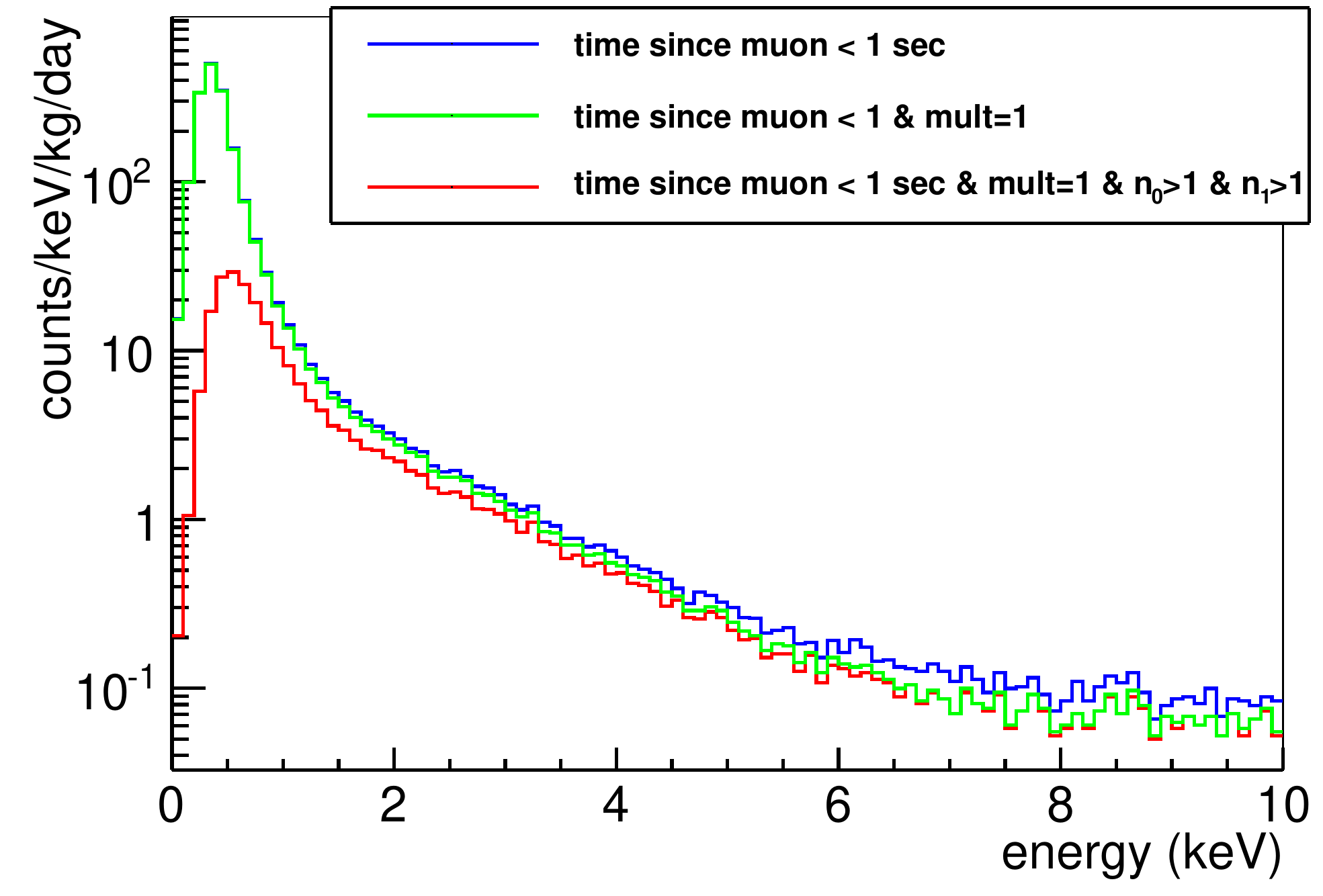}
\caption{Energy distribution (below 10~keV) of muon related events along the ANAIS-112 unblinded data: events occurring less than one second after a trigger of the Veto System (blue line), same but adding the anticoincidence condition (green line) and finally requiring also 
more than one phe at every PMT (red line). Colors referenced are available in the online version of the paper. 
}
\label{fig:muonrelated}
\end{figure}

Unfortunately, we cannot reject all the muon related events in that way. Figure~\ref{fig:muon2} top panel shows the total trigger rate in a period of 1000~s, %250~s, 
with a time binning of 0.1~s. After one of the muons crossing ANAIS-112 set-up, identified by the Veto System, a very relevant increase in trigger rate is observed in the bottom panel of Figure~\ref{fig:muon2}, which lasts for more than one second. 
Something similar is observed for very large energy depositions in one crystal and these events are not removed by the filter explained above. 
However, these events are easily eliminated by imposing an analysis threshold requiring more than one peak in the pulse at every PMT of a given ANAIS module. 
This analysis threshold is very soft, implying that at least 4 photoelectrons (phe) are required to produce an effective trigger (2 at each PMT). 
According to the light collection of ANAIS-112 modules reported in Section~\ref{sec:performance}, it should not compromise being able to register events at 1~keV or even below. 
Moreover, more stringent analysis threshold will be introduced in Section~\ref{subsec:psd} and more information of the implications of this filter will be given there. In conclusion, we are quite sure we are removing muon related events from ANAIS-112 data. 
The energy distribution of the events removed using the 1~s after a muon filter is shown in Figure~\ref{fig:muonrelated} for the ANAIS-112 unblinded data (see Section \ref{sec:dataAnalysis}). It is clear these events would contribute to the very low energy region if they were not rejected.

\subsection{Pulse shape cut}
\label{subsec:psd}

In a scintillation detector there are events at low energy 
whose origin is neither the scintillating crystal, nor the PMT dark current, but other light-emitting mechanisms 
in the PMTs structure, the optical windows used to couple the crystal to the PMT, or even the optical grease or silicone pads used in that coupling. Unlike dark current events, which are effectively removed by the trigger in logical AND between the two PMTs, the latter events can dominate the trigger rate 
below 10~keV and they have to be rejected prior to any dark matter analysis. On the other hand, slow scintillation components
(up to several hundreds of ms) have been measured in NaI(Tl) crystals after large energy 
depositions ($\alpha$ or $\beta/\gamma/\mu$)~\cite{2013OptMa..36..316C}.
These slow components can result in fake low energy events 
in the usual NaI(Tl) scintillation time scale if they are not properly identified as pulse tails.
\par
\begin{figure*}[htbp]
\centering
\includegraphics[width=0.40\textwidth]{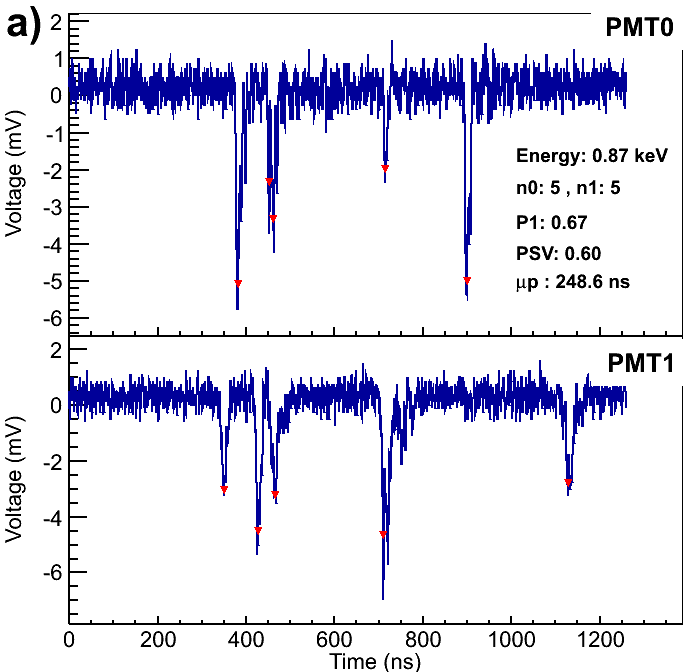}
\includegraphics[width=0.40\textwidth]{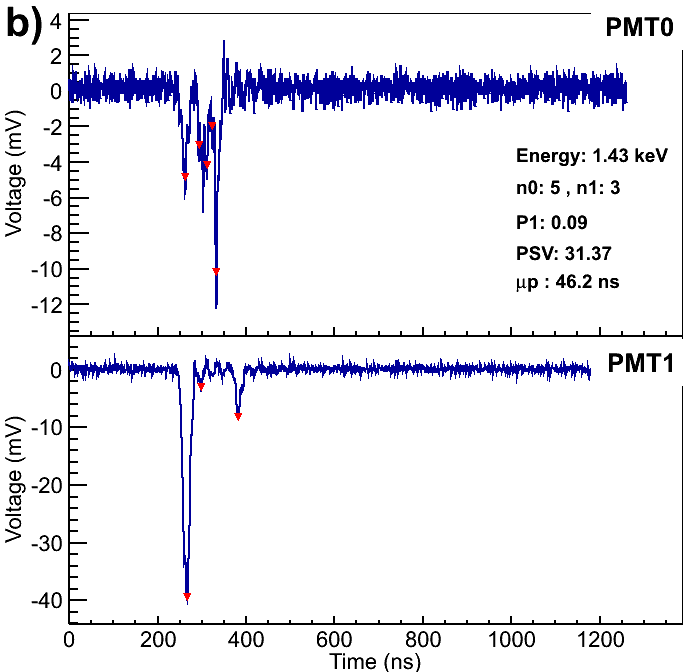}
\includegraphics[width=0.40\textwidth]{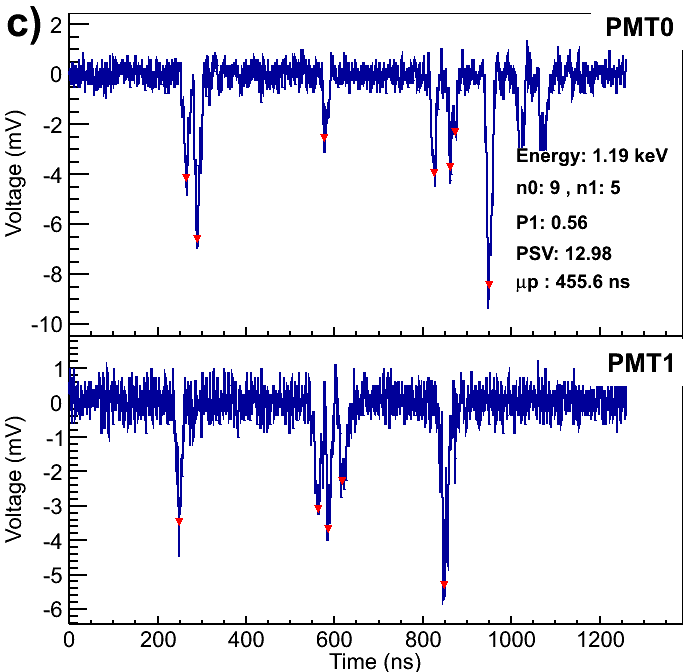}
\includegraphics[width=0.40\textwidth]{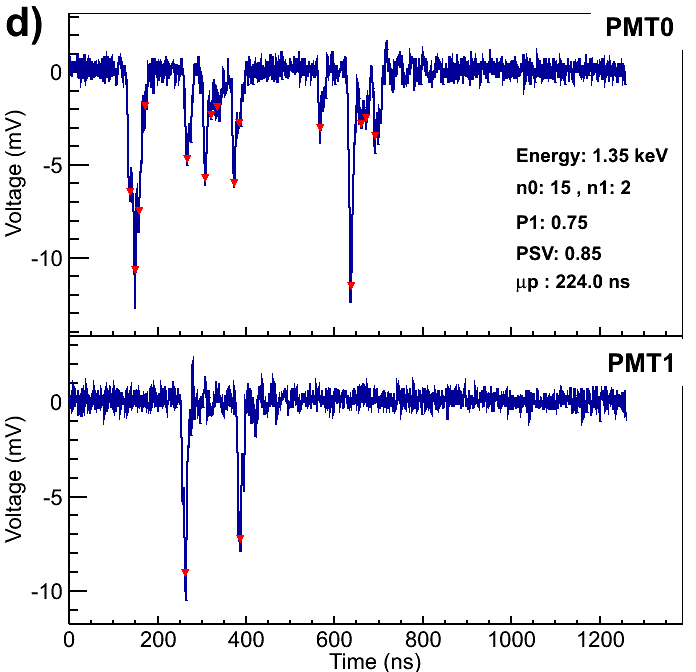}
\caption{Examples of low energy digitized pulses. The two traces at each panel 
correspond to the two PMT signals, while the legend display the energy of the event, number of peaks detected by the algorithm
at every trace (n$_0$ and n$_1$), and PSA parameters {\Puno}, {\PSVar} and $\mu_p$. Red triangles: phe identified by 
the peak-searching algorithm. Panel (a): 
0.87~keV bulk scintillation event from 
{\Na} decay selected by coincidence with a 1274.5~keV deposition in another crystal. Panel (b): 
fast event, likely a Cherenkov light emission in one PMT that is seen in the opposite PMT. Panel (c): 
long $\mu_p$ event, possibly caused by phosphorescence in the NaI(Tl) crystal. Panel (d):
low energy asymmetric event.}
\label{fig:pulseExamples}
\end{figure*}

The most common light production mechanism in the PMT is Cherenkov emission 
in the glass generated by radioactive contamination in the PMT itself or in the environment \cite{2014OptMa..36.1408A}.
These events are very fast, in fact they should be dominated by the temporal behaviour of the PMT response (tens of ns range), and then, they should be easily discriminated 
by pulse-shape analysis (PSA) with respect to NaI(Tl) scintillation events (hundreds of ns range). 
For the sake of illustration we show in (a) panel of Figure~\ref{fig:pulseExamples} 
a low energy bulk scintillation event from {\Na} decay and in panel (b) a fast pulse coming from the PMT. In order to discriminate this kind of events, we follow \cite{2008NIMPA.592..297B} and define the PSA parameter
%%%%%%%%%%%%%%%%%%%%%%%%%%%%%%%%%%%
\begin{equation}\label{eq:p1}
%\Puno=\frac{\int_{100~ns}^{600~ns}(S_0(t)+S_1(t)dt}{\int_{0}^{600~ns}(S_0(t)+S_1(t))dt} ,
\Puno=\frac{\sum_{t=100~ns}^{t=600~ns}(S_0(t)+S_1(t))}{\sum_{t=0}^{t=600~ns}(S_0(t)+S_1(t))} ,
\end{equation}
%%%%%%%%%%%%%%%%%%%%%%%%%%%%%%%%%%%
where $S_0(t)$  $(S_1(t))$ is the pulse amplitude at time $t$ after the trigger position for waveform $S_0$  $(S_1)$. 
For NaI(Tl) scintillation pulses (dominated by the scintillation time of $\sim$230~ns) 
this P$_1$ parameter has a value around 0.65,
while for fast scintillation (Cherenkov-like) or single phe is usually below 0.2.
\par 
This parameter also removes very efficiently random coincidences between individual dark current phe, however 
it is not  useful to discriminate fake low energy events produced 
by long phosphorescence in the crystal or pulse tails (see  Figure~\ref{fig:pulseExamples} panel (c)). 
To achieve this scope we use the logarithm of the 
mean time of the distribution of the individual phe arrival times in the digitized window~\cite{Kim:2018wcl}, {\LFM}, where $\mu_p$ is evaluated from 
the software trigger position until the end of the pulse as
%%%%%%%%%%%%%%%%%%%%%%%%%%%%%%
\begin{equation}\label{eq:fm}
\mu_p=\frac{\sum_{p}A_pt_p}{\sum_{p}A_p} ,
\end{equation}
%%%%%%%%%%%%%%%%%%%%%%%%%%%%%%
being $A_p$ and $t_p$ the amplitude and time of the phe identified in the pulse trace, respectively. 
For the phe identification we use the same peak identification algorithm developed for the 
SER construction (see Section~\ref{sec:performance}).

\par

\begin{figure}[t]
\centering
\includegraphics[width=0.42\textwidth]{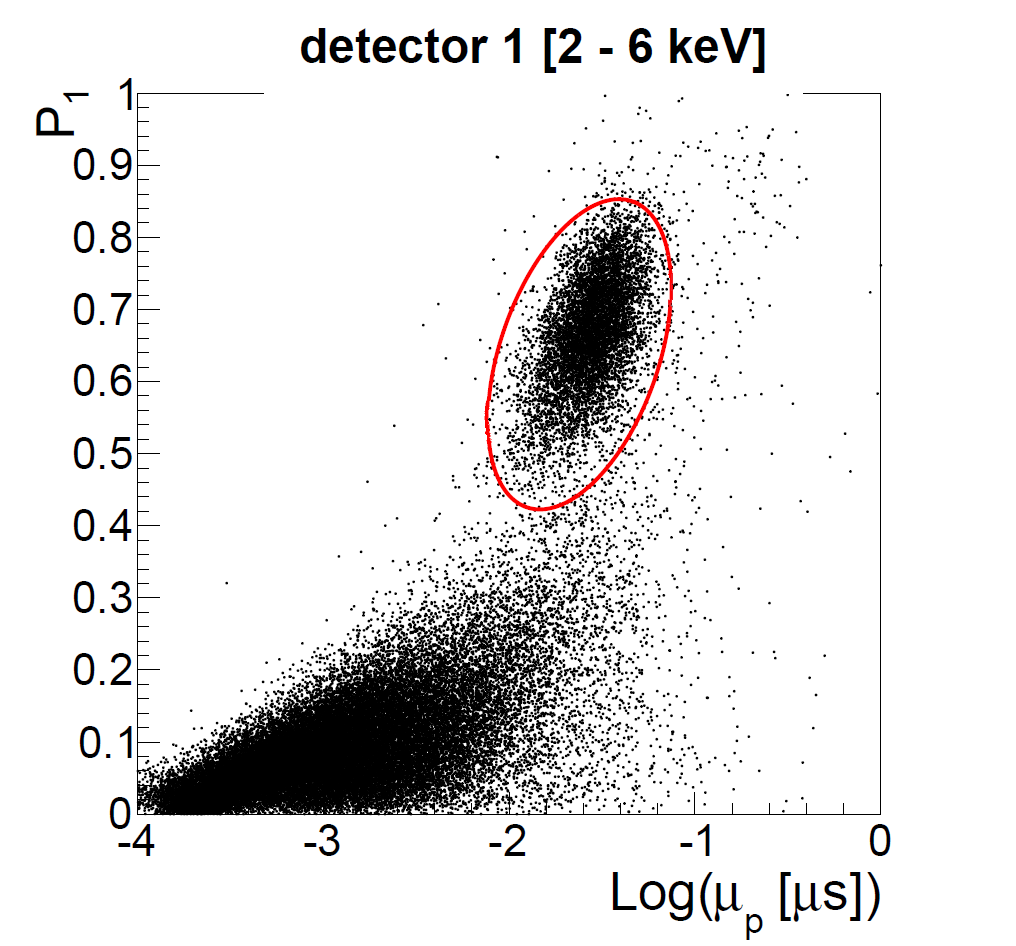}
\caption{Distribution of the background events of the $\sim$10\% unblinded data of detector D1 with energies between 2 and 6~keV
in the ({\Puno},{\LFM}) plane. Bulk scintillation events are clustered around (0.6,-1.5).
The red line encloses the events selected by the {\PSVar} cut in this detector.}
\label{fig:bkgMV}
\end{figure}

In Figure~\ref{fig:bkgMV} we show the distribution of these shape parameters for the low energy 
events between [2-6]~keV of detector D1 for the $\sim$10\% unblinded data. Bulk scintillation 
events are clustered around (${{\Puno}=0.65}$, ${{\LFM}=-1.5)}$, while most background events at these energies correspond
to very fast (likely Cherenkov) events with {\Puno}<0.1.
\par
Both parameters are correlated, so we perform a bivariate analysis, 
assuming that $\bm x = \begin{pmatrix} \Puno  \\ \LFM \end{pmatrix}$ follows a 2-dimensional Gaussian distribution 
with mean $\bm\mu$ and covariance matrix $V$.
$\bm{\mu}$ and $V$ are calculated between [1-2]~keV from 
the 0.87~keV bulk scintillation population from 
{\Na} decay selected by coincidence with a 1274.5~keV deposition in another crystal. 
We define 

\begin{equation}\label{eq:PSVar} 
\PSVar=(\bm{x}-\bm{\mu})^T V^{-1} (\bm{x}-\bm{\mu}).
\end{equation}

\par
For any given value of the parameter {\PSVar}$_{cut}$, the inequality {\PSVar}<{\PSVar}$_{cut}$ represents
an ellipsoid centered at $\bm \mu$, and such that 
the probability that an event $\bm x$ lies outside the ellipsoid  
is given by $1 - F_{\chi^2}(\PSVar_{cut})$, where $F_{\chi^2}$ is the cumulative $\chi^2$ 
distribution with 2 degrees of freedom~\cite{PhysRevD.98.030001}.
We select {\PSVar}$_{cut}$=3 (red line in Figure~\ref{fig:bkgMV}), so the efficiency of the cut at [1-2]~keV 
is 77.7\%, but as the distribution of the PSA parameters depends on energy, the efficiency of the cut in other energy regions has to be calculated, 
as we explain in detail in next section.

\par
\begin{figure*}[b]
\centering
\includegraphics[width=0.85\textwidth]{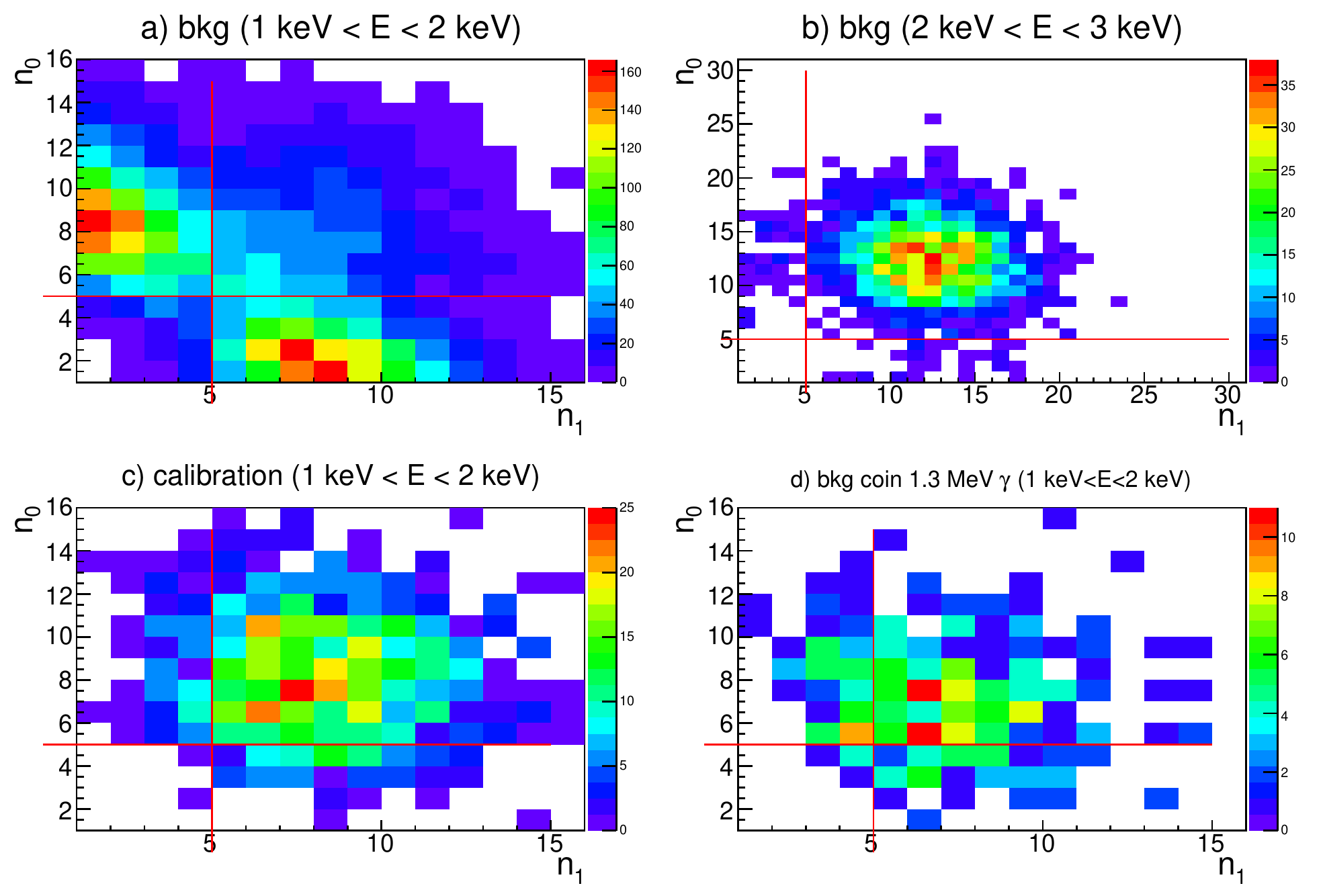}
\caption{Distribution in the (n$_0$, n$_1$) plane (i.e., number of peaks detected by the 
algorithm for each of the two PMT signals from the same module) for: (a) $\sim$10\% unblinded data background events with 
energy between 1 and 2~keV; 
(b): same but for events with energy between 2 and 3~keV; (c) $^{109}$Cd calibration events with energy between
1 and 2~keV; (d): bulk scintillation 0.87~keV events from {\Na} decay selected by coincidence with a 1274.5~keV 
deposition in another crystal (whole 1 year statistics).}
\label{fig:asy}
\end{figure*}

Below 2~keV, we observe a different population of background events whose main characteristic is 
a strong asymmetry in the energy partition among both PMTs (see an example in Figure~\ref{fig:pulseExamples} panel (d)). 
These events, which do not have the timing characteristics of afterpulses or pulse tails, are also
observed by  the COSINE collaboration \cite{Adhikari:2017esn}, and could be related to 
the light emission observed in PMTs by other experiments \cite{Akimov:2015cta,Li:2015qhq}.
They can be easily 
identified by the difference in the number of peaks that our algorithm finds in the 
signals of the two PMTs of every detector ($n_0$ and $n_1$). In 
the panel (a) of Figure~\ref{fig:asy} we show the distribution of $n_0$ and $n_1$ 
for background events corresponding to the $\sim$10\% unblinded data with energies between 1 and 2~keV.
We observe that most of them feature a very asymmetric distribution, with few peaks (one or two) in one PMT
and more than five in the other one\footnote{Note that, with more than 7~phe/keV/PMT, the probability of having 1 or 2 phe in one PMT at 1~keV is around 3\%.}.
These events, only visible below 2~keV in dark matter runs (see Figure~\ref{fig:asy} panel (b)), are not present neither
in the $^{109}$ Cd calibration runs (Figure~\ref{fig:asy} panel (c)), 
nor in the 0.87~keV bulk scintillation population from {\Na} (Figure~\ref{fig:asy} panel (d)), 
so they are presumably fake light events produced at or near the PMTs. 
\par
This asymmetric population is not distinguishable from the bulk scintillation one by the PSV parameter,
so to reject them we use a cut based on the number of peaks detected at every 
PMT: we require an event to have more than four peaks detected by our algorithm at every PMT signal
(see red lines in Figure~\ref{fig:asy}). 
The efficiency of this cut is calculated on calibration runs as explained in next section.

\section{Efficiency}
\label{sec:eff}

We can factorize the total detection efficiency of ANAIS-112 as the product of
the trigger efficiency (probability that an event is triggered by the DAQ system),
and the efficiency of the PSV and asymmetry cuts.
For the muon vetoed events, we do not need an efficiency, but we calculate a 1~s dead time for every muon triggering the veto system, that is added to the 
total dead time of the DAQ system and then, discounted from the total live time. 
The efficiency of the anticoincidence cut can be safely taken as one, 
given that, with a total rate below 5~Hz, the probability of an accidental 
coincidence between two detectors in the digitization window (1.2~$\mu$s) is negligible.

\subsection{Trigger efficiency}
\label{subsec:triggEff}

As explained in Section~\ref{sec:DAQ}, the hardware threshold is low enough to efficiently trigger
at the phe level each PMT. Nevertheless, at very low energy the coincidence requirement 
reduces the efficiency of the trigger, as very few phe are collected and the mean 
time interval between them increases.
\par
\begin{figure}[htbp]
\centering
\includegraphics[width=0.5\textwidth]{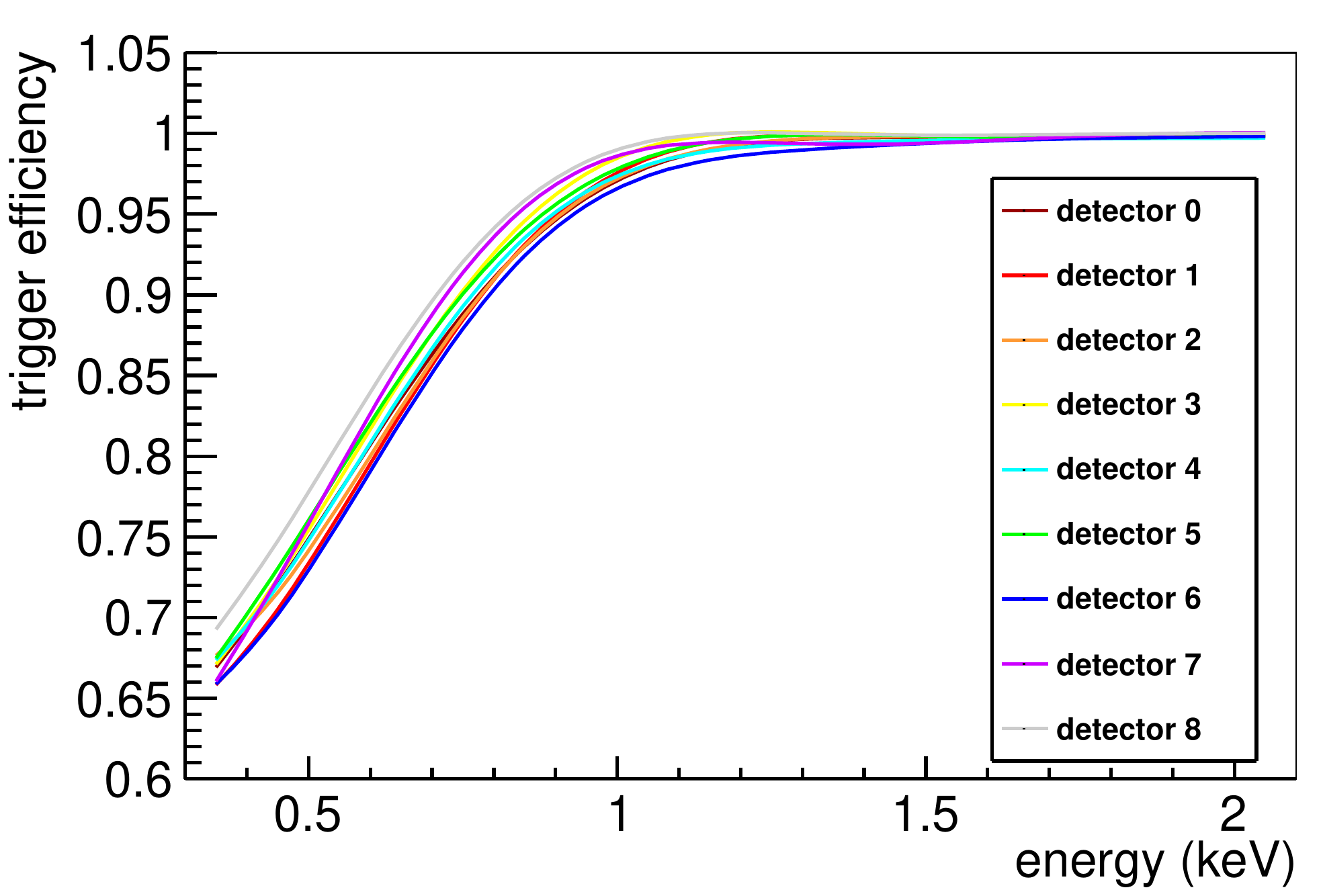}
\caption{Trigger efficiency for every detector, calculated from the measured light collected by a Monte Carlo technique. Colors referenced are available in the online version of the paper. 
}
\label{fig:triggEff}
\end{figure}

For ANAIS-112 we do not expect a significant trigger efficiency reduction above 1~keV, 
given the high light collection and large coincidence window (200~ns). 
Trigger efficiency was experimentally  measured for previous ANAIS prototypes in set-ups having two or three modules and a slightly different readout: all the modules traces were stored independently on the module triggering. {\Na} and {\K} high energy events allowed to identify real coincidence events with a very low energy deposition in one of the modules, and by selecting those effectively triggering our DAQ to calculate the trigger efficiency~\cite{Cuesta:2014vna}. Very high trigger efficiency was observed then, and as far as the same DAQ hardware/software is used in ANAIS-112, we do not expect differences in trigger efficiency in the present set-up. 
To quantify the trigger efficiency we use a Monte Carlo technique:
for every detector we simulate 2$\times10^4$ events of random energy $E$ between 0 and 10~keV. 
For each simulated event, the number of phe in a PMT is selected according to a Poisson distribution 
with mean given by $E\times$LC$_{[det, PMT]}$, where LC$_{[det,PMT]}$ is the measured light collection 
for that detector and PMT (see Table~\ref{tab:lightyield}).
The phe are simulated as Gaussian peaks of widths between 5 and 6~ns, depending on the detector.
The arrival time of every phe is sampled from an exponential PDF 
with $\tau$=230~ns, as for NaI(Tl), and their amplitudes are 
sampled from the SER amplitude distribution of the corresponding PMT. Then, we calculate the trigger efficiency 
as a function of energy
as the number of events having the first phe identified in every PMT above the trigger threshold and within 200~ns divided by the total number of simulated events generated for that energy. 
The result is displayed in Figure~\ref{fig:triggEff}. As expected, the efficiency is larger than 98\% down to 1~keV. The small differences among the detectors are only due to the spread in light collection and phe width. This MC will be very useful in following sections to check the efficiencies of PSV and asymmetry cuts, which are more difficult to check experimentally than trigger efficiency and that could be more affected by systematics.

\subsection{PSV cut efficiency}
\label{subsec:psaEff}

From Eq.~\ref{eq:PSVar}, the definition of the {\PSVar} variable provides
the efficiency for every cut value, but it is only valid in the energy region
in which the means and covariances are calculated, i.e., between 1 and 2~keV.
Instead of changing the variable definition for every energy range, we keep 
the same definition and calculate the efficiency of the cut as a function
of energy. In order to do so we use the {\Na} and {\K} low energy populations
selected in coincidence with a 1274.5~keV or 1460.8~keV $\gamma$ in another 
detector. 
\par
\begin{figure*}[ht] 
\centering
\includegraphics[width=0.38\textwidth]{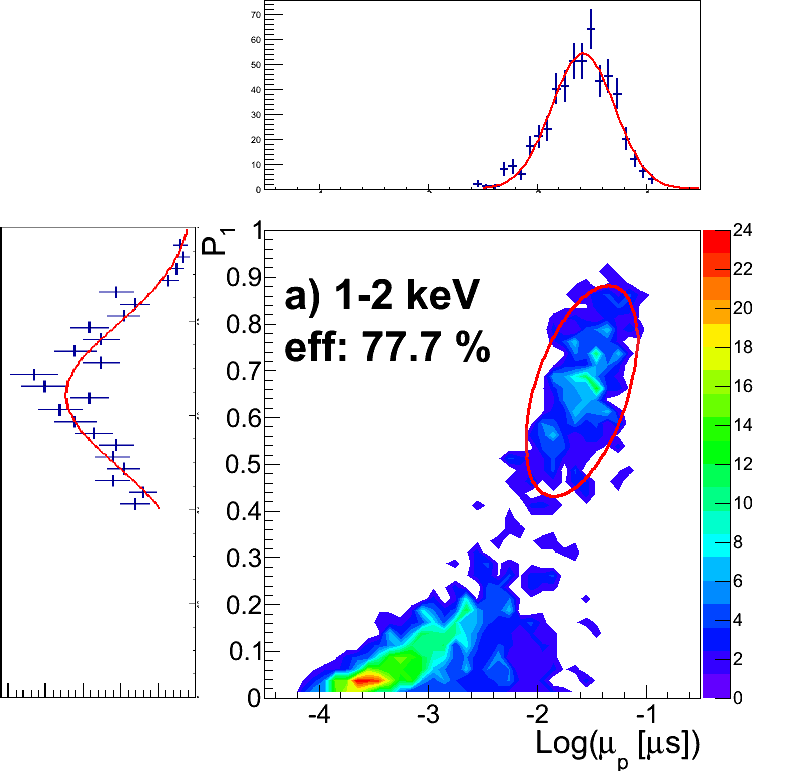}
\includegraphics[width=0.58\textwidth]{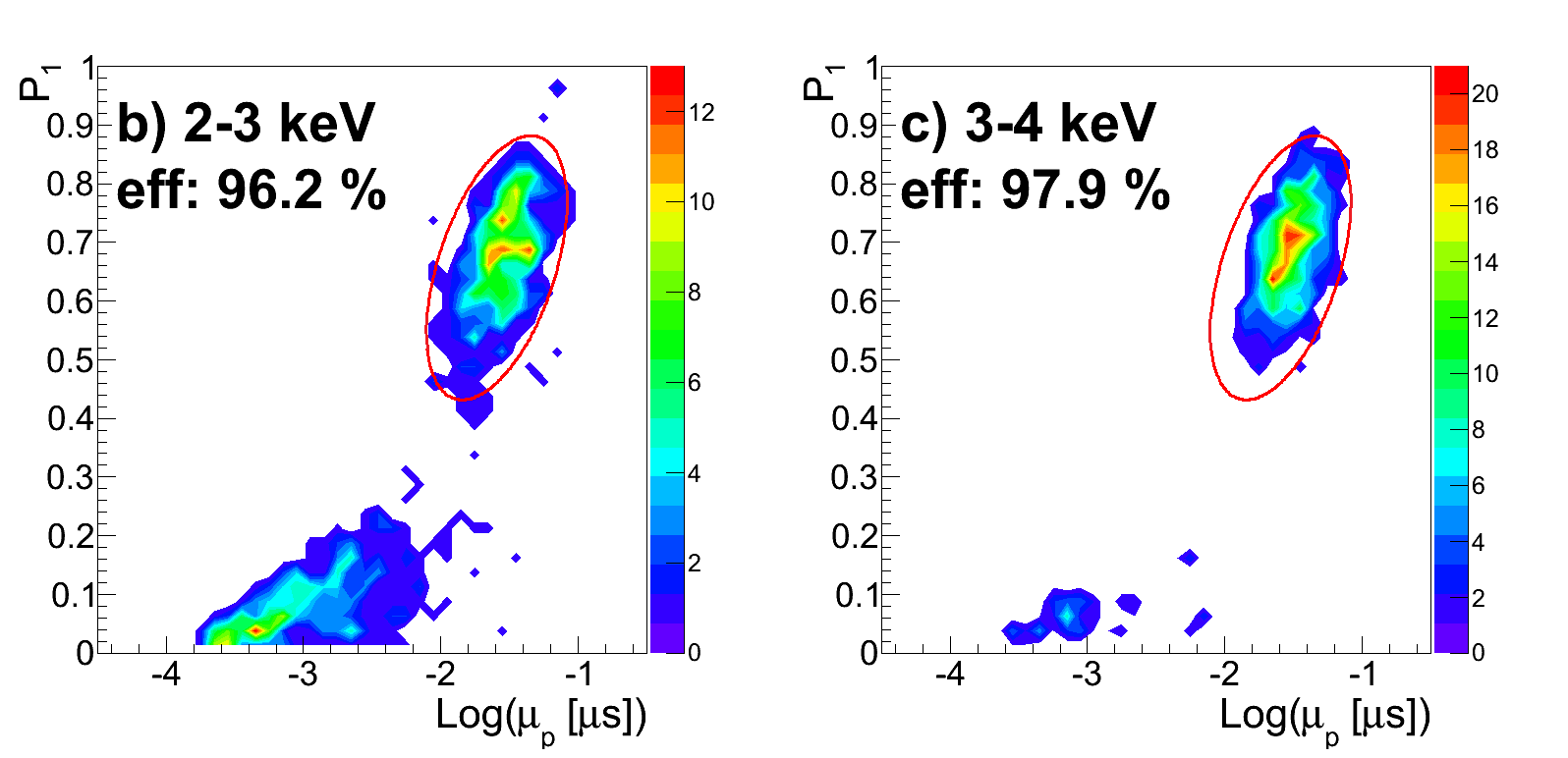}
\caption{Distribution of the {\Na/\K} events at low energy selected by coincidence with  
a 1274.5~keV or 1460.8~keV $\gamma$ in another detector in the ({\Puno}, {\LFM}) plane.
The three panels correspond to 1-2, 2-3 and 3-4~keV energy regions. The red line corresponds to {\PSVar}=3. The {\PSVar} variable is defined in the [1-2]~keV range and the 
cut is fixed for an efficiency of 77.7\% in that energy region. Then the efficiency is 
calculated for the other energy ranges as explained in the text. In panel (a) we show the projections of {\Puno} and {\LFM} variables in left and top panels, respectively, and the fitting to single Gaussian lineshapes, considering only {\Puno}>0.4 events, which supports the bivariate analysis carried out. }  
\label{fig:mvEffCalc}
\end{figure*}
\begin{figure}[htbp]
\centering
\includegraphics[width=0.5\textwidth]{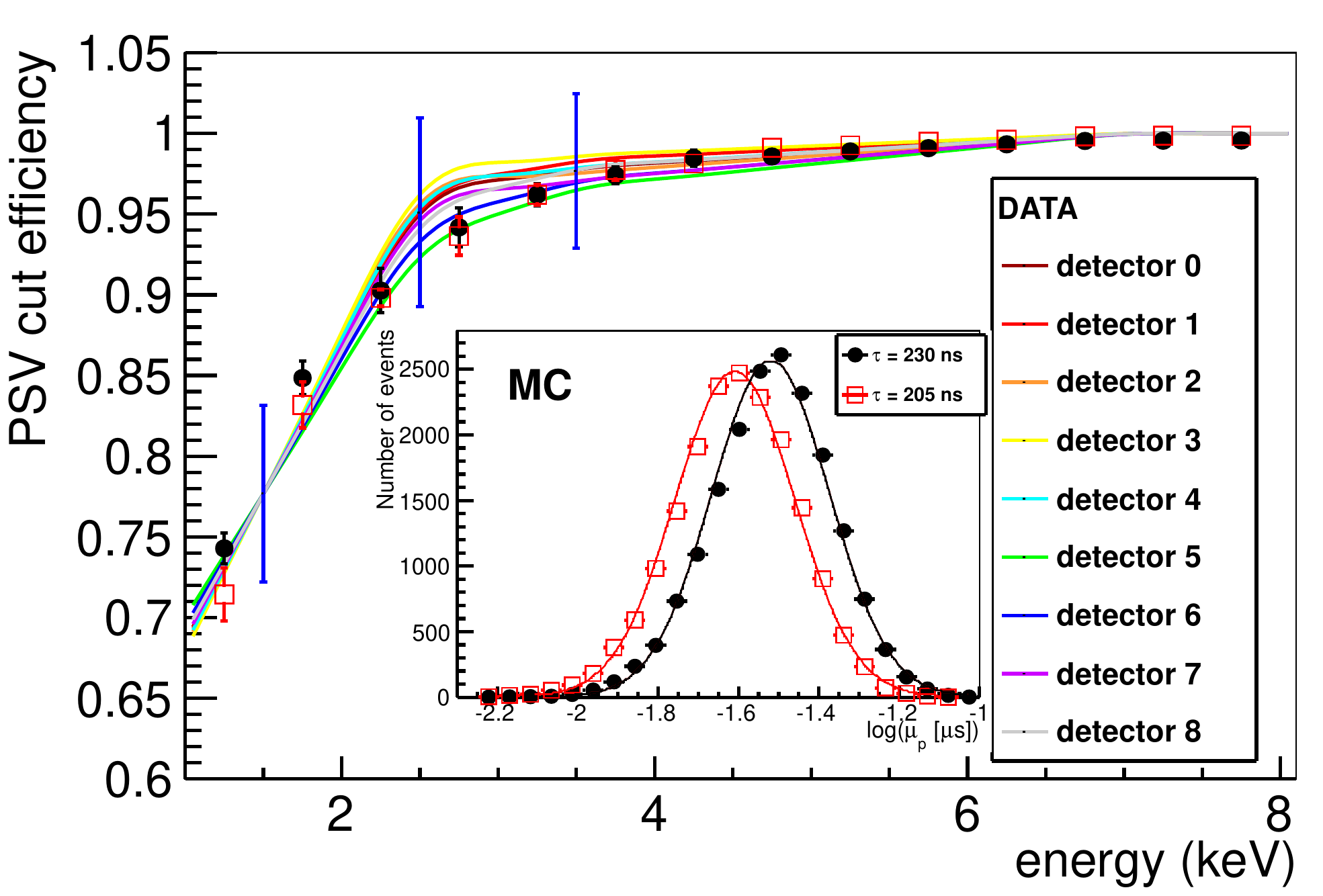}
\caption{Continuous lines: PSV cut efficiency for every detector, calculated from the distribution 
of the {\Na/\K} bulk events selected in coincidence with a high energy $\gamma$. The average of the statistical uncertainties for these efficiencies is shown in blue (see text for more information).
Closed black dots: PSV cut efficiency calculated from a MC simulation of the 
scintillation pulses with $\tau$=230~ns. Red open squares: the same, but for a MC simulation 
with $\tau$=205~ns. Inset: distribution in {\LFM} of the 3-4~keV events of both MC simulations. Colors referenced are available in the online version of the paper. }
\label{fig:mvEff}
\end{figure}

Figure~\ref{fig:mvEffCalc} shows the distribution of these 
low energy events of detector D1 in the PSA parameters 
plane for three energy ranges. In the first one (1-2~keV) we calculate
 $\bm\mu$ and $V$ and fix {\PSVarcut}=3 (red ellipse) in order to 
have a 77.7\% efficiency at that energy. This calculation is done for every detector independently. 
As we have commented in Section~\ref{sec:lecal}, the {\Na} and {\K} populations are polluted with 
fast (Cherenkov-like) events in the PMT in coincidence with a $\gamma$ in one of the crystals~\cite{Amare:2018ndh}. 
Therefore, to avoid these events we require {\Puno}>0.4  before calculating $\bm\mu$ and $V$ (see projections along the P$_1$ and {\LFM} axis in Figure~\ref{fig:mvEffCalc} panel (a)).
Then, we compute the efficiency in 1~keV bins
for larger energies as the number of {\Na/\K} events in the ellipsoid 
divided by the total number of {\Na/\K} events having {\Puno}>0.4. Results are 
shown in Figure~\ref{fig:mvEff} (continuous lines). Results are very similar for all the ANAIS-112 modules,
with efficiency increasing from 77.7\% at 1.5~keV (fixed by the {\PSVarcut} choice) up to 
95\% at 2.5~keV and 98\% at 3.5~keV. The statistical uncertainties in the efficiency, due to the small number of {\Na/\K} coincident events in each module, are of the order of  5\% for all the modules. Although we calculate these uncertainties individually and propagate them conveniently to estimate the final uncertainties in the total efficiencies, they have been averaged to be shown in Figure~\ref{fig:mvEff}.
\par

This procedure to estimate acceptance efficiencies is not valid for energies larger than $\sim$4~keV, as there are very few events selected by the coincidence condition (see Figure~\ref{fig:coin}).
In order to extend the efficiency calculation up to 10~keV and 
double-check the results,
we use the MC simulation described in Section~\ref{subsec:triggEff}.
To ensure consistency between the MC and the data, first we check that the distributions of the PSA 
parameters are compatible. The distribution in {\LFM} is shown in the inset of Figure~\ref{fig:mvEff}
(black dots), to be compared with the upper panel in Figure~\ref{fig:mvEffCalc}. 
With the simulated events we follow the same method described before, i.e., 
we compute $\bm\mu$ and $V$ between 1 and 2~keV
and construct the ellipse {\PSVar}=3. Then we calculate the efficiency at higher energies as the number of 
events inside the ellipse divided by the total number of simulated events. The result
is displayed in Figure~\ref{fig:mvEff} as black closed dots, showing an excellent agreement with 
the efficiencies calculated on the {\Na/$^{40}$K} populations. 
\par 
Finally, it is worth noting that our sample population for the efficiency calculation
is originated from electron recoils (ER), 
while for most DM models the expected signal comes from nuclear recoils (NR).
It is well known that ER and NR
feature slightly different scintillation constants in NaI(Tl) at very low energy. 
For example, in \cite{Kim:2018wcl} the KIMs collaboration
has measured a variation in the {\LFM} parameter of 0.08 between the mean values for ER and NR at 2-4~keV, which 
roughly corresponds to a factor 0.9 in $\tau_{scint}$. 
We simulate a MC-NR population with a scintillation constant 
$\tau_{NR}\sim$0.9$\times\tau_{ER}\approx$205~ns, and calculate the efficiency for the NR population
when the {\PSVar} variable is computed on the ER one. Results are shown in Figure~\ref{fig:mvEff} as open red squares. The largest difference in the efficiency is 0.03 in the region from 1 to 2~keV, that we will take 
as a systematic error in our efficiency estimate.

\subsection{Asymmetry cut efficiency}
\label{subsec:asyEff}

\begin{figure}[h]
\centering
\includegraphics[width=0.5\textwidth]{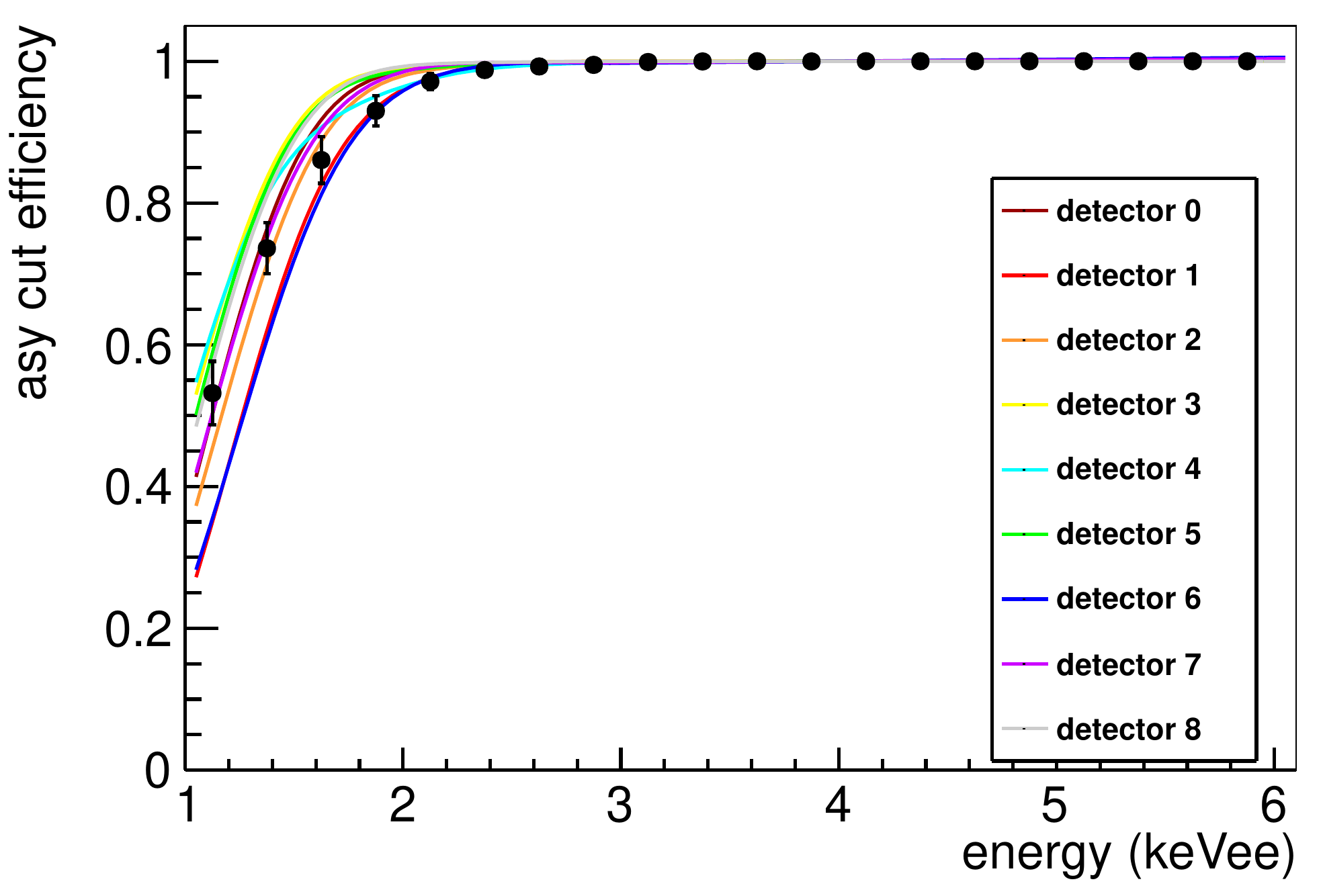}
\caption{Continuous lines: asymmetry cut efficiency for every detector, calculated in calibration runs.
Black dots: the same, but calculated with the MC simulation. Colors referenced are available in the online version of the paper. 
}
\label{fig:asyEff}
\end{figure}

We evaluate the efficiency of the asymmetry cut (more than 4 phe identified by the algorithm at every PMT) 
using all the $^{109}$Cd calibration events accumulated along the whole year calibration runs to have enough events in the 1-2~keV energy region. In order to do so we count the number of events surviving 
the cut and divide by the total number of events, as a function of energy. Results are displayed in Figure~\ref{fig:asyEff}. The efficiency is equal to one down to 2~keV, and then it decreases steeply down to 
0.3-0.55 at 1~keV (depending on the detector). We have chosen this energy, 1~keV, as common analysis threshold for all the modules, considering the acceptance efficiency of this filtering, which is the most restrictive. 
The result is in good agreement with expectations from Poisson statistics, taking into account 
that the peak-detecting algorithm cannot identify all the phe, as they overlap in the waveform. 
We have estimated the phe-finding efficiency of the algorithm  with the MC simulation described 
in Section~\ref{subsec:triggEff}: it properly identifies between 70 and 80\% of the phe at 1~keV, 
and then the efficiency decreases 
linearly with energy (as the probability of overlap increases) down to 40-50\% at 5~keV.
We show also in Figure~\ref{fig:asyEff} the asymmetry cut efficiency calculated from the MC, averaged 
for all detectors (black dots).

\begin{figure}[b]
\centering
\includegraphics[width=0.50\textwidth]{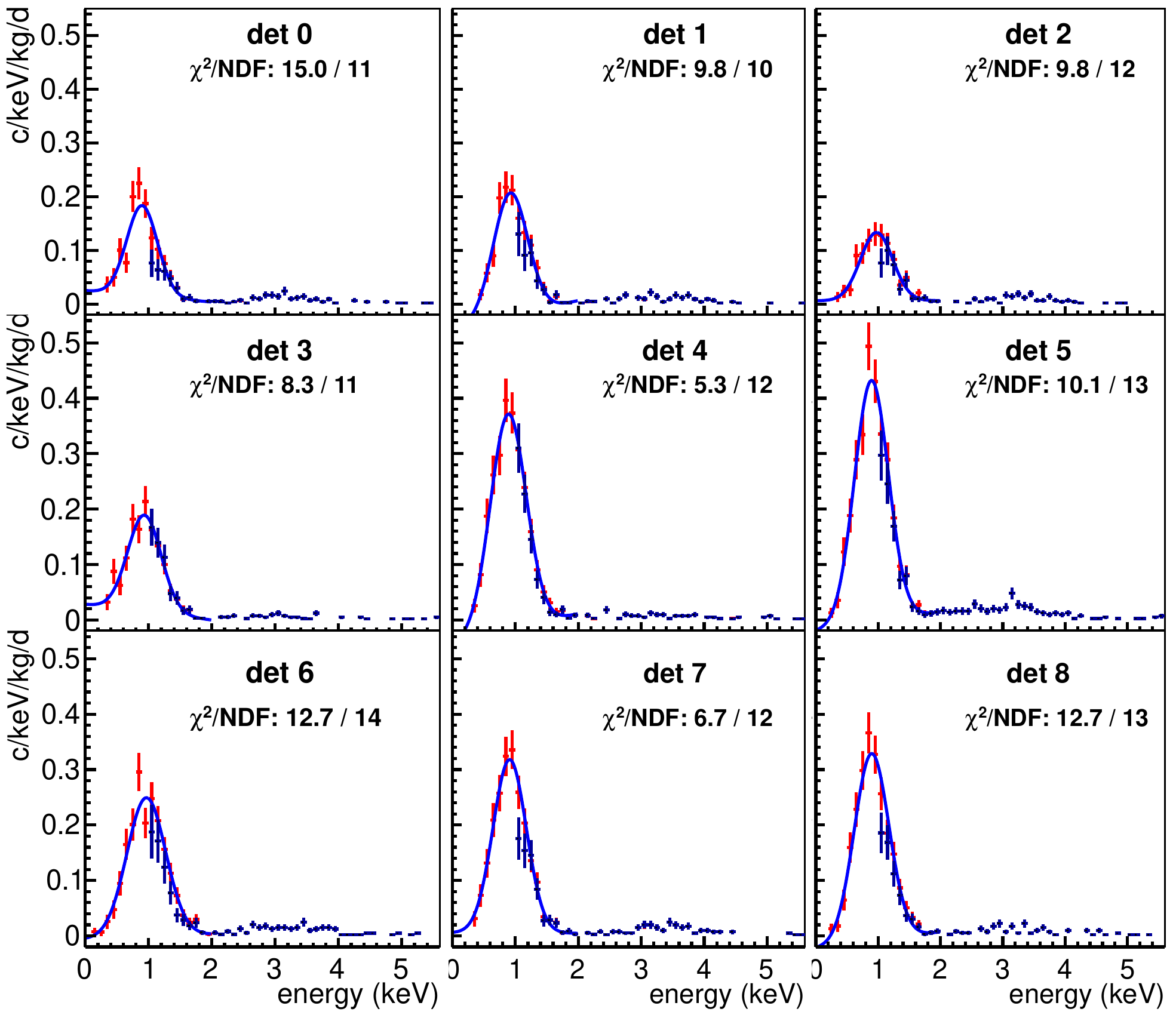}
\caption{Low energy spectra of the nine modules in coincidence
with a high energy gamma between 1215 and 1335~keV in a second module.
Data correspond to the full first year of ANAIS-112.
Red points: Data selected by the PSV cut and corrected with the corresponding efficiency (PSV and trigger).
Black points: In addition to the previous one, we apply the asymmetry cut and correct with the 
corresponding efficiency. Blue line: gaussian fit to the red points. The chi-squared of the fit is also displayed. Colors referenced are available in the online version of the paper. 
}
\label{fig:NaAsyEff}
\end{figure}

We also check the consistency of the procedure with the {\Na} and {\K} populations selected 
by coincidence. These events are not polluted with the asymmetric population, as we have shown 
in Figure~\ref{fig:asy}.
Figure~\ref{fig:NaAsyEff} compares for each detector the coincidence low energy spectra obtained after 
applying the PSV cut and correcting for the corresponding efficiency (red points) with the one obtained
after applying also the asymmetry cut and correcting for the corresponding efficiency (black points).
The agreement is excellent for all detectors down to 1.2~keV. 
Below this energy there is a discrepancy for some of the detectors
(mainly D0, D1 and D6) up to  20\% at 1~keV. 
We take into account this effect as a systematic error in the asymmetry cut efficiency.

\subsection{Total detection efficiency}
\label{subsec:totalEff}

\begin{figure*}[th]
\centering
\includegraphics[width=0.75\textwidth]{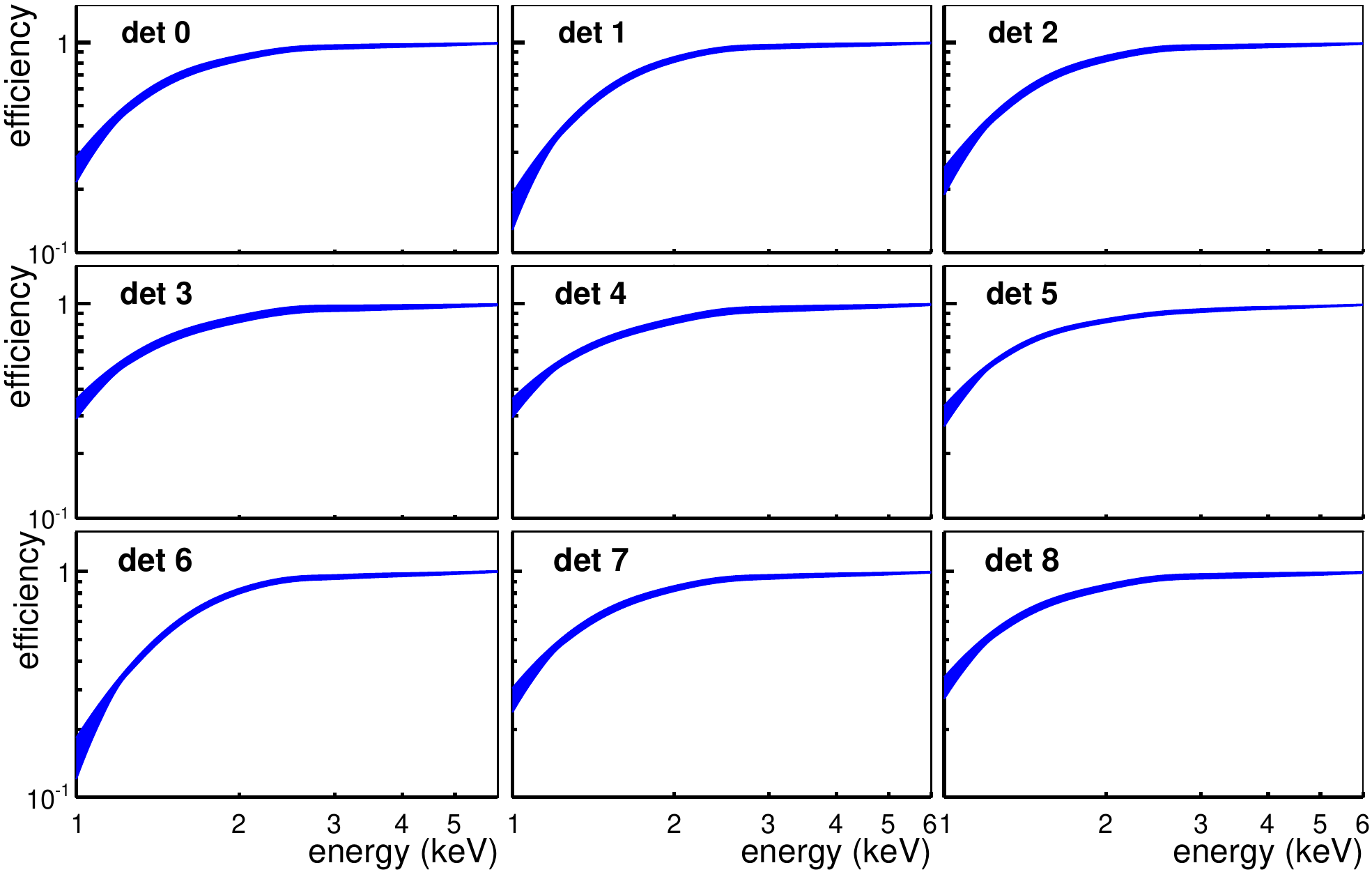}
\caption{Total efficiency, obtained as the product of the trigger, pulse shape 
cut and asymmetry cut efficiencies. The statistical and systematic errors estimated for all these efficiencies, as described in Section~\ref{sec:eff}, have been combined in quadrature and are shown in the width of the lines. }
\label{fig:totalEff}
\end{figure*}

Finally we multiply the three efficiencies to obtain the total efficiency for the selection of dark matter compatible events in every ANAIS-112
detector, that is shown in Figure~\ref{fig:totalEff}. 
The estimated systematic errors in these efficiencies, previously commented, are shown in the width of the lines. 
At 1~keV the efficiency 
ranges from 0.15 to 0.35, depending on the detector, increases up to 0.8 at 2~keV for all the ANAIS-112 modules and is nearly 1 at 4~keV.

\section{Low energy spectrum}
\label{sec:low_energy_spectrum}

\begin{figure*}[bh]
\centering
\includegraphics[width=0.76\textwidth]{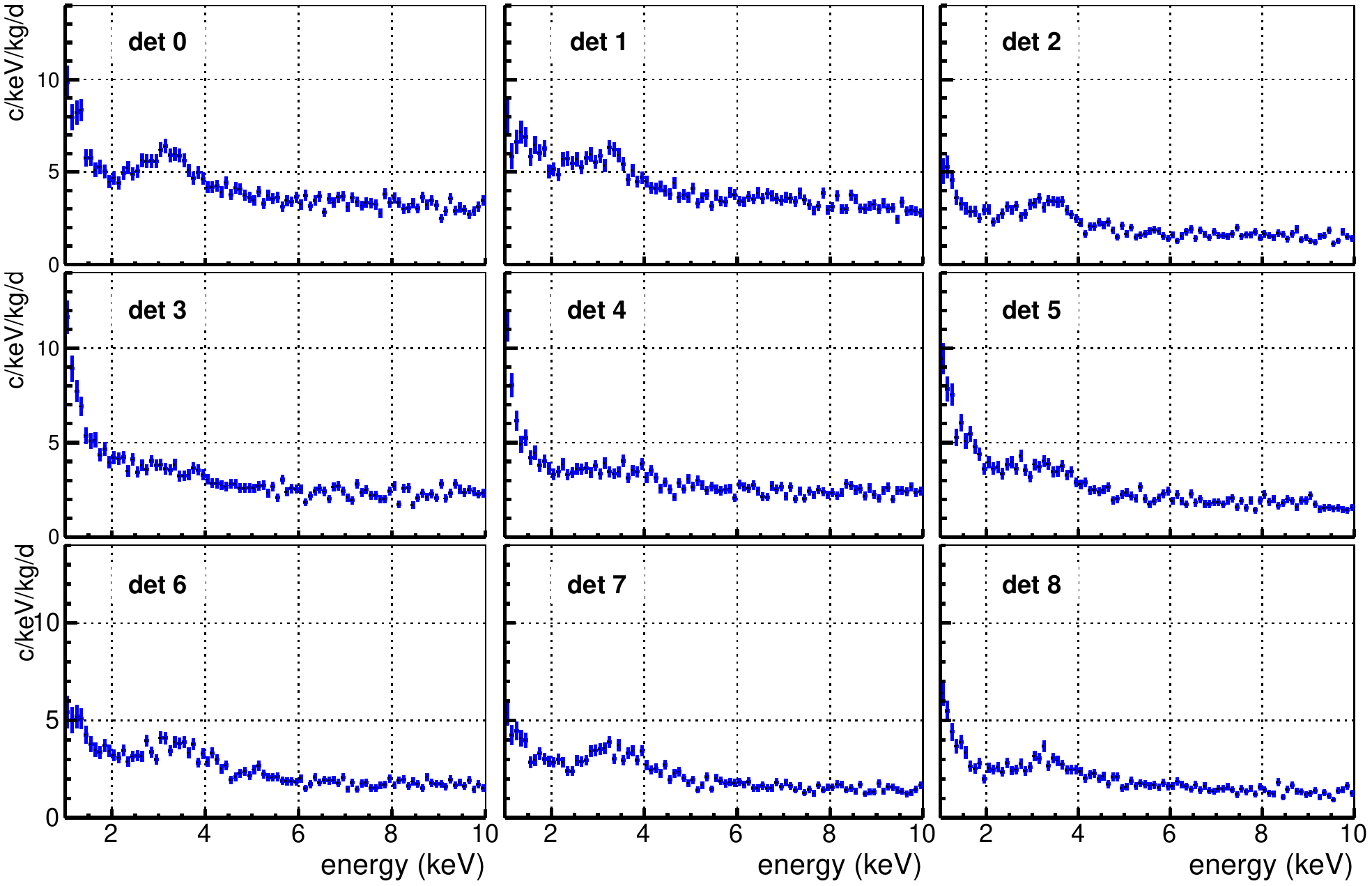}
\caption{Anticoincidence energy spectrum measured in the ROI for each detector, corresponding to the $\sim$10\% unblinded data.
The {\K} peak at $3.2$~keV is clearly visible.
}
\label{fig:bkglow}
\end{figure*}

Figure~\ref{fig:bkglow} shows the energy spectra in the ROI corresponding to the $\sim$10\% unblinded data for each of the detectors in anticoincidence (single hit events)
after event selection and efficiency correction.
A detailed analysis of the different background contributions, based on MC simulations of quantified radioactive contaminants, 
is reported in the companion paper~\cite{Amare:2018ndh}.
The background in the ROI is dominated by the internal contamination of the NaI(Tl) crystals themselves, 
and varies depending on the detector as do their powder purification procedures, crystal growing protocols and starting time of 
underground storage. The main internal background contributions come from {\K}, {\Pb} (contaminants present in the starting powder, and presumably entering into the powder purification or growing procedures through $^{222}$Rn, respectively), and 
{\Na} and $^3$H (having cosmogenic origin).
In all detectors the {\K} peak at $3.2$~keV is clearly visible. The background level at 2~keV ranges from 2 to 5~counts/keV/kg/day,
depending on the detector, and then increases up to 5-8~c/keV/kg/day at 1~keV. Corresponding values for the single-hit total spectrum of DAMA/LIBRA phase 2 are 1 and 1.8~counts/keV/kg/day, respectively~\cite{Bernabei:2018yyw}.
\par
\begin{figure}[t]
\centering
\includegraphics[width=0.5\textwidth]{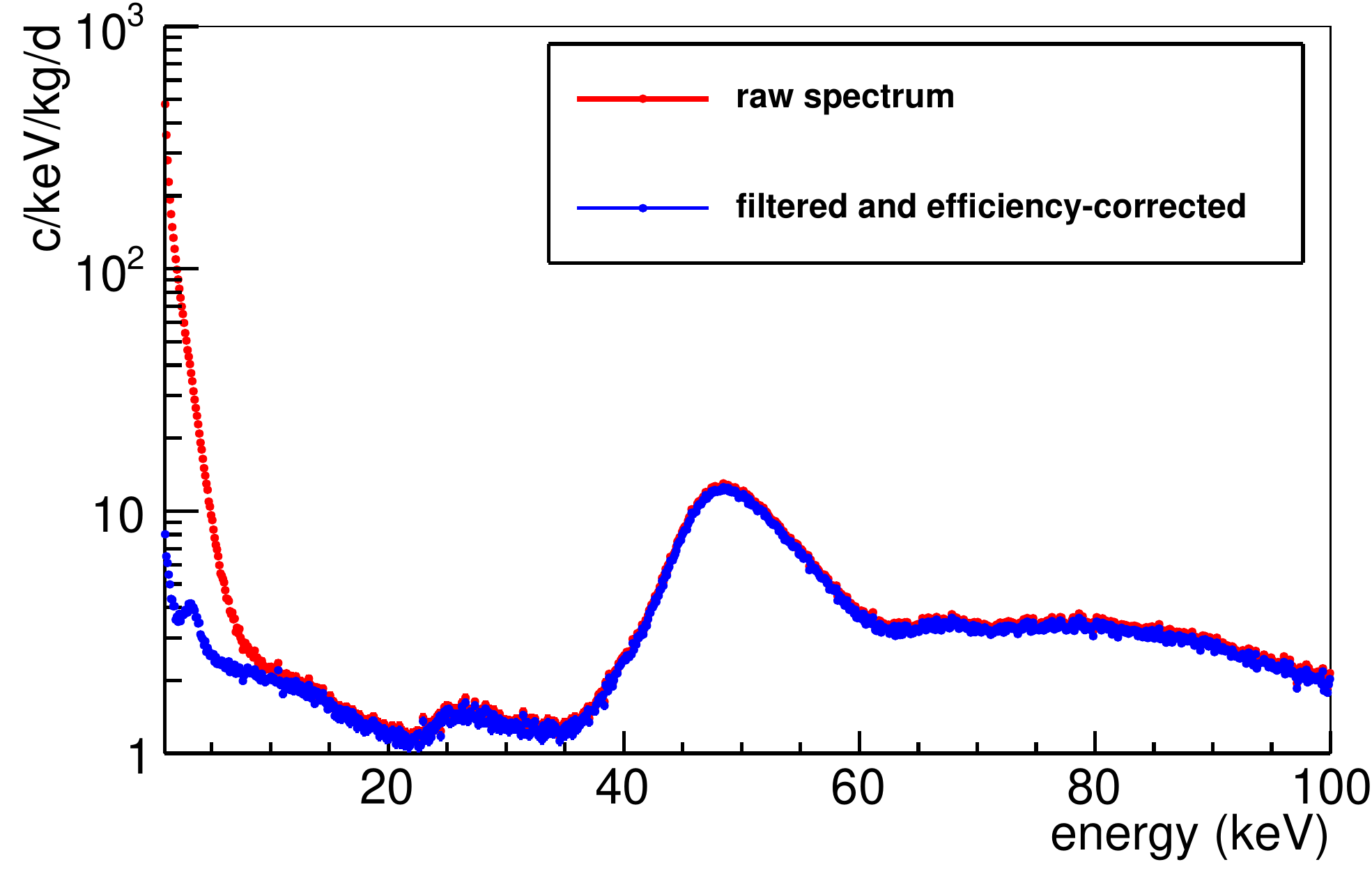}
\caption{Red points: total anticoincidence energy spectrum (all detectors) corresponding to the $\sim$10\% unblinded data in the energy region from 0 to 100~keV. 
Blue points: same, but after applying the event selection criteria and correcting for efficiency. Main features identified in this spectrum correspond to the decay of isotopes found in the crystals: $^{210}$Pb (around 50~keV), $^{109}$Cd and $^{113}$Sn (from 20 to 30~keV) of cosmogenic origin, and $^{40}$K (around 3~keV). Colors referenced are available in the online version of the paper. }
\label{fig:bkgmedium}
\end{figure}

 
We also display in Figure~\ref{fig:bkgmedium} (blue points) the total anticoincidence background (adding up all 9 detectors) in the energy region below 100~keV for 
the unblinded data. This energy region is also dominated by the internal contamination of the crystal, 
with a prominent peak at $\sim$50~keV coming from  {\Pb}. The small peak at $\sim$26~keV originated by $^{109}$Cd and $^{113}$Sn cosmogenic isotopes (T$_{1/2}$=461.9 and 115.1~days, respectively) is also visible. 
Above 100~keV (see Figure~\ref{fig:heSpc}) the background is dominated by the PMTs radioactive contamination.
In order to illustrate the magnitude of the rejected populations, 
we display in the same plot (Figure~\ref{fig:bkgmedium} red points) the data before any filtering procedure (apart from the 
anticoincidence requirement). As we have commented before, non-bulk scintillation events clearly
dominates the rate in the ROI and their correct identification is a key point in this kind of experiments.

\section{ANAIS-112 stability}
\label{sec:stability}

The annual modulation analysis requires to have under control all the environmental conditions which could produce any modulation in the detectors performance. Because of that, many of the environmental parameters that could affect ANAIS-112 setup have been monitored along this first year of data taking in order to evaluate the possible influence in any modulation which could be observed. In this section we also present an study of stability for several background events populations which can be taken as control samples in order to discard a possible relationship among a modulation in the ROI for dark matter analysis and the known sources of background. 

\begin{figure}[htbp]
\centering
\includegraphics[width=0.53\textwidth]{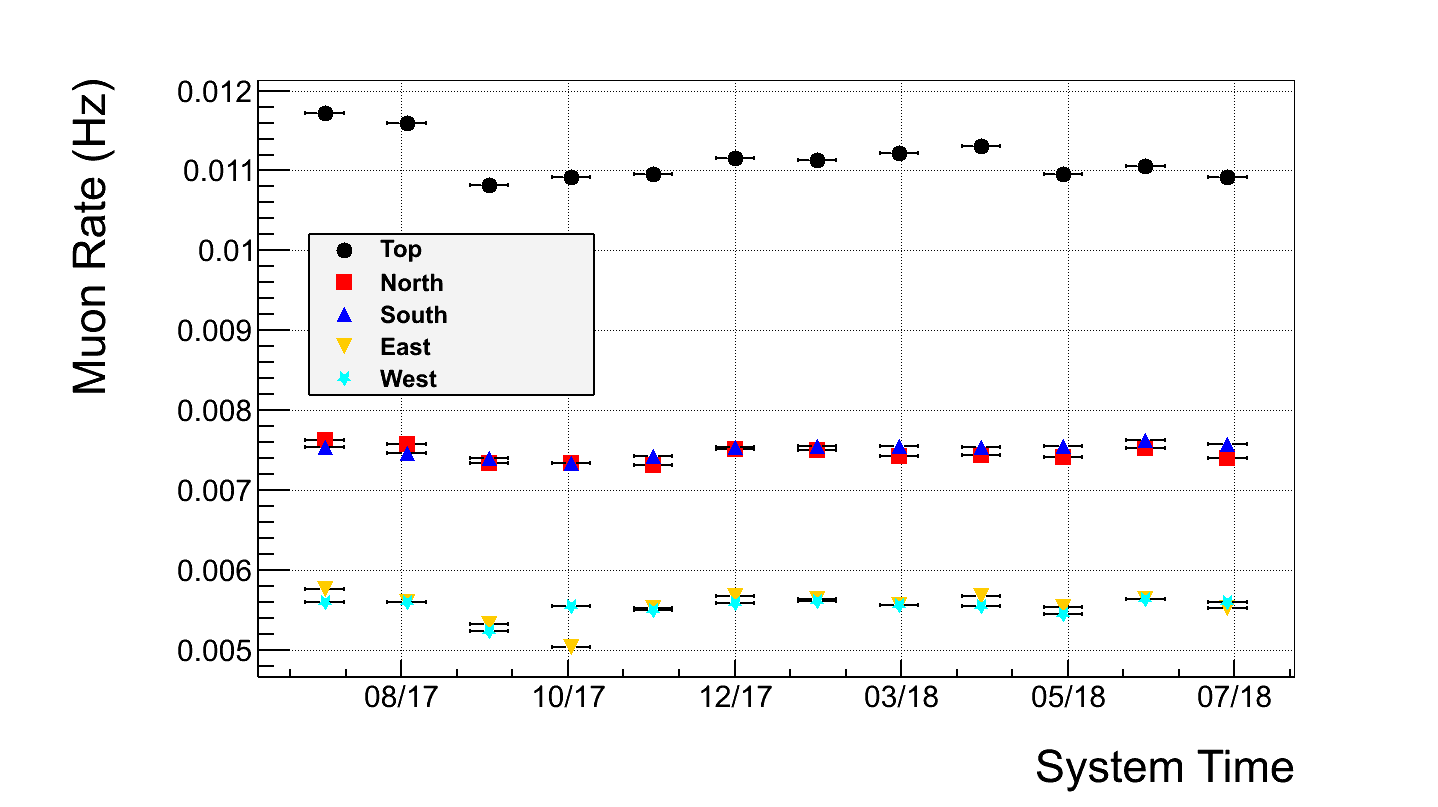}
\includegraphics[width=0.53\textwidth]{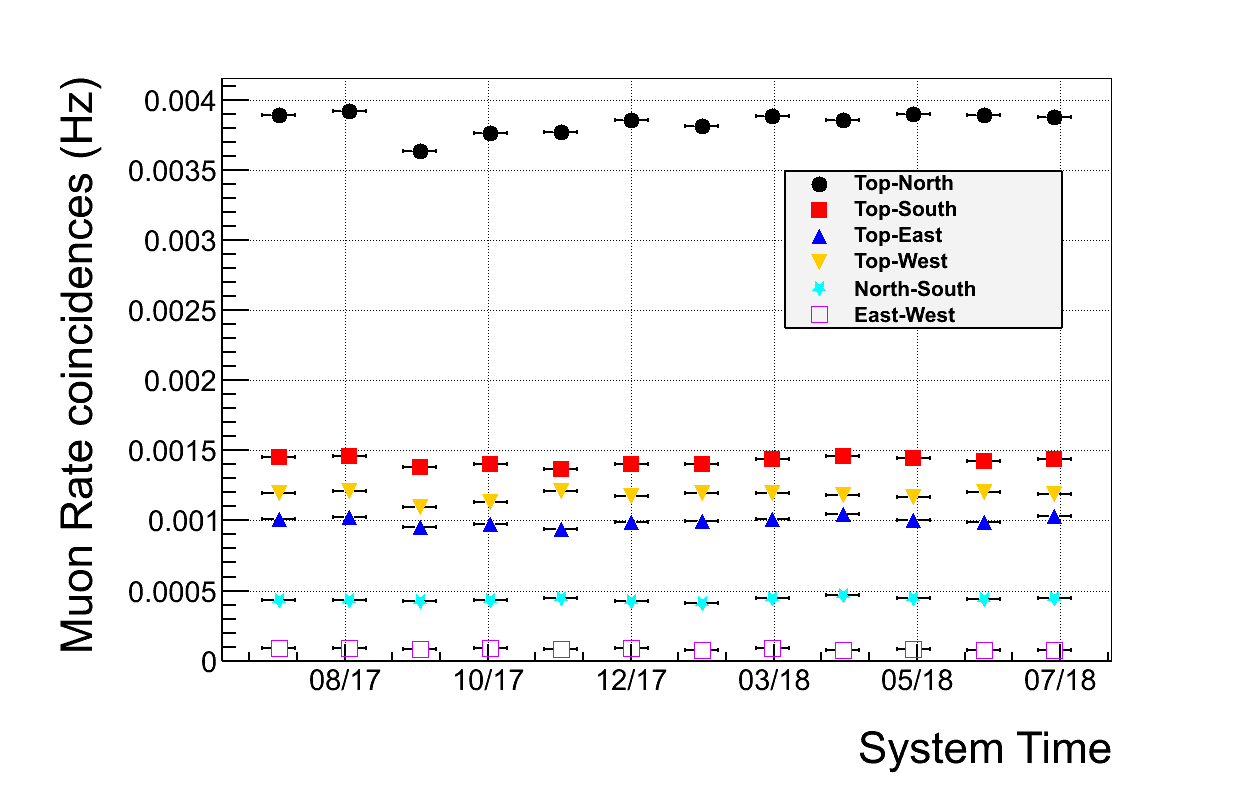}
\caption{Evolution of the muon rate. Upper panel: rate on each side of the scintillator veto system. 
Lower panel: muon coincidences rate between different sides of the scintillator Veto System. 
Rates are calculated per month. See text for more information. Colors referenced are available in the online version of the paper.  
}
\label{fig:muonrate_month}
\end{figure}

Muon interaction rate onsite is being continuously monitored (see Section~\ref{sec:experimental}). 
Figure~\ref{fig:muonrate_month} shows the muon rate measured by each side of the Veto System (rate after PSA), and the muon coincidence rate between different sides, expressed both in muons/second, and calculated on a monthly basis. We had a problem with the QDC module used for the PSA required to identify muons in the scintillator vetoes, which produced a wrong identification of muons a few days in October-November 2017. Most of the data could be recovered, but for a few of the plastic scintillators, being noticed as an apparent decrease in muon rate in the East side of the scintillator Veto System at the end of October 2017.

\begin{figure}[htbp]
\centering
\includegraphics[width=0.5\textwidth]{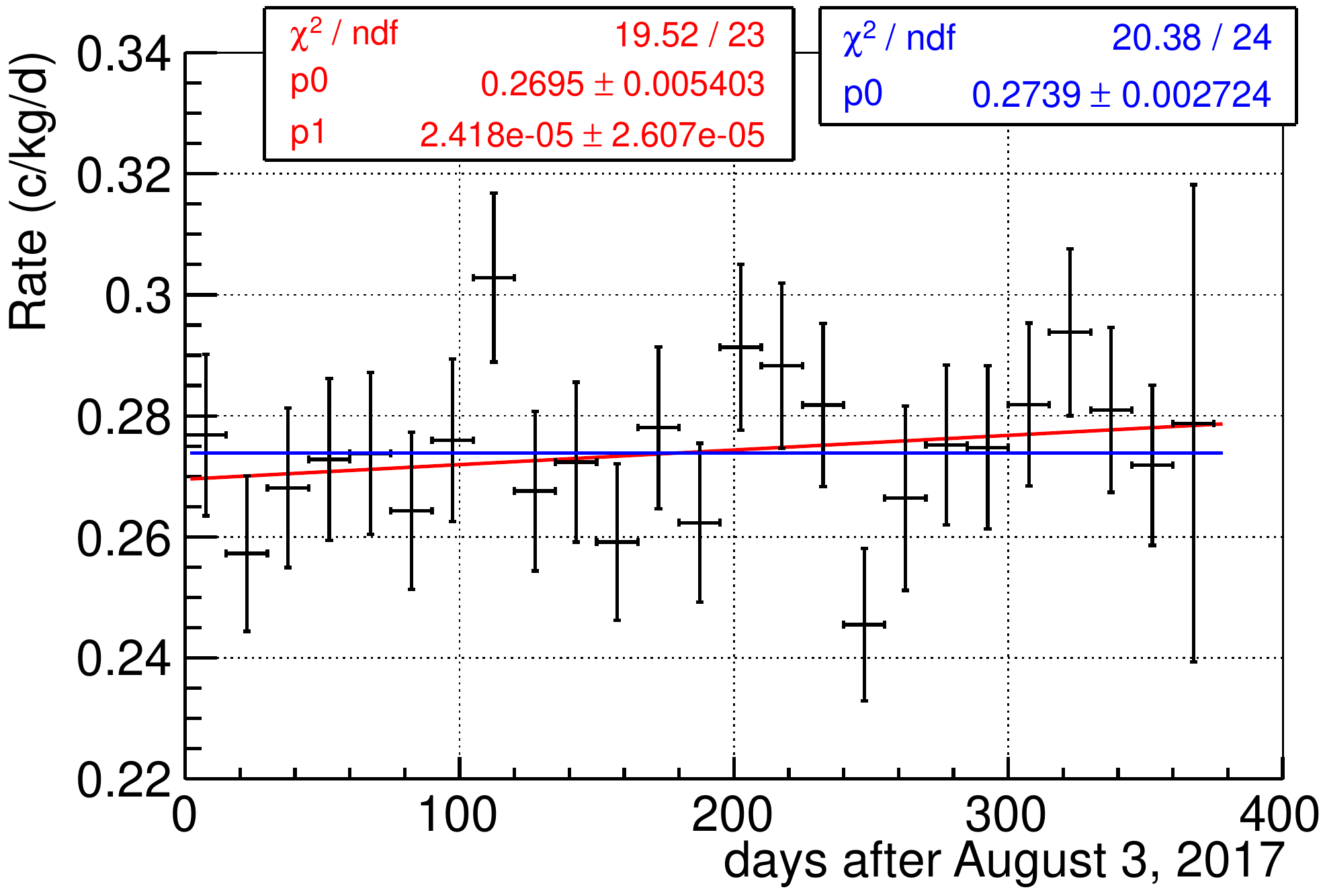}
\caption{Evolution of the rate of coincidences with a high energy gamma between 1400 and 1510~keV 
in a second module, after filtering (PSV only) and efficiency correction in the energy range from 2 to 5~keV, 
corresponding to events attributed to $^{40}$K decays in the crystal bulk. Rates are averaged every 10~days and 
fitted to a constant rate (blue) or to a straight line (red). The results of the fits are also shown in the plot 
with the corresponding color. In both cases compatibility with constant rate is obtained. Colors referenced are available in the online version of the paper. 
}
\label{fig:KRate}
\end{figure}

\begin{figure}[htbp]
\centering
\includegraphics[width=0.5\textwidth]{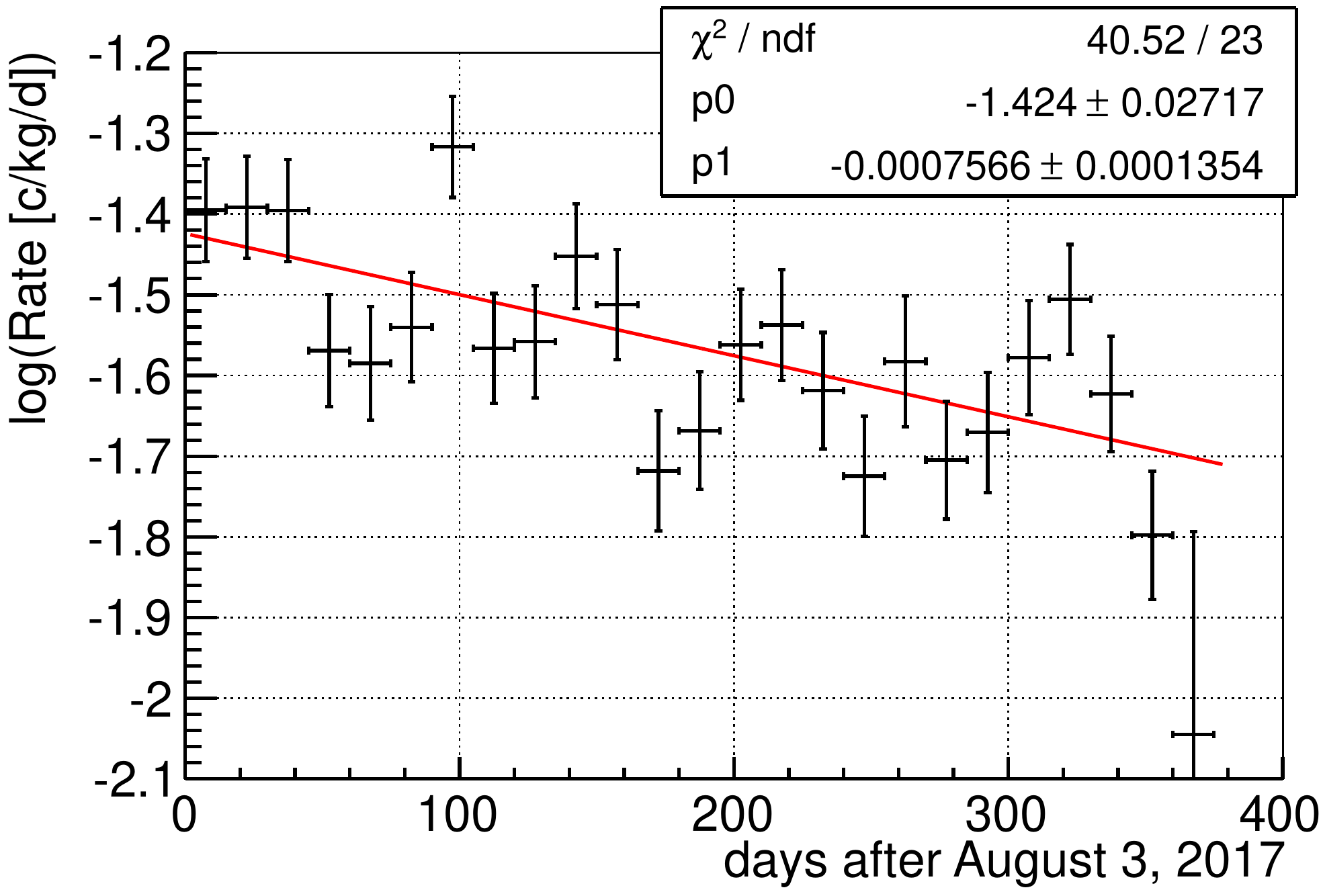}
\caption{Evolution of the natural logarithm of the rate of coincidences with a high energy gamma between 1215 and 1335~keV 
in a second module, after filtering (PSV only) and efficiency correction in the energy range below 2~keV
corresponding to events attributed to $^{22}$Na decays in the crystal bulk (T$_{1/2}$=2.6~y). Rates are averaged every 10~days and 
fitted to a straight line (red). The events rate is decreasing as expected with a half life of 2.5$\pm$0.4~y.}
\label{fig:NaRate}
\end{figure}

Figure~\ref{fig:KRate} shows the evolution of the rate of coincidences with 
high energy gammas in a second module attributed to the decay of $^{40}$K in the NaI 
bulk averaged every 10~days. As in Figure~\ref{fig:NaK}, we apply the PSV cut and correct with the corresponding efficiency.
The rate has been fitted to a constant rate (in blue) and to a straight line (in red). 
In both cases compatibility with constant rate is obtained. 
We have performed the same study with the $^{22}$Na population selected by coincidence, which rate is expected to decrease
with a half-life of 2.6~y.  Figure~\ref{fig:NaRate} shows the evolution of the natural logarithm of the $^{22}$Na events selected by the 
coincidence requirement, with the same filtering as the $^{40}$K population. We fit to a straight line, obtaining a decay
constant of (0.00076$\pm$0.00014)~d$^{-1}$ corresponding to a half life of 2.5$\pm$0.4~y. 
These populations of events in the ROI would allow in the long term to check that 
the efficiencies of our filtering are stable or they show some modulation. 

\begin{figure*}[b]
\centering
\includegraphics[width=1\textwidth]{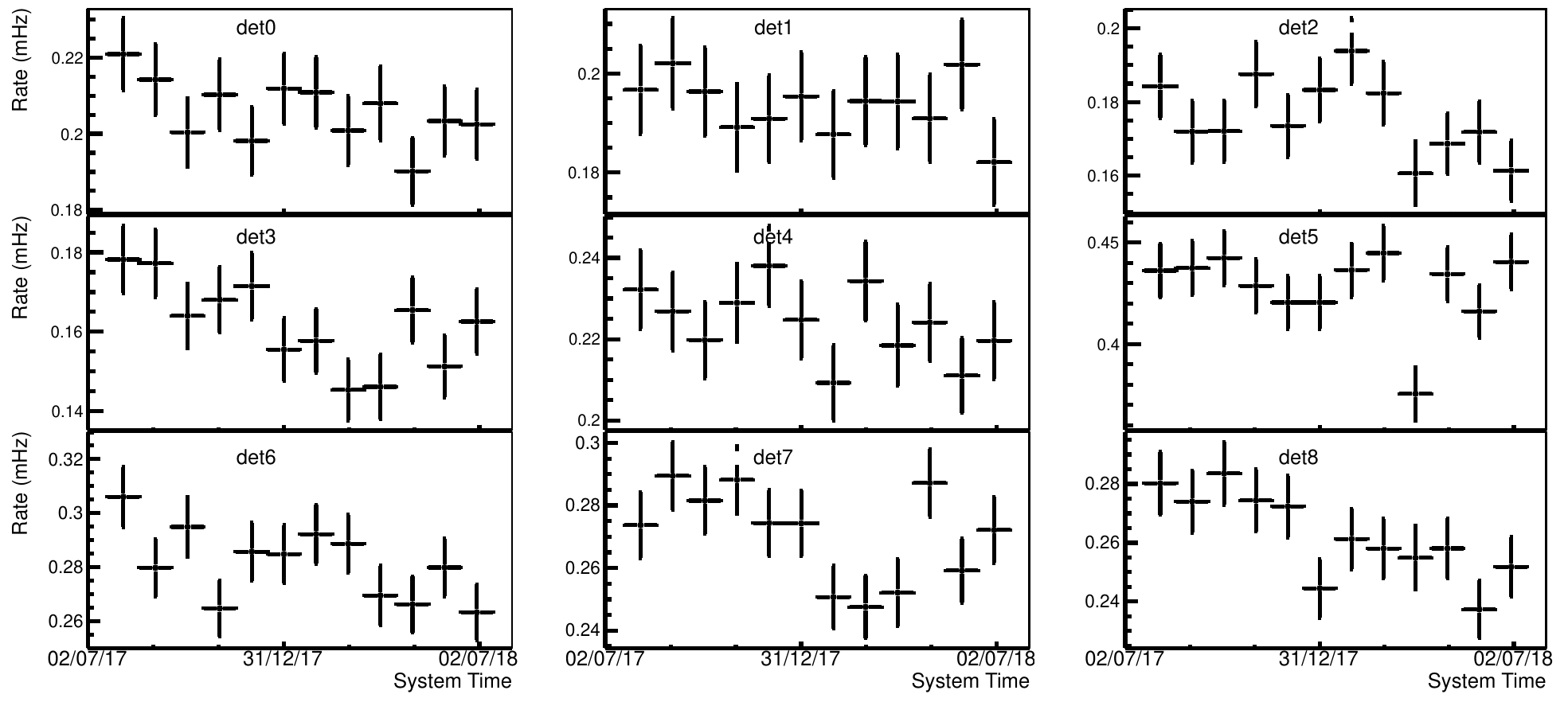}
\caption{Evolution of the rate of events producing an energy deposition below 10~keV in the detector studied and passing the filtering procedure described in Section \ref{sec:eventSel}, which are in coincidence with any energy deposition in a second ANAIS-112 module. Coincident events are those triggering the DAQ system in 1 $\mu$s window after the first trigger. Rates are calculated per month. See text for more information.
}
\label{fig:coinrate}
\end{figure*}

In Figure~\ref{fig:coinrate} we show the rate of events in every module which have an energy deposition below 10~keV, pass the filtering procedure explained in Section~\ref{sec:eventSel}, and are in coincidence with any energy deposition in a second ANAIS-112 module. Coincident events are defined as those triggering the ANAIS-112 DAQ system in 1 $\mu$s window after the first trigger (see Section~\ref{sec:DAQ}).

\begin{figure*}[htbp]
\centering
\includegraphics[width=0.95\textwidth]{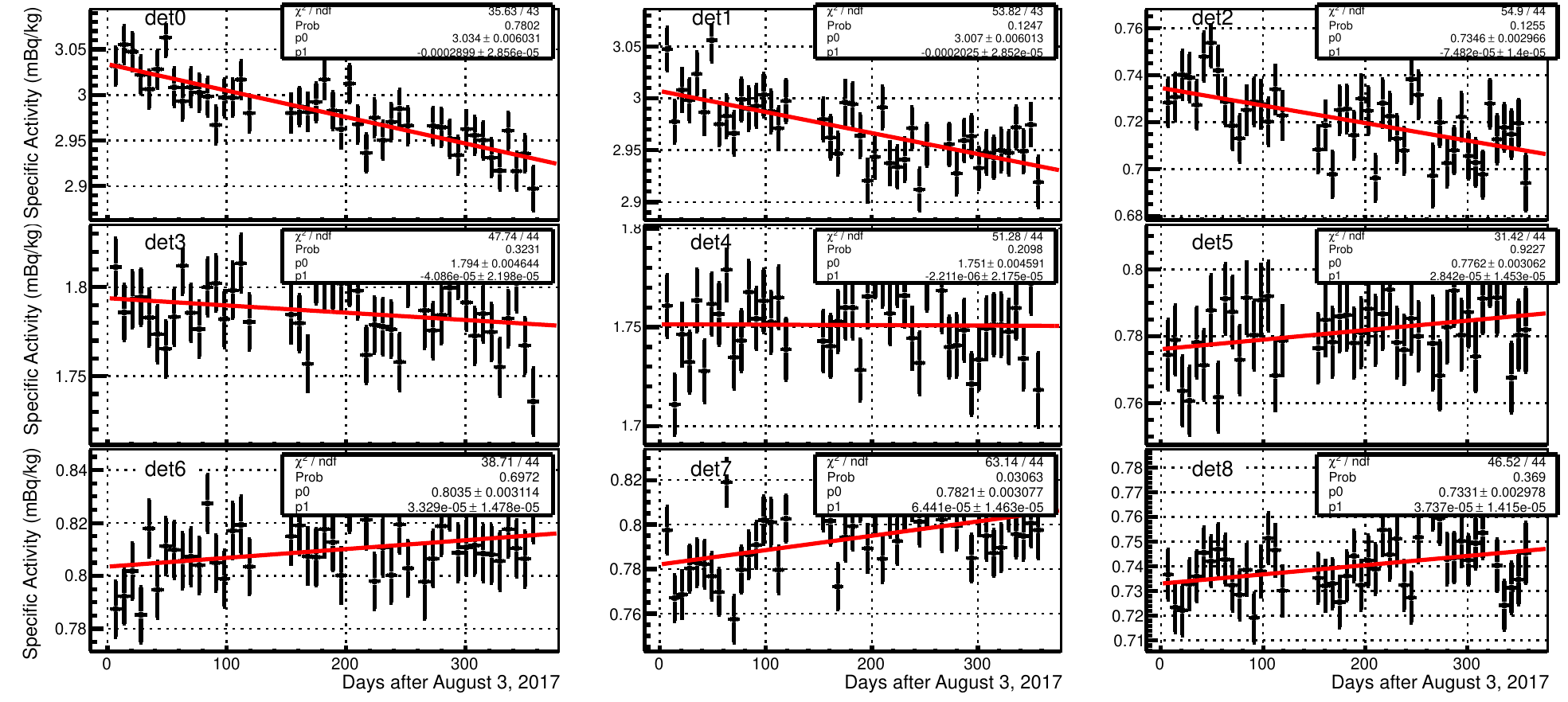}
\caption{Evolution of the specific activity of alpha events attributed mainly to $^{210}$Po decay in the crystal bulk. Alpha events are selected by PSA (see section~\ref{sec:hecal} and Figure~\ref{fig:heLin}). Rates are averaged per week and linearly fitted. It can be observed that modules arriving first at Canfranc (D0, D1, D2 and D3) show a decaying alpha rate because equilibrium in the chain was reached and $^{210}$Pb decay time constant rules the $^{210}$Po decay; for modules arriving later at Canfranc, rate of alpha events is either constant or slightly increasing because $^{210}$Po activity is still being built.  
}
\label{fig:alfas1}
\end{figure*}

Figure~\ref{fig:alfas1} shows the evolution of the alpha specific activity derived from the rate of alpha events identified by PSA in every module along the first year of ANAIS-112. It can be observed that in D0 and D1 modules (which arrived in 2012 at the LSC), this activity is clearly decreasing with the expected $^{210}$Pb progenitor half-life, whereas in the case of the rest of modules the alpha rate is either stable or still slightly increasing, as $^{210}$Po activity is still being built before reaching equilibrium in the decay chain. 

\begin{figure*}[htbp]
\centering
\includegraphics[width=0.95\textwidth]{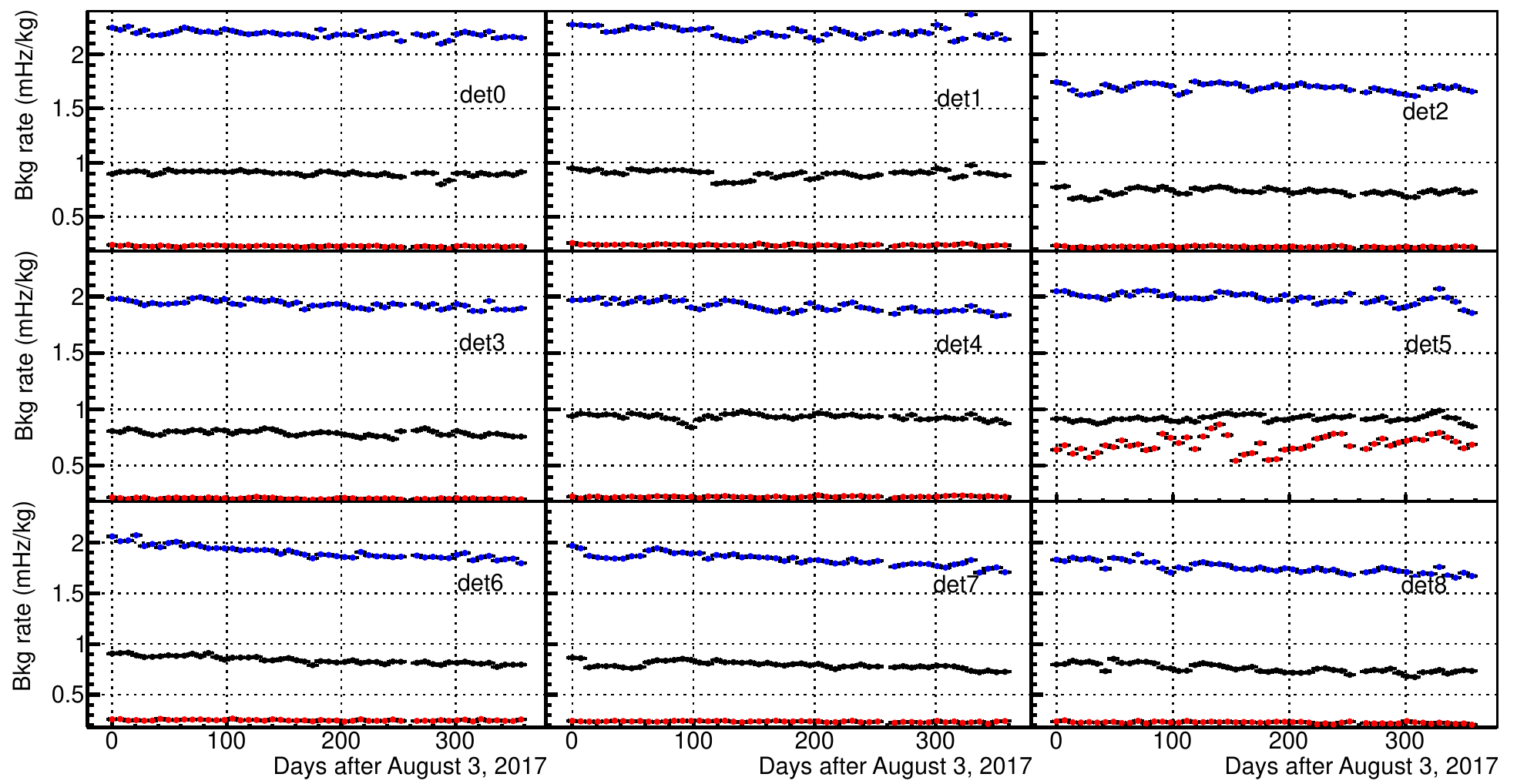}
\caption{Evolution of the rate of events attributed to interactions of background gammas in the crystal bulk in three different energy windows (in black around the 609.3~keV line (from 559 to 659~keV), in red around the 1460.8~keV line (from 1410 to 1510~keV) and in blue in the region from 300 to 450~keV). Rates are averaged per week. Colors referenced are available in the online version of the paper. 
}
\label{fig:bkg_evol}
\end{figure*}

Figure~\ref{fig:bkg_evol} shows the evolution of the rate of events attributed to gamma background in three different energy windows. It has to be commented that D6, D7 and D8 have still cosmogenic activated isotopes decaying in the crystal bulk, because they arrived later at LSC. This can be clearly observed in Figure~\ref{fig:bkg_evol}.

\begin{figure}[b]
\centering
\includegraphics[width=0.5\textwidth]{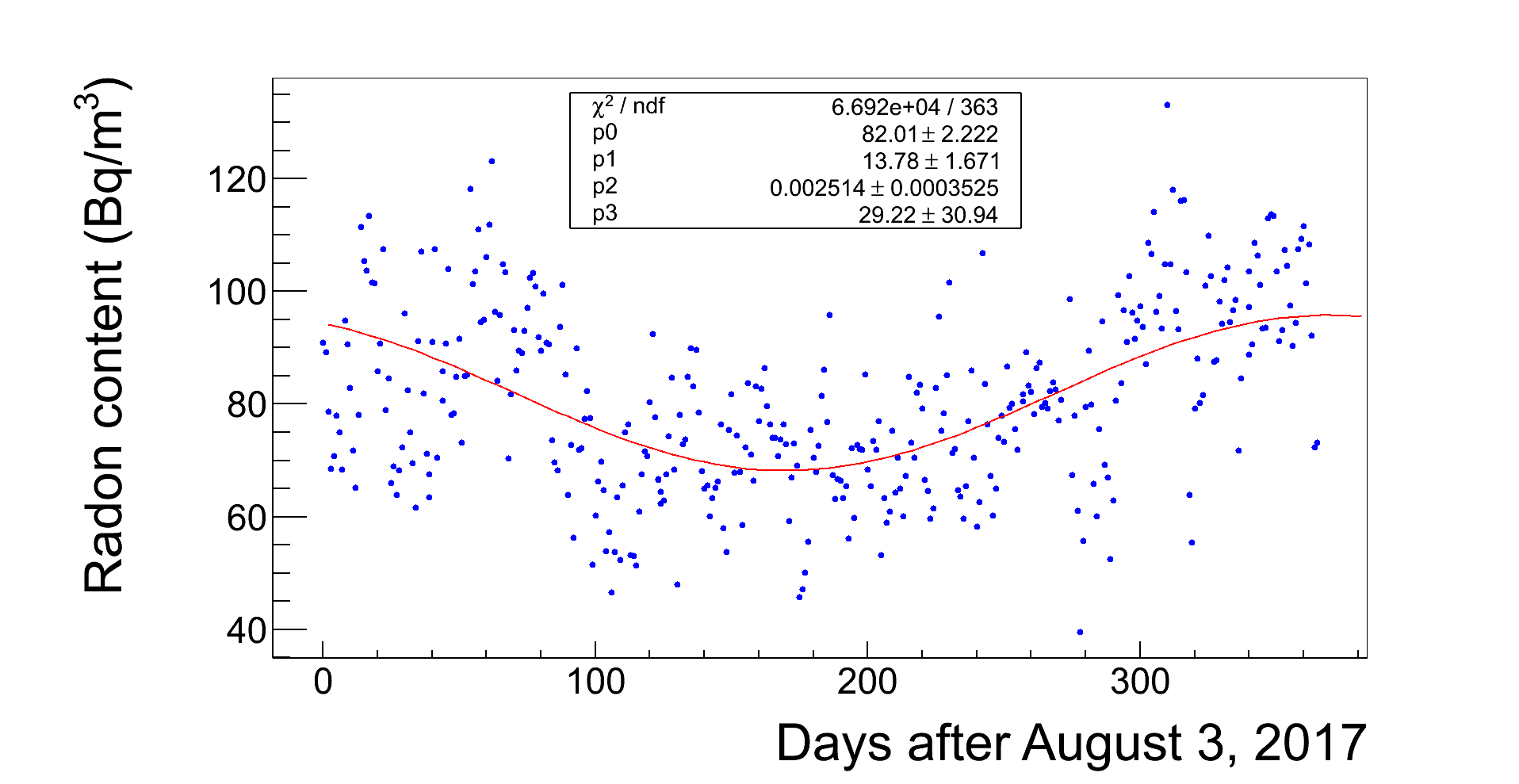}
\caption{Evolution of the radon content in the LSC hall B air in the ANAIS hut outside the shielding. Rates are calculated per day and fitted to a cosinoidal dependence.
}
\label{fig:Rnevol_fit}
\end{figure}

\begin{figure}[htbp]
\centering
\includegraphics[width=0.5\textwidth]{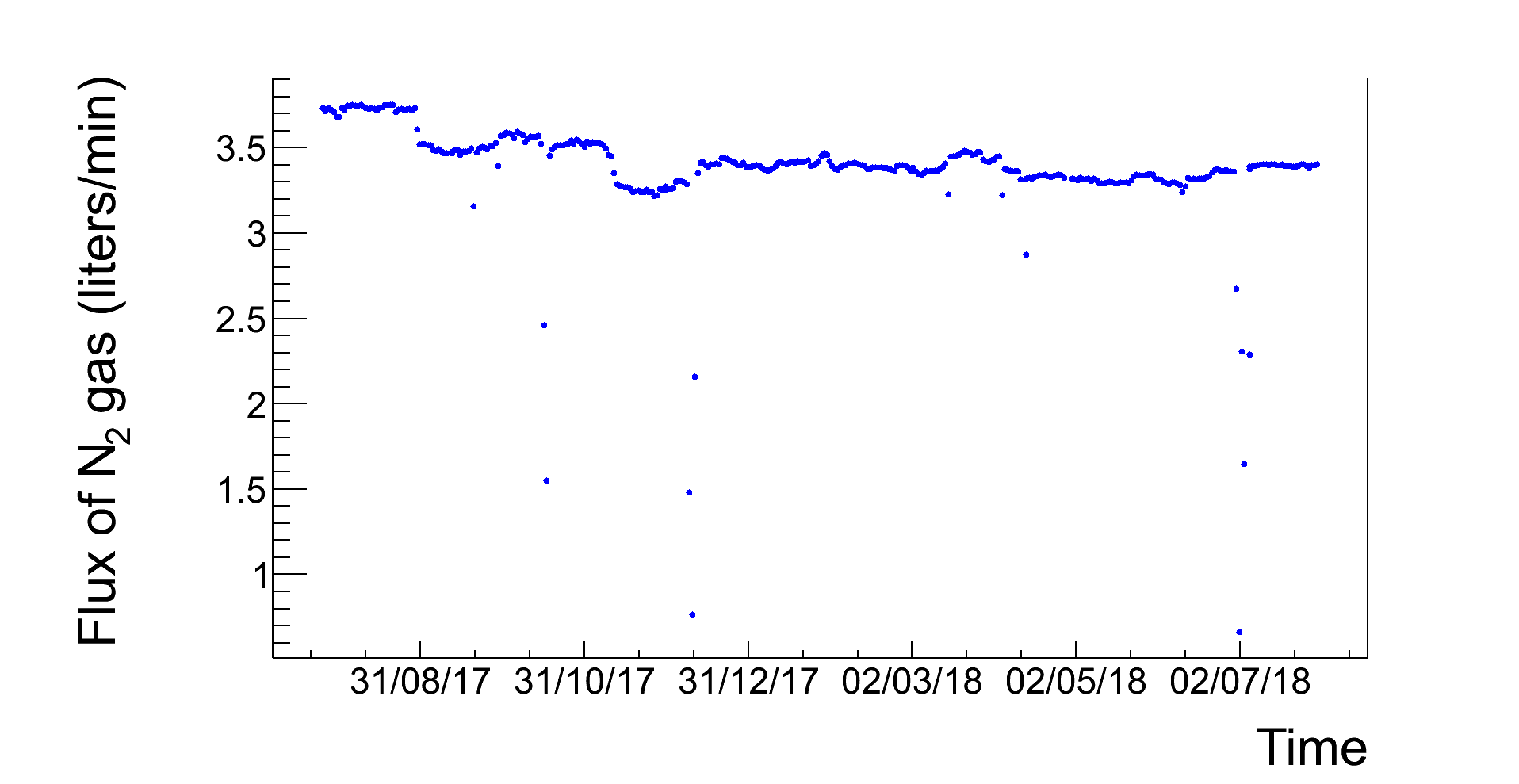}
\caption{Evolution of the nitrogen gas flux entering into the ANAIS-112 shielding to prevent radon intrusion, averaged per day.
}
\label{fig:N2Flux_evol}
\end{figure}

\begin{figure}[htbp]
\centering
\includegraphics[width=0.5\textwidth]{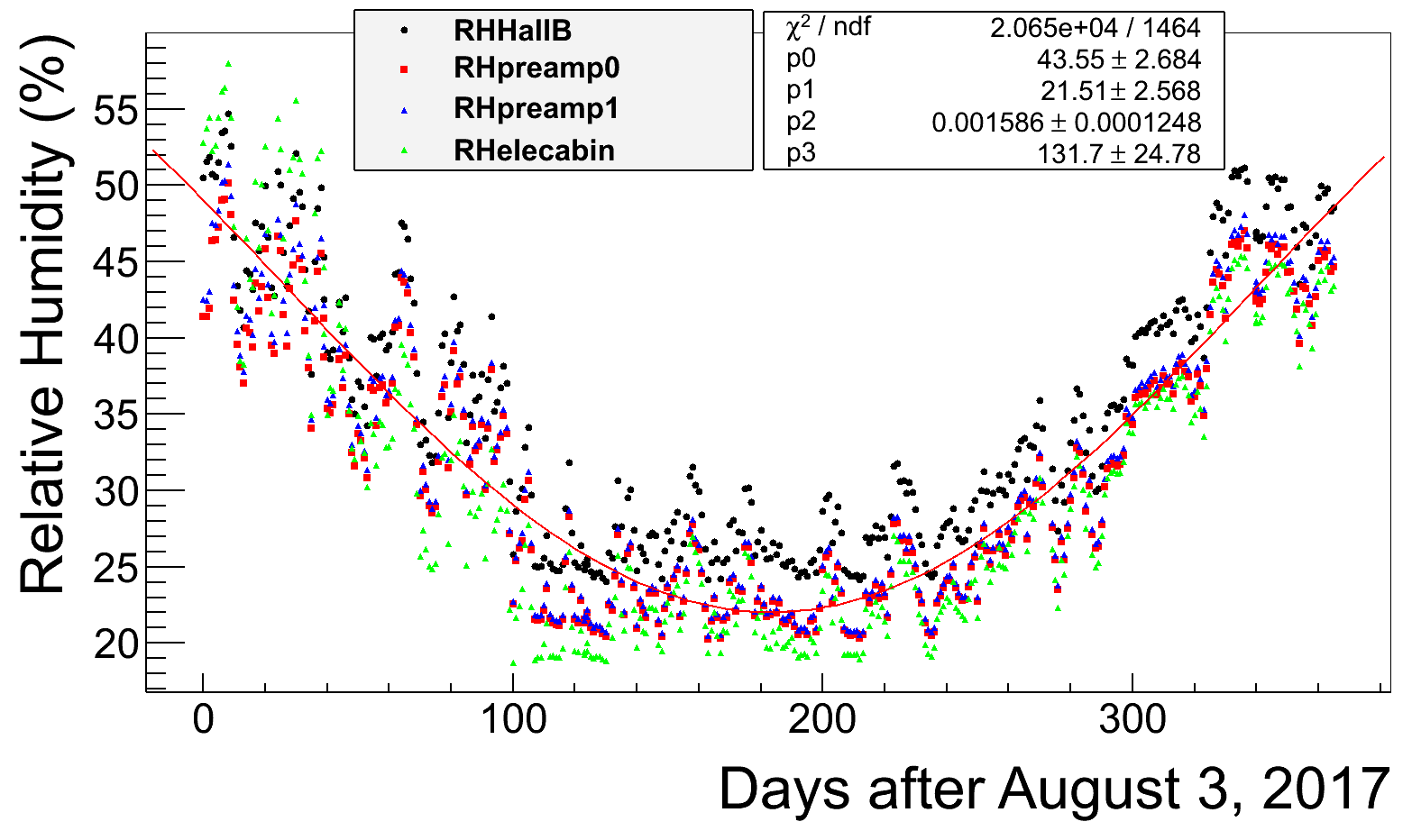}
\caption{Evolution of the Relative Humidity inside the ANAIS hut (RHHallB), at the preamplifiers positions (RHpreamp0, RHpreamp1) and at the ANAIS-112 electronics air-conditioned room (RHelecabin). All of them have been averaged per day and fitted to a cosinoidal dependence. Colors referenced are available in the online version of the paper. 
}
\label{fig:RH_evol_fit}
\end{figure}

Monitoring of environmental parameters has been continuously ongoing along ANAIS-112 dark matter run data taking: N$_2$ gas flux entering into the shielding; temperatures at electronics, inner shielding, 
ANAIS hut, preamplifiers, etc.; radon content in laboratory air (inside ANAIS hut); 
relative humidity in several places; HV supply for every PMT; electronics voltage supply; etc. 
All the data are saved every few minutes and alarms have been set on the most relevant parameters 
sending an alarm message through Telegram. Very few incidents have set on the ANAIS-112 alarms along the 
first year of data taking, only the radon-free nitrogen gas flux entering into the shielding which has been 
reduced down to zero three times in the data taking because of exhausting the liquid nitrogen dewar, 
as already explained in Section~\ref{sec:experimental}. 
Figure~\ref{fig:Rnevol_fit} shows the evolution of the radon content in the 
air of the ANAIS hut (but outside the shielding) along the first year of data taking. 
It has to be remarked that radon content inside the shielding is expected to be about 
two orders of magnitude lower than that thanks to the radon-free nitrogen gas overpressure. 
We show in Figure~\ref{fig:N2Flux_evol} 
the time behaviour of the nitrogen gas flux entering into the ANAIS-112 shielding and 
in \ref{fig:RH_evol_fit} 
the relative humidity in different positions (inside the ANAIS hut and at the electronics air conditioned space). 
There is a clear correlation between relative humidity and the radon content in the lab air
presented in Figure~\ref{fig:Rnevol_fit}. 
For both a fit of the data to a function \( f(t)=p_0+p_1 \cdot \cos(2\pi p_2(t+p_3)) \) is performed, showing clear differences in the best fit 
for the period of the oscillation (400 $\pm$ 56~d vs 630 $\pm$ 50~d). % and the phase (29 $\pm$ 30 vs 132 $\pm$ 25). 
Further information on this correlation will be obtained by accumulating more years of data.  

\begin{figure}[htbp]
\centering
\includegraphics[width=0.5\textwidth]{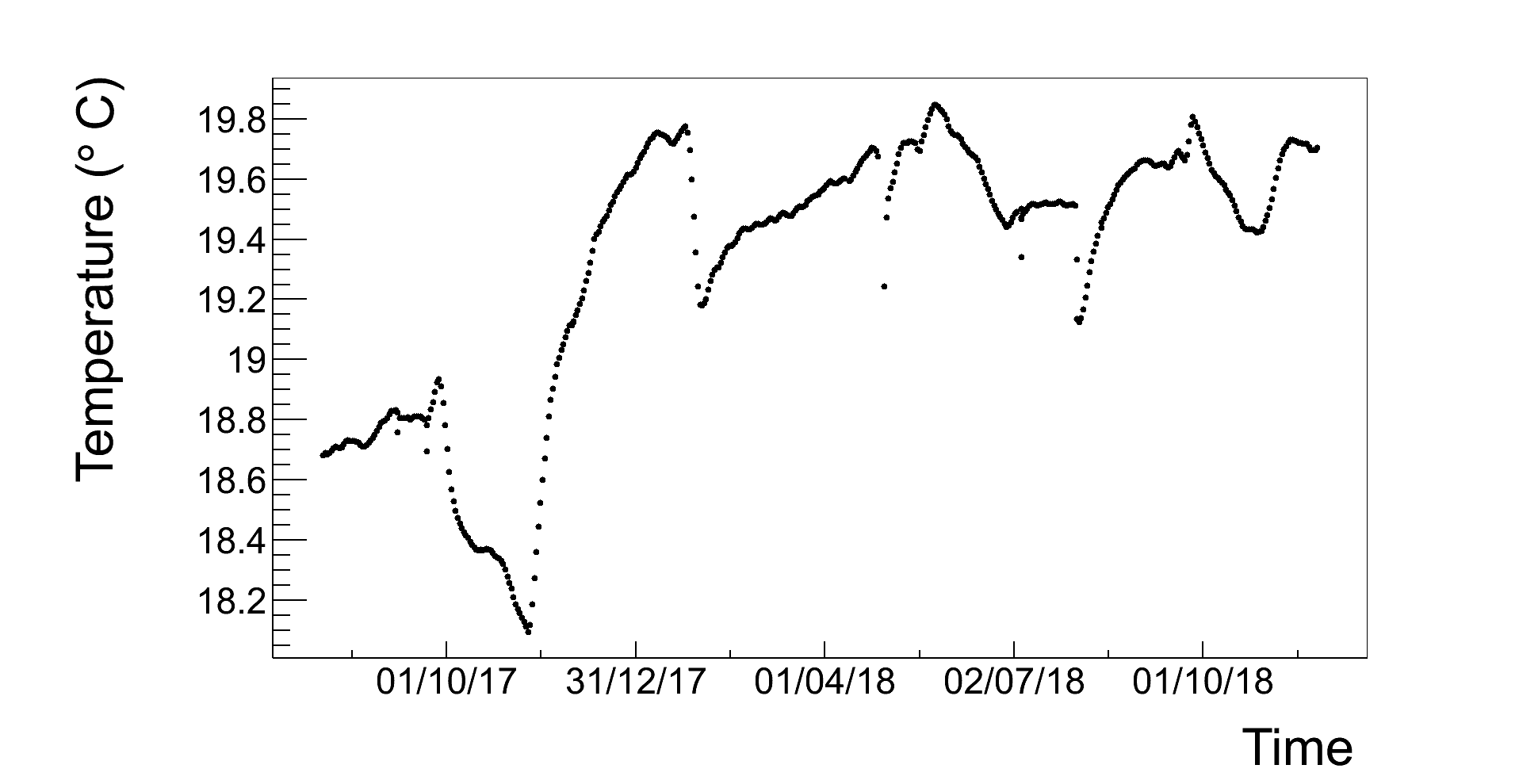}
\caption{Evolution of the temperature inside the ANAIS-112 shielding (Tint), averaged per day. Tint follows the external conditions at LSC hall B. In the first year of operation, it varies from 18.09 to 19.85$^\circ$C, having a mean value of 19.24$^\circ$C and standard deviation of 0.48$^\circ$C. LSC temperature regulation is based on sensors placed in hall A, and then, temperature at hall B is affected, for instance by the operation of ArDM experiment at hall A, which introduces an additional heat source and forces the system to cool down, which is observed in the figure in November 2017. Efforts of the LSC staff to improve the hall B temperature control are ongoing, and along 2018 variations in temperature at hall B have been much smaller. No correlation among Tint and ANAIS-112 trigger rate has been observed along the first year of operation.
}
\label{fig:Tevol1}
\end{figure}

\begin{figure}[htbp]
\centering
\includegraphics[width=0.5\textwidth]{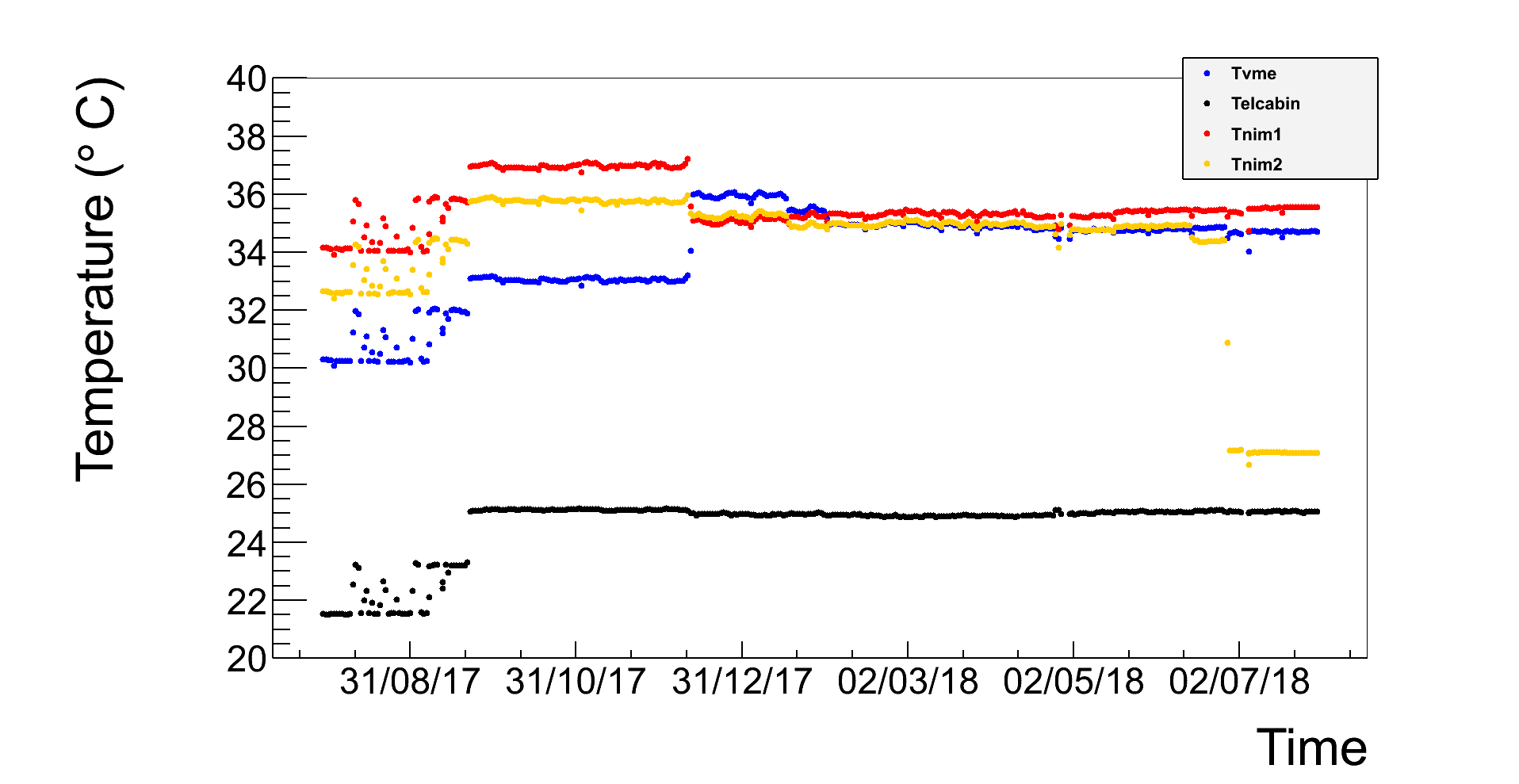}
\caption{Evolution of the temperatures inside the ANAIS-112 electronics air-conditioned room (Telcabin), and at different positions of the electronics racks (Tvme, Tnim1, Tnim2). All of them have been averaged per day. Colors referenced are available in the online version of the paper. 
}
\label{fig:Tevol3}
\end{figure}

Figures~\ref{fig:Tevol1} and \ref{fig:Tevol3} show the evolution in time along the first year of 
ANAIS-112 data taking of the temperature inside the ANAIS-112 shielding (Tint), the temperature at the 
electronics air-conditioned space (Telcabin), and the temperatures at VME and NIM racks placed inside that 
conditioned space (Tvme, Tnim1, Tnim2). The temperature at the detectors position, 
Tint, follows the external conditions at the LSC hall B. It varies from 18.09 to 19.85$^\circ$C, 
having a mean value of 19.24$^\circ$C and standard deviation of 0.48$^\circ$C. 
The temperature of the preamplifiers follows exactly the same trend, while on the other hand, 
temperature at the VME and NIM electronics racks is fully decoupled from hall B temperature. The latter has been quite stable along the first year of data taking but for sudden changes (see Fig.~\ref{fig:Tevol3}), which are due to adjustments in the configuration of the temperature control and accidental shift of the temperature sensors while maintenance works.

\section{Summary}
\label{sec:summary}

ANAIS-112 experiment has been taking data at the Canfranc Undeground Laboratory for more than one year, and the experiment goal is to accumulate five years of data in stable conditions in order to test the DAMA/LIBRA modulation signal in a model independent way. 
An excellent duty cycle has been achieved, accumulating 341.72 days of data with the nine NaI(Tl) modules (12.5 kg each). 
ANAIS-112 modules have shown a very high light collection per unit of energy deposited, about 15~phe/keV in 8 out of 9 modules. In this article, the performance of the ANAIS-112 experiment along the first year of data taking has been reviewed, as well as the analysis techniques developed for low energy calibration, in particular, selection of events corresponding to interactions attributable to dark matter, and the corresponding efficiencies have been thoroughly described and estimated. Background events populations in the ROI, from {\K} and {\Na} decays in the crystal bulk, are well tagged and used to control any modulation in the operation of the detectors that could mimic the signal searched for. Environmental parameters (muon flux, radon content in the laboratory air, temperature at the detectors positions and electronics, etc.) are being continuously monitored. 
ANAIS-112 will be able to test DAMA/LIBRA signal at 3$\sigma$ level in five years of data taking in the 
achieved experimental conditions, having a large discovery potential if dark matter particles are 
responsible for such signal. Estimates of the ANAIS-112 sensitivity prospects have been 
updated~\cite{Coarasa:2018qzs}
with the background level corresponding to the unblinded data and the efficiencies reported in this work.

\section*{Acknowledgments}

This work has been financially supported by the Spanish Ministerio de Econom{\'\i}a y Competitividad and the European Regional Development Fund (MINECO-FEDER) under grants No. FPA2014-55986-P and FPA2017-83133-P, the Consolider-Ingenio 2010 Programme under grants MultiDark CSD2009-00064 and CPAN CSD2007-00042 and the Gobierno de Arag{\'o}n and the European Social Fund (Group in Nuclear and Astroparticle Physics and I.~Coarasa predoctoral grant). 
We thank the support of the Spanish Red Consolider MultiDark FPA2017-90566-REDC. 
We acknowledge the technical support from LSC and GIFNA staff.
We also would like to acknowledge the use of Servicio General de Apoyo a la Investigaci{\'o}n-SAI, Universidad de Zaragoza.
Professor J.A. Villar passed away in August, 2017. Deeply in sorrow, we all thank
his dedicated work and kindness.
\bigskip

\bibliographystyle{spphys}

\end{document}